\documentclass[3p]{elsarticle}
\usepackage[hidelinks]{hyperref}

\usepackage[utf8]{inputenc}

\usepackage[table]{xcolor}
\hypersetup{
    colorlinks,
    linkcolor={red!50!black},
    citecolor={blue!50!black},
    urlcolor={blue!80!black}
}

\usepackage{amsmath,amssymb}
\usepackage{savesym}
\savesymbol{amalg}
\restoresymbol{ABXF}{amalg}
\usepackage{mathtools}
\usepackage[electronic]{ifsym}
\usepackage{bm}
\usepackage{longtable}
\usepackage{stackrel}
\usepackage[center,bf,tight]{subfigure}
\usepackage[textwidth=4cm,textsize=footnotesize,disable]{todonotes}
\usepackage{tikz}
\usetikzlibrary{decorations.pathmorphing}
\usetikzlibrary{decorations.pathreplacing}

\usepackage{stmaryrd}

\usepackage{enumitem}
\usepackage{amsthm}

\usepackage{soul}
\usepackage{cancel}

\usepackage{lineno}

\usepackage{multirow}

\newtheorem{proposition}{Proposition}
\newtheorem{lemma}{Lemma}
\newtheorem{theorem}{Theorem}
\newtheorem{corollary}{Corollary}
\theoremstyle{definition}
\newtheorem{remark}{Remark}
\newtheorem{definition}{Definition}

\theoremstyle{plain}
\newtheorem{claim}{Claim}
\newtheorem*{claim*}{Claim}

\newcommand{\Nat}{\mathbb{N}_{> 0}}
\newcommand{\Natgez}{\mathbb{N}_{\geq 0}}

\newcommand{\Real}{\mathbb{R}}
\newcommand{\Realgez}{\mathbb{R}_{\geq 0}}

\newcommand{\pair}[2]{(#1, #2)}

\newcommand{\A}{\mathcal{A}}

\newcommand{\x}{\mathbf{x}}

\makeatletter
\def\Eqlfill@{\arrowfill@\Relbar\Relbar\Relbar}
\newcommand{\longmodels}[1][]{\,|\!\!\!\ext@arrow 0359\Eqlfill@{#1}}
\makeatother

\newcommand{\set}[1]{\{ #1 \}}

\newcommand{\todoMR}[2]{\todo[color=red!50,#1]{\textbf{MR:}#2}}

\newcommand{\todoPL}[2]{\todo[color=yellow!50,#1]{\textbf{PL:}#2}}

\def\NOTRaisingEdge{{%
    \setbox0\hbox{\RaisingEdge}%
    \rlap{\hbox to \wd0{\hss/\hss}}\box0
}}

\def\NOTFallingEdge{{%
    \setbox0\hbox{\FallingEdge}%
    \rlap{\hbox to \wd0{\hss/\hss}}\box0
}}

\def\NOTShortPulseHigh{{%
    \setbox0\hbox{\ShortPulseHigh}%
    \rlap{\hbox to \wd0{\hss/\hss}}\box0
}}

\def\NOTShortPulseLow{{%
    \setbox0\hbox{\ShortPulseLow}%
    \rlap{\hbox to \wd0{\hss/\hss}}\box0
}}

  \DeclareFontFamily{U}  {MnSymbolA}{}
  \DeclareSymbolFont{MnSyA}         {U}  {MnSymbolA}{m}{n}
  \SetSymbolFont{MnSyA}       {bold}{U}  {MnSymbolA}{b}{n}
\DeclareFontShape{U}{MnSymbolA}{m}{n}{
    <-6>  MnSymbolA5
   <6-7>  MnSymbolA6
   <7-8>  MnSymbolA7
   <8-9>  MnSymbolA8
   <9-10> MnSymbolA9
  <10-12> MnSymbolA10
  <12->   MnSymbolA12}{}
\DeclareFontShape{U}{MnSymbolA}{b}{n}{
    <-6>  MnSymbolA-Bold5
   <6-7>  MnSymbolA-Bold6
   <7-8>  MnSymbolA-Bold7
   <8-9>  MnSymbolA-Bold8
   <9-10> MnSymbolA-Bold9
  <10-12> MnSymbolA-Bold10
  <12->   MnSymbolA-Bold12}{}
  \DeclareMathSymbol{\leftfilledspoon}{\mathrel}{MnSyA}{114}
  \DeclareMathSymbol{\leftspoon}{\mathrel}{MnSyA}{106}
  \DeclareMathSymbol{\upfilledspoon}{\mathrel}{MnSyA}{113}
  \DeclareMathSymbol{\upspoon}{\mathrel}{MnSyA}{105}
  \DeclareMathSymbol{\leftfree}{\mathrel}{MnSyA}{130}

  \DeclareFontFamily{U}  {MnSymbolB}{}
  \DeclareSymbolFont{MnSyB}         {U}  {MnSymbolB}{m}{n}
  \SetSymbolFont{MnSyB}       {bold}{U}  {MnSymbolB}{b}{n}
\DeclareFontShape{U}{MnSymbolB}{m}{n}{
    <-6>  MnSymbolB5
   <6-7>  MnSymbolB6
   <7-8>  MnSymbolB7
   <8-9>  MnSymbolB8
   <9-10> MnSymbolB9
  <10-12> MnSymbolB10
  <12->   MnSymbolB12}{}
\DeclareFontShape{U}{MnSymbolB}{b}{n}{
    <-6>  MnSymbolB-Bold5
   <6-7>  MnSymbolB-Bold6
   <7-8>  MnSymbolB-Bold7
   <8-9>  MnSymbolB-Bold8
   <9-10> MnSymbolB-Bold9
  <10-12> MnSymbolB-Bold10
  <12->   MnSymbolB-Bold12}{}
  \DeclareMathSymbol{\nleftfilledspoon}{\mathrel}{MnSyB}{114}

\DeclareRobustCommand{\shortf}[3] %
{\ensuremath{%

	 \ifthenelse{\not \equal{#3}{} }  {\stackrel{#2}{#1}_{#3}}  {\stackrel{#2}{#1}}
  }%
}%

\DeclareRobustCommand{\LogOp}[3] %
{\ensuremath{%

	 \ifthenelse{\not \equal{#2}{}}  {\mathbf{#1}_{#2}}  {\mathbf{#1}}

	 \ifthenelse{\not \equal{#3}{}}  {\!\left({#3}\right)} {}

  }%
}%

\DeclareRobustCommand{\LogOpPast}[3] %
{\ensuremath{%

	 \ifthenelse{\not \equal{#2}{}}  {\overleftarrow{\mathbf{#1}}_{#2}}  {\mathbf{#1}}

	 \ifthenelse{\not \equal{#3}{}}  {\!\left({#3}\right)} {}

  }%
}%

\DeclareRobustCommand{\LogOpInfix}[4] %
{\ensuremath{%

	 \ifthenelse{\not \equal{#3}{}}  {#3}  {}

	 \ifthenelse{\not \equal{#2}{}}  {\mathbf{#1}_{#2}}  {\mathbf{#1}}

	 \ifthenelse{\not \equal{#4}{}}  {#4} {}

  }%
}%

\newcommand{\fract}[1]{\mathit{frac(#1)}}

\newcommand{\cmax}{C}
\newcommand{\const}{c}
\newcommand{\param}{\mu}
\newcommand{\intprm}{m}
\newcommand{\floorprm}{n}
\newcommand{\halffrac}{\frac{\floorprm}{2}}
\newcommand{\cnf}{\eta}

\newcommand{\ZR}{{\underline{0}}}
\newcommand{\ZA}{{\underline{0\ell}}}
\newcommand{\ZB}{{\underline{\ell}}}
\newcommand{\ZC}{{\underline{\ell\widehat{\ell}}}}
\newcommand{\ZD}{{\underline{\widehat{\ell}}}}
\newcommand{\ZE}{{\underline{\widehat{\ell}1}}}
\newcommand{\ZZ}{{\underline{\iota}}}

\newcommand{\onerst}{one-reset}
\newcommand{\crtonerst}{critical}

\newtagform{Alph}[]()
\newtagform{roman}[]()
\newtagform{scroman}[][]

\newcommand{\orssym}{{\xi}}

\newcommand{\runsym}{{\rho}}

\newcommand{\Low}{Dn}
\newcommand{\LowK}[1]{\ensuremath{\Low}(#1)}

\newcommand{\leftend}[1]{\textit{left}(#1)}
\newcommand{\rightend}[1]{\textit{right}(#1)}
\newcommand{\SetZZ}{\ensuremath{S_\ZZ}}
\newcommand{\ellgenpos}{\ensuremath{\chi^+}}
\newcommand{\ellgenneg}{\ensuremath{\chi^-}}
\newcommand{\widiota}{\ensuremath{w_{\SetZZ}}}

\newcommand{\crtind}[1]{\ensuremath{c(#1)}}

\begin{document}

\begin{frontmatter}

\title{On Decidability Timed Automata with 2 Parametric Clocks}

\author[pdm]{Marcello M. Bersani}
\ead{marcellomaria.bersani@polimi.it}
\author[pdm]{Matteo Rossi}
\ead{matteo.rossi@polimi.it}
\author[pdm,cnr]{Pierluigi {San Pietro}}
\ead{pierluigi.sanpietro@polimi.it}

\address[pdm]{Dipartimento di Elettronica Informazione e Bioingegneria, Politecnico di Milano,\\Piazza Leonardo da Vinci 32, Milano, Italy}
\address[cnr]{CNR IEIIT-MI, Milano, Italy}

\begin{abstract}
In this paper, we introduce a restriction of Timed Automata (TA), called non-resetting test Timed Automata (nrtTA). 
An nrtTA does not allow to test and reset the same clock on the same transition. The model has the same expressive power of TA, but it may require one more clock than an  TA to recognize the same language. 
We consider the parametric version of nrtTA, where one parameter can appear in clock guards of transitions. 
The focus of this draft is to prove that the $\omega$-language emptiness problem for 2-clock parametric nrtTA is decidable. 
This result can be compared with the parametric version of TA, where the emptiness problem for 2-clock TA with one parameter is not known to be decidable. Our result, however, extends the known decidability of the case of TA with one clock and one parameter from finite words to infinite words.
\end{abstract}


\end{frontmatter}

\section{Introduction}\label{section-intro}

In this draft paper, we introduce non-resetting test Timed Automata (nrtTA, namely TA with the additional constraint that the same clock cannot be tested and reset in the same transition. This family is as expressive as the family of ``traditional'' TA, since, as shown in Section~\ref{section-nrtTA}, any TA with $k>0$ clocks can be simulated by a nrtTA with $k+1$ clocks. However, a nrtTA with $k+1$ clocks is more expressive than a TA with $k$ clocks.

Here we focus on parametric nrtTA, where one parameter, denoted by $\param$ and whose value is not determined a priori, can be used in clock guards, e.g., with constraints of the form $x<\param$, $x=\param$, etc., where $x$ is a clock. If the parameter occurs in a guard, then no constant can appear in the same guard, i.e., guards such as $\param>2$ or $x<\param+1$ are not allowed. 
Parametric TA is a widely used formalism (see~\cite{DBLP:journals/sttt/Andre19} for a thorough survey.)

Our main result is that $\omega$-language emptiness is decidable for nrtTA with two clocks and one parameter. This extends the known fact that the emptiness of 1-clock-1-parameter TA is decidable in two directions; first, 2-clock-1-parameter nrtTA are more expressive than 1-clock-1-parameter TA; second, decidability of the latter was only proven over finite languages.

This draft is organized as follows: Section~\ref{section-languages} shortly summarizes the definition of TA; Section~\ref{section-nrtTA} introduces the nrtTA model; Section~\ref{section-decTA} proves the main result.

\section{Timed Automata}\label{section-languages}

In this section, we recall the basic definitions of Timed Automata. %

Le $\Sigma$ be a finite alphabet.
A {\em timed $\omega$-word} (sometimes called simply \emph{timed word}) over $\Sigma$ is a pair $\pair{\pi}{\tau}$ where $\pi: \Nat\to \Sigma$
and the {\em timed sequence} $\tau$ is a monotonic function $\tau: \Nat \to \Realgez$ such that, for all $i>0$, $\tau(i)<\tau(i+1)$ holds (strong monotonicity).
The value $\tau(i)$ is called the \emph{timestamp} at position $i$, $i\in\Nat$. 

Let $X$ be a finite set of clocks with values in $\Realgez$.
 $\Gamma(X)$ is the set of clock constraints $\gamma$ over $X$ defined by the syntax
 $\gamma := x \sim \const \mid
 \neg\gamma \mid \gamma\wedge\gamma$,
 where $\sim \in \set{<,=}$, $x,y \in X$ and $\const \in \Natgez$.
 A clock valuation
 is a function $v:X\to\Realgez$.
 We write $v \models \gamma$ when the clock valuation satisfies $\gamma$.
 For $t \in \Realgez$, $v + t$ denotes the clock valuation mapping each clock $x$ to value $v(x)+t$---i.e., $(v + t) (x) = v(x)+t$ for all $x \in X$.
 
A \emph{Timed Automaton} \cite{Alur&Dill94} is a tuple $\A=(\Sigma, Q, T, q_0, B)$ where
 $Q$ is a finite set of control states,
 $q_0 \in Q$ is the initial state, 
 $B\subseteq Q$ is a subset of control states (corresponding to a B\"uchi condition) and
 $T \subseteq Q \times Q \times \Gamma(X)\times \Sigma \times 2^{X}$
  is a set of transitions.
 Thus, a transition has the form 
 $q\xrightarrow{\gamma,a,S} q'$ where $q,q'\in Q$, 
 $\gamma$ is a clock constraint of $\Gamma(X)$, $a\in \Sigma$, %
 and $S$ is a set of clocks to be reset.
 Two transitions $q\xrightarrow{\gamma,a,S} q'$
 and $p \xrightarrow{\gamma',b,P} p'$
 of $T$ 
 are \emph{consecutive} when $q' = p$.
 A pair $(q, v)$, where $q\in Q$ and $v:X\to\Realgez$ is a clock valuation, is a \emph{configuration} of $\A$.
 A \emph{run} $\rho$ of $\A$ over a timed $\omega$-word $(\pi,\tau) \in (\Sigma \times \Realgez)^\omega$ is an infinite sequence of configurations $(q_{i_0}, v_0) \xrightarrow[\tau(1)]{\pi(1)} (q_{i_1}, v_1) \xrightarrow[\tau(2)]{\pi(2)} (q_{i_2}, v_2)\dots,$
 satisfying the following three constraints: 
\begin{itemize}
	\item $q_{i_0}=q_0$;
	\item $q_{i_0}\xrightarrow{\gamma_1,\pi(1),S_1} q_{i_1} \xrightarrow{\gamma_2,\pi(2),S_2} q_{i_2} \dots$ is a sequence of consecutive transitions and, for all $i > 0$, $v_{i-1}+\tau(i)-\tau(i-1) \models \gamma_i$ (conventionally $\tau(0) = 0$);
	\item for all $x\in X$,  $v_0(x)=0$ and for all $i>0$ either $v_{i}(x)=0$, if $x \in S_i$, or $v_{i}(x)=v_{i-1}(x) + \tau(i)-\tau(i-1)$ otherwise.
\end{itemize}
 
 Let $\mathit{inf}(\rho)$ be the set of control states $q\in Q$ such that $q = q_{i_j}$ for infinitely many positions $j\geq 0$ of $\rho$. 
 A run is \emph{accepting} when $\mathit{inf}(\rho)\cap B \not = \emptyset$---i.e., when a B\"uchi condition holds.
 In the rest of this paper, we indicate with $\cmax_{\mathcal{A}}$ the maximum constant appearing in the guards and invariants of $\mathcal{A}$; when no ambiguity can arise, we shorten $\cmax_{\mathcal{A}}$ simply as $\cmax$.

\subsection*{Extending Timed Automata with Parameters}

We extend TA by allowing for comparisons with constant parameters. 
More precisely, a \emph{parametric TA} is a tuple $\A=(\Sigma, Q, T, q_0, B, P)$, where $P$ is a set of parameters.
The set of clock constraints $\Gamma(X)$ now includes formulae of the form $x \sim \param$ (where $\sim \in \set{<,=}$).
We introduce a mapping $\mathcal{I} : P \rightarrow \Real$ that associates a real number with each parameter of set $P$.
We write $v, \mathcal{I} \models \gamma$ to indicate that constraint $\gamma$ is satisfied by clock valuation $v$ given parameter evaluation $\mathcal{I}$.
When we want to highlight the number of parameters of $\A$ we will say that it is an $n$-parametric automaton, with $n = |P|$.
A \emph{parametric run} $\rho$  %
over a timed word $\pair{\pi}{\tau}$ with parameter evaluation $\mathcal{I}$ is an infinite sequence of configurations $(q_{i_0}, v_0) \xrightarrow[\tau(1)]{\pi(1)} (q_{i_1}, v_1) \xrightarrow[\tau(2)]{\pi(2)} (q_{i_2}, v_2)\dots,$ that satisfies the constraints introduced above, with the only difference that now, for all $i > 0$, $v_{i-1}+\tau(i)-\tau(i-1), \mathcal{I} \models \gamma_i$ must hold.

When the set $P$ of parameters is a singleton, we call the TA \emph{1-parametric}.

\section{Non-Resetting Test Timed Automata}
\label{section-nrtTA}

In this section, we consider a syntactic restriction on TA, for which it is not possible to test and reset the same clock on the same transition; we call this restriction \emph{non-resetting test TA} (which we abbreviate in nrtTA).
In particular, we first define nrtTA---and parametric nrtTA---and then study their expressiveness.
Then, Section \ref{section-decTA} studies their decidability.

\begin{definition}
Let $\A=(\Sigma, Q, T, q_0, B)$ be a TA, whose set of clocks is $X$.
For every transition $u \in T$ of $\A$ of the form $q_u\xrightarrow{\gamma_u,a_u,S_u} q'_u$, let $X(\gamma_u)$ be the set of clocks that appear in constraint $\gamma_u$.
We say that $\A$ is a \emph{non-resetting test Timed Automaton} (nrtTA for short)
if, for all $u \in T$, $X(\gamma_u) \cap S_u = \emptyset$ holds.
\end{definition}

\begin{figure}[!tp]
\centering
\subfigure[][]{\label{subfig:exnrtTA}}\includegraphics[scale=0.5]{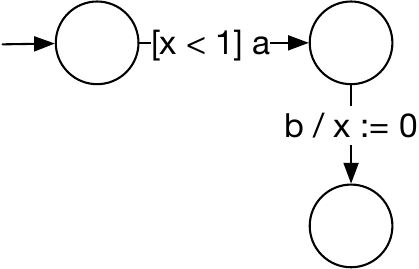}
\hspace{0.5cm}
\subfigure[][]{\label{subfig:exTA}}\includegraphics[scale=0.5]{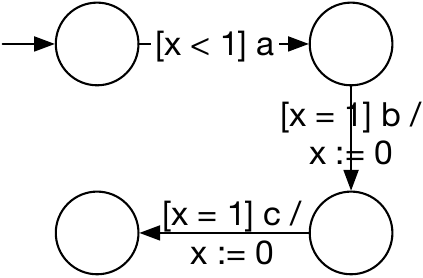}
\subfigure[][]{\label{subfig:exnrtTA2clock}}\includegraphics[scale=0.5]{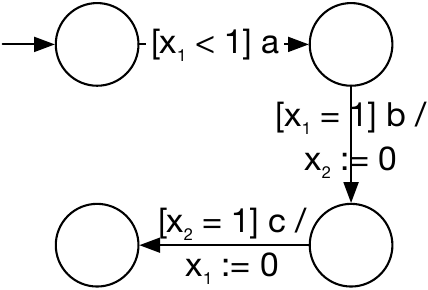}
\caption{\subref{subfig:exnrtTA} Fragment of nrtTA with 1 clock; \subref{subfig:exTA} fragment of TA with 1 clock; \subref{subfig:exnrtTA2clock} fragment of nrtTA with 2 clocks equivalent to the TA of \subref{subfig:exTA}.}
\label{fig:exnrtTA}
\end{figure}

The notion of nrtTA can of course be extended with parameters.
The notion of nrtTA was inspired by the CLTLoc logic \cite{BFMPRS12}, where clocks are handled in a similar way as TA, but in which at each position each clock has exactly one value---wheres in TA, when a clock $x$ is reset at the $i$-th position, it takes two values at the same time instant, the one before the reset, captured by $v_{i-1}(x)+\tau(i)-\tau(i-1)$, and 0 (which corresponds to $v_{i}(x)$).
Notice that, when a clock $x$ is reset, it cannot impose any constraint on the delay preceding the reset:
for example, in the fragment of nrtTA shown in Figure \ref{subfig:exnrtTA}, the time elapsed between the $a$ and the $b$ can be any.

A clock $x$ that can be tested and reset at the same time can be represented, in a nrtTA, with two clocks $x_1, x_2$ which are alternatively tested and reset.
Consider, for example, the fragment of TA of Figure \ref{subfig:exTA}, which uses one clock $x$ that on some transitions is both tested and reset, and which is such that $b$ must occur 1 time instant from the start of the run, and $c$ must occur after another time instant.
To build an equivalent nrtTA we can introduce two clocks $x_1$ and $x_2$, such that initially $x_1$ is used to represent the value of $x$, until its first reset (that is, constraints on $x$ are replaced with constraints on $x_1$); then, after
the first reset, and until the next reset of $x$, $x_2$ is used to represent the value of $x$, then $x_1$ again, and so on.
In this way, to represent the situation in which $x$ is both tested and reset in a transition, it is enough to test one of $x_1, x_2$ (depending on which one is currently representing $x$; keeping track of which clock is representing $x$ is trivially done through the states of the automaton), and reset the other clock.
For example, the nrtTA of Figure \ref{subfig:exnrtTA2clock} is equivalent to the TA of Figure \ref{subfig:exTA}.

\begin{figure}[!tp]
\centering
\subfigure[][]{\label{subfig:exTA2clock}}\includegraphics[scale=0.5]{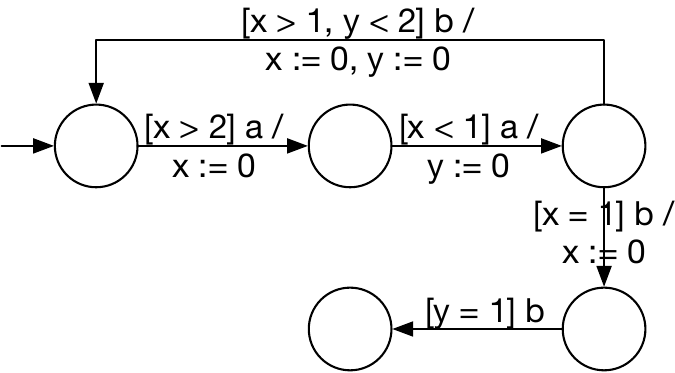}
\hspace{0.3cm}
\subfigure[][]{\label{subfig:exnrtTA3clock}}\includegraphics[scale=0.5]{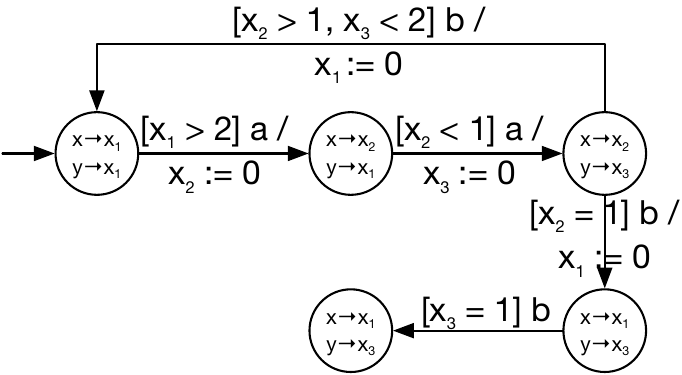}
\caption{\subref{subfig:exTA2clock} Fragment of TA with 2 clocks; \subref{subfig:exnrtTA3clock} fragment of nrtTA with 3 clocks equivalent to the TA of \subref{subfig:exTA2clock}.}
\label{fig:fromTAtonrtTA}
\end{figure}

It is trivial to see that the construction can be generalized to any number of clocks, so one can conclude that it is possible to simulate $n$ clocks of TA with $2n$ nrtTA clocks.
However, it can be shown that the number of nrtTA clock that are enough to simulate $n$ TA clocks is indeed smaller than $2n$ and equal to $n+1$.
Intuitively, we keep a mapping between TA clocks and nrtTA clocks that is used when testing the values of clocks; the mapping is updated when a TA clock is reset, using the ``spare'' ($n+1$-th) nrtTA clock.
Consider, for example, the TA of Figure \ref{subfig:exTA2clock}, which uses two clocks, $x$ and $y$; it is equivalent to the nrtTA of Figure \ref{subfig:exnrtTA3clock}, which uses 3 clocks, $x_1, x_2, x_3$.
The idea is that $x_1, x_2, x_3$ are used in a circular manner, depending on the next clock that is reset. For example, initially $x_1$ represents both $x$ and $y$ (as indicated in the label of the initial state in Figure \ref{subfig:exnrtTA3clock}); then, after the first reset of $x$, the first unused clock---$x_2$ in this case---is reset and from now on it represents $x$ (while $x_1$ still represents $y$); when $y$ is reset, the next unused clock---i.e., $x_3$---is reset, and now corresponds to $y$; and so on.
Notice that, if both clocks are reset at the same time (as in the transition that goes back to the initial state in Figure \ref{subfig:exTA2clock}), after the reset they are both mapped to the same clock $x_i$ ($x_1$ in the case of Figure \ref{subfig:exnrtTA3clock}).
Also, when a clock $x$ or $y$ needs to be tested, the corresponding $x_i$ is used in the guard, depending on the current mapping.

It is easy to see that the construction exemplified in Figure \ref{fig:fromTAtonrtTA} can be generalized to any number of clocks; the same construction works also when the timed automata are parametric. Hence, we have the following result.

\begin{lemma}
\label{lm:fromTAtonrtTA2}
For any (parametric) TA $\A$ whose set of clocks is $X$, there exists an equivalent (parametric) nrtTA $\A'$ whose set of clocks $X'$ has size $|X|+1$.
\end{lemma}

\section{Decidability of 2-clock Parametric nrtTA}
\label{section-decTA}

In this section we prove the following result.

\begin{theorem}
\label{thm:nrtTAdec}
The $\omega$-language emptiness problem is decidable for parametric nrtTA that have at most 2 clocks and 1 parameter.
\end{theorem}

To prove Theorem \ref{thm:nrtTAdec} we consider two cases:
\begin{enumerate}
\item 
the value of the parameter is greater than $2\cmax$ (Theorem \ref{thm:over2cmax}),

\item
and the one in which it is less than $2\cmax$ (Theorem \ref{thm:below2cmax}).
\end{enumerate}
(the case where the parameter is exactly equal to $2\cmax$ can be handled separately, as discussed in Remark \ref{rem:over2cmax}).
We start by considering the first case, for which we have the following result.
\todoMR{}{Introduzione sarebbe da espandere un po'}

\begin{theorem}
\label{thm:over2cmax}
Let $\A=(\Sigma, Q, T, q_0, B)$ be a parametric nrtTA with one parameter, whose set of clocks $X$ is such that $|X| = 2$.
There exists a value $\Xi > 2\cmax$ such that, for every real value $\bar{\param} > 2\cmax$ with $\bar{\param} \neq \Xi$,  
if there is a {parametric run} $\rho$ for $\A$ over a timed word $\pair{\pi}{\tau}$ with  parameter evaluation $\mathcal{I}(\param) = \bar{\param}$, 
then there is also a {parametric run} $\hat \rho$ for $\A$ over a timed word $\pair{\pi}{\hat \tau}$ such that $\hat{ \mathcal{I}}(\param) = \Xi$ holds.
\end{theorem}

\begin{proof}

Let $\Xi$ be any value greater than $1+\cmax(1+|Q|)$.
Let $X = \set{x,y}$.
Assume first that there is a {parametric run} $\rho$ for $\A$ over a timed word $\pair{\pi}{\tau}$ with  parameter evaluation $\mathcal{I}(\param) = \bar{\param}>2\cmax$, $\bar{\param} \neq \Xi$. 
For simplicity, in the following we ignore the input alphabet, i.e., $\Sigma$ can be assumed to be a singleton. The timed word can thus be represented just by the mapping $\tau$.
Run $\rho$ is a sequence of configurations $\rho= \cnf_0 \cnf_1 \cnf_2 \dots$,
where every configuration $\cnf_i$ is $(q_i,v_i)$.

We show we can modify $\tau$ into a $\hat \tau$ such that $\hat \rho$ over $\hat \tau$ is a {parametric run} for $\A$ with parameter evaluation $\hat{\mathcal{I}}(\param) = \Xi$.

For clock $x$, consider a finite factor of run $\rho$, denoted as $\cnf_{h} \cnf_{h+1} \dots \cnf_{k}$, $0\le h <k$. This factor is called a {\em simple $\param$-increasing} sequence of $\rho$ for $x$, if:
\begin{enumerate}
	\item $v_h(y)=0$,
	\item  for all $j$, $h<j\le k$, $0<v_j(y)\le v_j(x)<\bar\param$.
\end{enumerate}

 The {\em duration} of the sequence is the difference $v_k(x)-v_h(x) = v_k(y)$.
  
 A  simple $\param$-increasing sequence $\cnf_{h}  \dots \cnf_{k}$ for $x$ is called {\em complete} if it is not a factor of a longer simple $\param$-increasing sequence for $x$---i.e, $\cnf_{h}  \dots \cnf_{k+1}$ is not a simple $\param$-increasing sequence, or, in other words, in $\cnf_{k+1}$  a clock 
 is reset  or $x\ge\param$ holds.
 The only case where a simple $\param$-increasing sequence $\cnf_{h}  \dots \cnf_{k}$
 cannot be extended to be complete
is when the sequence  is a prefix of an infinite sequence $\cnf_{h} \cnf_{h+1} \dots$ where both clocks are smaller than the parameter. 
 Therefore, the run must be Zeno (otherwise it would require an infinite value for the parameter). Thus, 
 for every $\epsilon>0$, the infinite sequence starting in $\cnf_h$
 can  be shrunk so  that the distance $\tau(n)-\tau(h)$ between positions $h$ and $n$, for every
 $n>h$, is less than $\cmax +\epsilon$.
 We call this sequence a {\em Zeno $\param$-increasing sequence} for $x$.
  
  Symmetrical definitions hold for clock $y$.

The concatenation of complete
$\param$-increasing sequences for $x$ is called a \emph{$\param$-increasing}
sequence of $\rho$ for $x$.    
A  $\param$-increasing sequence $\cnf_{h}  \dots \cnf_{k}$  for $x$ is called {\em maximal} if it is not a factor of another $\param$-increasing sequence for $x$---e.g., $\cnf_{k+1}$ is  such that $v_{k+1}(x)=0$ or $v_{k+1}(x)\ge \bar\param$. 
The duration of the sequence is  $v_k(x)-v_h(x)$ (notice that $x$ is not necessarily reset in $\cnf_h$). The {\em rank} of the sequence is the number of configurations $(q_i,v_i)$
such that $v_i(x)>\cmax$ and $v_i(y)=0$, $q_i \in Q$ hold.

There are two cases: $\bar{\param} > \Xi$ or $\bar{\param} < \Xi$.

 \textbf{Let $\Xi<\bar{\param}$ hold}.
Consider all maximal $\param$-increasing sequences in the infinite run $\rho$ for clock $x$ and those defined analogously for clock $y$. 
Given one maximal $\param$-increasing sequence, we notice that if its rank is greater than the number $|Q|$ of states, 
then it is possible to build a new run $\tilde\rho$ for $\A$ over a timed word $\tilde\tau$ such that every maximal $\param$-increasing sequence of $\tilde\rho$  has rank at most
$|Q|$.
In fact, if the rank is greater than $|Q|$, the maximal sequence must be of the form:

$$
\cnf_{i_1} \dots \cnf_{i_h-1}\cnf_{i_h} \dots \cnf_{i_n} \cnf_{i_n+1} \dots \cnf_{i_k},
$$ 
such that $q_{i_h}=q_{i_n}$, $v_{i_h}(y)=v_{i_n}(y)=0$ and $v_{i_h}(x)>\cmax, v_{i_n}(x)>\cmax$. Replace the above maximal sequence in $\rho$ with the shorter (but still maximal), $\param$-increasing sequence:

$$
\cnf_{i_1} \dots \cnf_{i_h-1}\cnf_{i_n} \cnf_{i_n+1} \dots \cnf_{i_k},
$$
obtaining a new infinite sequence $\tilde\rho$. Since both $v_{i_h}(x),v_{i_n}(x)$
 are greater than $\cmax$ and $v_{i_h}(y)=v_{i_n}(y)=0$ hold, then we can define a new timed word $\tilde\tau$ identical to $\tau$ but lacking the portion from  $i_h$ to $i_n$ and letting $\tilde\tau(j+i_h)=\tau(j+i_n)$ for all $j\ge 0$. 
 The sequence  $\tilde\rho$ is  still a run of $\A$ over the timed word $\tilde\tau$. 

By repeatedly applying this procedure (considering both clocks $x$ and $y$)
we can obtain an infinite sequence, still called  $\tilde\rho$, of configurations of $\A$ over a timed word $\tilde\tau$ such that it has no maximal $\param$-increasing sequence of rank greater than $|Q|$.
It is clear that $\tilde\rho$ it is still a run of $\A$ over $\tilde\tau$.

We now buld a sequence $\rho' = \cnf'_0 \cnf'_1 \cnf'_2 \dots$ where the duration of a maximal sequence may be smaller than the corresponding one in $\tilde\rho$. 
In general, $\rho'$ will not be a run, since it may not verify the constraints comparing the clocks with the parameter, but it will respect all other constraints, i.e, the ``classical regions''.

We claim that, for any value $\epsilon>0$, we can shrink the duration of every complete $\param$-increasing sequence $\cnf_{h}  \dots \cnf_{k}$ to be less than $\cmax + \epsilon$, respecting the classical regions. In fact, if 
 the duration is greater than $\cmax + \epsilon$, then $v_k(y)>\cmax$ holds: let $h\le j<k$
 be the rightmost position in the sequence such that $v_j(y)\le \cmax$.   
We can  decrease the duration from $j$ to $j+1\le k$ so that $v_{j+1}(y)= \cmax + \epsilon'$ for any $\epsilon'>0$.  
If $k>j+1$ holds we can clearly shrink the duration between position $j+1$ and position $k$ down to any arbitrarily small value $\epsilon''$: 
 $v_{k}(y)=\cmax+\epsilon'+\epsilon''$ (since both clocks will be greater than $\cmax$ in each one of those positions anyway). 
Just let $\epsilon= \epsilon'+\epsilon''$. Notice that, if the duration of the original sequence
was greater than $\cmax + \epsilon$, then it is still greater than $\cmax$ after shrinking, so the evaluation of all guards in $\cnf_k$  is not changed. 
The shrinking of the duration requires to reduce also one or more of the values of timestamps $\tau(h+1), \dots, \tau(k)$;
 for the moment being, however, we do not change the timestamp $\tau(k+1)$, thus $\cnf_{k+1}$ is unchanged. 
 
Therefore, every complete, or Zeno, $\param$-increasing sequence $\cnf_{h}  \dots \cnf_{k}$ for $x$ can be assumed in the following to have duration less than $\cmax +\epsilon$---i.e., $v_k(x) -v_h(x)<\cmax+\epsilon$ holds, hence
 clock $x$ is always increased less than  $\cmax+\epsilon$ compared to its value in the leftmost position.
An immediate consequence  is that in every position of a complete $\param$-increasing sequence for $x$, including the rightmost position (if any),
clock $y$ is always less than $\cmax+\epsilon$.

The duration of a maximal $\param$-increasing sequence %
of rank $n$ can thus similarly be assumed to be less than $n(\cmax + \epsilon)$ (by reducing the duration between each complete sequence), still respecting the classical regions.
Since no maximal $\param$-increasing sequence of $\rho'$ has rank  greater than $|Q|$,
the longest duration of a maximal sequence
is less than $|Q|(\cmax +\epsilon)$, i.e., it is less than $\Xi - \cmax = 1+\cmax|Q|$.

We now need to show that we can transform $\rho'= \cnf'_0 \cnf'_1 \cnf'_2 \dots$ into a run $\hat\rho=\widehat{\cnf_0}\widehat{\cnf_1}\widehat{\cnf_2}\dots$
of $\A$ over a timed word $\hat\tau$ when $I(\param)=\Xi$.

It is easy to notice that in each configuration $\cnf'_i=(q_i,v_i')$, for every clock $z \in \{x,y\}$, if $v_{i}(z)\le \cmax$ holds,
then 
$v'_i(z)=v_i(z)$ holds, else $v'_i(z)>\cmax$ does. Define a timed word $\tau'$ such that $\rho'$ is well defined over $\tau'$. 
As already remarked, the only constraints that may not have the same value in $\cnf_i, \cnf'_i$ are the comparisons with the parameter $\param$. This may require adjustments to timestamps and clocks' values in $\rho', \tau'$, thus defining new sequences $\hat{\rho}, \hat{\tau}$.

We build $\hat \rho, \hat \tau$ based on $\tilde \rho$, $\tilde \tau$, $\rho'$, $\tau'$ (where $\tilde{\rho}= \cnf_0 \cnf_1 \cnf_2 \dots$) by induction on the position $h\ge 0$. The induction hypothesis is: 

\begin{itemize}
\item[(*)]
the same set of constraints, including those over the parameter, is verified in $\cnf_{j}$,
for all $j < h$, with $\mathcal{I}(\param)=\Xi$, over $\tilde\tau(j)$
and in $\widehat{\cnf}_j$ over $\hat \tau (j)$. 
In particular, if a clock $x$ (and symmetrically for $y$) is such that its value in $\cnf_j$ is less than or equal to $\cmax$, then it has the same value in $\hat \cnf_j$; also, every complete $\param$-increasing sequence of configurations (for $x$ or for $y$, up to $h-1$)
has duration less than $\cmax + \epsilon$.
\end{itemize}

The base case is obvious (just let $\hat \cnf_0 = \cnf'_0=\cnf_0)$.
Let $\cnf_h=(q_h, v_h)$ be a configuration of $\tilde\rho$, with $h > 0$. We define $\hat \cnf_h=(q_h,\hat v_h)$. The induction step considers the various cases of comparison with the parameter.   
\begin{enumerate}
	\item  If $\cnf_{h}$ verifies the constraint $x \geq {\param}\land y \geq {\param}$
	(hence, also $x>{\cmax}\land y>{\cmax}$), then let, for instance, $\hat \tau(h) = \hat \tau(h-1) + \Xi +1$, $\hat v_h (x)= \hat v_{h-1}(x) +\Xi + 1$, $ \hat v_h (y) = \hat v_{h-1}(y) +\Xi + 1$.
	All constraints in $\widehat{\cnf}_h$ have thus the same truth value as in $\cnf_h$ by considering $\mathcal{I'}(\param)=\Xi<\bar\param$ instead of $\mathcal{I}(\param)=\bar\param$.
	
	\item If $\cnf_{h}$ verifies the constraint $x<{\param}\land y\ge{\param}$, then, since clocks can only be incremented or reset, position $h$ %
	must be preceded in $\tilde\rho$ by a complete $\param$-increasing sequence $\sigma_y$ for $y$, namely  $\sigma_y= {\cnf}_j \dots {\cnf}_{j+n}$, $0\le n\le h-1-j$. 
	By induction hypothesis, there is a complete $\param$-increasing sequence $\hat \sigma_y$ for $y$ in $\hat\rho$,  namely 
	$\hat \sigma_y= \widehat{\cnf}_j \dots \widehat{\cnf}_{j+n}$.
	This sequence may possibly be followed by a few configurations 
	$\widehat{\cnf}_{j+n+1} \dots \widehat{\cnf}_{h-1}$,
	where $y$ is greater than $\Xi$ and $x$ is less than $\Xi$,
	with in this case
	$n < h-1-j$. For simplicity, we ignore those configurations, i.e.,  let $n=h-1-j$. %
	
	First, we notice that, from the discussion above, $\hat v_{j-1}(y)\le \Xi-\cmax-\epsilon$ holds, since $\sigma_y$ is the rightmost complete sequence of a maximal sequence of total rank less than $|Q|$. 
	In the following, we adjust, if necessary, $\hat \tau(j), \hat v_j(y)$ and $\hat v_{j+i}(x)$ (for all $1 \leq i \leq n+1$) so that the remaining values of $\hat \tau(j+i) = \hat \tau(j) + \hat v_{j+i}(x)$ and $\hat v_{j+i}(y) = \hat v_{j}(y)+\hat v_{j+i}(x)$ are such that the desired constraints on $x$ and $y$ hold.
	For all $i$, $1 \le i \le n+1$, such that  $v_{j+i}(x)\le \cmax$ let first $\hat v_{j+i}(x) = v_{j+i}(x)$ hold.  
	\begin{enumerate}[label={\roman*)}]
		\item 	
		Case $v_{j+n+1}(x)\le \cmax$:
		to allow for $v_{j}(y)+ v_{j+n+1}(x)\geq\bar{\param}>2\cmax$ to hold, it is thus necessary that $v_j(y)>\cmax$ also holds,
		hence also $\hat v_{j}(y)$ must be greater than $\cmax$. Therefore the increment of $\hat\tau(j)$ over $\hat\tau(j-1)$ can be as large as needed, since $\A$ is an nrtTA, because there cannot be any constraint on the value of $x$ at the moment of its reset in position $j$.
		\todoMR{}{da rivedere questa spiegazione, però, qualcosa sul fatto che sono nrtTA va detto.}
		To obtain $\hat v_j(y)+\hat v_{j+n}(x)<\Xi$ and $\hat v_j(y)+\hat v_{j+n+1}(x)\geq\Xi$, let $\hat v_j(y) = \Xi-\hat v_{j+n+1}(x)+ \varepsilon$,
		for $0 \leq \varepsilon < \hat v_{j+n+1}(x)-\hat v_{j+n}(x)$. Since obviously $\hat v_j(y)\geq \Xi-\cmax>\hat v_{j-1}(y)$ holds,
		the timestamp $\hat \tau(j)$ can be correctly defined  as  $\hat\tau(j-1)+\hat v_j(y)- \hat v_{j-1}(y)$.
		
		\item 
		Case $v_{j+n+1}(x)> \cmax$. We  need to distinguish two subcases, depending on $v_{j}(y)$ being greater or smaller than $\cmax$. 
		
		If $v_{j}(y)> \cmax$, then by induction hypothesis $\hat v_{j+n}(x)<\cmax +\epsilon$ holds; the duration between positions $j+n$ and $j+n+1$ can be made as small as necessary to guarantee that $\hat v_{j+n+1}(x)<\cmax +\epsilon$ holds and the same constraints on $x$ and $y$ hold in $v_{j+n+1}$ and $\hat v_{j+n+1}$.
		Notice that, if $\hat v_{j}(y)<\Xi-\cmax-\epsilon$ holds, then neither $\hat v_{j+n}(y) = \hat v_{j}(y) + \hat v_{j+n}(x) \geq \Xi$ nor $\hat v_h(y) = \hat v_j(y)+\hat v_{j+n+1}(x)\geq \Xi$ can hold.
		In this case, we must redefine timestamp $\hat \tau(j)$---thus, $\hat v_j(y)=\hat \tau(j) - \hat \tau(j-1) +\hat v_{j-1}(y)$---so that $\hat v_j(y)+\hat v_{j+n+1}(x) \geq\Xi$ holds.
		This is possible because, by induction hypothesis, $\hat v_{j}(y)> \cmax$ and $\hat v_{j}(x) = 0$ hold, hence, since $\A$ is an nrtTA, the distance $\hat \tau(j) - \hat \tau(j-1)$ between points $j-1$ and $j$ can be arbitrary.\\
		If $v_{j}(y)\le \cmax$, then we define $\hat v_{j}(y) = v_{j}(y)$.
		Let $\Delta = \bar{\param}-v_{j+n+1}(x)$.
		Notice that $\Xi > \cmax \geq v_j(y) = v_j(y) - v_j(x) = v_{j+n+1}(y) - v_{j+n+1}(x) > \Delta$ hold, since $x$ and $y$ advance with the same rate and $v_j(x) = 0$, $v_{j+n+1}(y) > \bar{\param} > v_{j+n+1}(x)$ hold by hypothesis.
		Since $\hat v_{j+n+1}(x)$ must be greater than $\cmax$ and $\A$ is an nrtTA, we can define $\hat v_{j+n+1}(x)=\Xi -\Delta > \cmax$
		(since $\Xi > 2\cmax$ holds by hypothesis), hence  the ``distance'' of $x$ from $\Xi$ when in configuration $\cnf_{j+n+1}$ is still $\Delta$. 
		This will be useful in point 4 of the proof.
		Of course, $\hat v_{j+n+1}(x)<\Xi, \hat v_{j+n+1}(y)>\Xi$ hold, since $\hat v_{j+n+1}(y) = \hat v_{j}(y) + \hat v_{j+n+1}(x)$ (notice that $\hat v_j(y) = v_j(y) > \Delta$ and $\hat v_{j+n+1}(x)=\Xi -\Delta$
		hold).
		
	\end{enumerate}

        \item The case where $\cnf_{h}$  verifies the constraint $y<{\param}\land x\ge{\param}$ is symmetrical to the previous one. 
        
	\item If $\cnf_{h}$ verifies the constraint $x<{\param}\land y<{\param}$, then $\cnf_h$ is a part of a $\param$-increasing sequence for, say, clock $x$.
	 
We assume that  $\cnf_{h}$ is the {\em leftmost} position of a maximal $\param$-increasing sequence $\sigma_x$ for clock $x$, corresponding to sequence $\sigma'_x$ in $\rho'$ (hence, $v_h(y) = v'_h(y) = 0$; on the other hand, either $v'_h(x) = 0$ or $v'_h(x) > 0$ hold, where the latter case occurs if $\sigma_x$ is preceded by a $\param$-increasing sequence for $y$).
We build a corresponding maximal sequence $\hat\sigma_x$ for clock $x$ in the following way (where $i$ is such that $\cnf'_{h+i}$ is a configuration of $\sigma'_x$):
(a) we define $\hat \tau(h+i) = \hat \tau(h) + \tau'(h+i) - \tau'(h)$;
(b) $\hat v_{h+i}(y) = 0$ holds if, and only if, $v'_{h+i}(y) = 0$ holds (recall that clock $x$ is never reset along $\sigma'_x$, $\hat \sigma_x$, except possibly for the first position $h$);
and (c) $\hat \cnf_{h}$ and $\hat \tau({h})$
are suitably defined---as described below---so that the induction hypothesis is verified for all positions in the maximal sequence.
We show that, as long as the induction hypothesis holds for the previous configurations, then it also holds in $\widehat{\cnf}_h$ and in every configuration in the sequence $\hat\sigma_x$. 
Thus, it is not necessary to consider positions different from the first one in a maximal sequence.

 We consider the various possibilities for configuration $\cnf_{h-1}$.

\begin{enumerate}[label={\roman*)}]

\item
Case $v_{h-1}(x) \geq \bar{\param}$.
Then $\hat{v}_h(x) = v'_h(x) = 0$ must hold, hence, since $\A$ is an nrtTA, the delay $\hat{\tau}(h) - \hat{\tau}(h-1)$ can be arbitrary, and the values of $\hat{\tau}(h+i)$ defined as above
obviously allow us to verify the induction hypothesis in every position of $\hat \sigma_x$.

\item 
Case $v_{h-1}(x)<\bar{\param}$ and $v_{h-1}(y)<\bar{\param}$.
Position ${h-1}$ is the end point of a maximal $\param$-increasing
sequence  $\sigma_y$ for $y$ in $\tilde\rho$, which by induction hypothesis must correspond to a $\param$-increasing sequence for $y$ in $\hat\rho$,
(it cannot be a $\param$-increasing sequence for $x$ since $\sigma_x$ is maximal).
We compute the value of clock $x$ in $\widehat{\cnf}_h$ (of course, clock $y$ is 0 in both $\widehat{\cnf}_h$ and $\cnf_{h}$).
If $v_h(x) \leq \cmax$ holds, then we simply define $\hat{\tau}(h) = \hat{\tau}(h-1) + \tau'(h) - \tau'(h-1)$ so that $\hat{v}_h(x) = {v}_h(x)$ holds and the induction hypothesis is satisfied.
If, instead, $v_h(x) > \cmax$ holds, since $h-1$ is the rightmost position of a maximal $\param$-increasing sequence for the other clock $y$, then
by induction hypothesis  the value of $x$ in $\widehat{\cnf}_{h-1}$ is less than $\cmax +\epsilon$; thus, we can define the value of $x$ in $\widehat{\cnf}_{h}$ such that it is also less than $\cmax +\epsilon$, 
since $x$ must be reset at the beginning of the rightmost $\param$-increasing complete sequence for $y$ in $\sigma_y$---i.e., it is also less than $\Xi$; notice that---as in point 2.ii above---the duration from position $h-1$ to $h$ can be defined to be as small as necessary to make $\hat{v}_{h}(x) < \cmax + \epsilon$ hold.
Therefore, the largest value that clock $x$ can assume in every configuration of the maximal sequence $\hat \sigma_x$ is less than $(\Xi - \cmax -\epsilon) + (\cmax + \epsilon)=\Xi$.
In addition, by Condition (a) above, the same constraints on $x$ and $y$ hold along $\sigma_x$ and $\hat{\sigma}_x$, no matter if $v_x(x) \leq \cmax$ or $v_x(x) > \cmax$ hold.

\item
Case $v_{h-1}(x)< \bar{\param}$ and $v_{h-1}(y) \geq \bar{\param}$.
Since clocks can only be incremented or reset, position $h-1$ must be preceded by a $\param$-increasing sequence $\sigma_y= \widehat{\cnf}_{j} \dots \widehat{\cnf}_{j+n}$ for $y$, with $j+n<h-1$,
possibly followed by a few configurations $\widehat{\cnf}_{j+n+1} \dots \widehat{\cnf}_{h-1}$ where $y$ is greater than $\Xi$ and $x$ is less than $\Xi$, with $j+n +1\le h-1$. For simplicity let $j+n+1=h-1$.
We notice that, since $\A$ is an nrtTA and clock $y$ is reset at position $h$, the distance $\hat{\tau}(h) - \hat{\tau}(h-1)$ can be chosen arbitrarily as long as the same clock constraints hold for ${v}_h(x)$ and $\hat{v}_h(x)$ (and, if $v_h(x) \leq \cmax$ holds, ${v}_h(x) = \hat{v}_h(x)$ also holds).
In addition, when previously dealing with such sequence in case 2.ii,
one of the following two conditions held:
\begin{enumerate}[label={\Roman*)}]
\item 
$\hat v_{j+n+1}(x) < \cmax+\epsilon$, or

\item
$\hat v_{j+n+1}(x)=\Xi -\Delta$

\end{enumerate}
where $0 < \Delta= \bar{\param} -v_{j+n+1}(x) < \cmax$.
We show that, from each condition (I) or (II)
it descends that in $\hat \sigma_x$ both clocks are less than $\Xi$.\\
If Condition (I) holds, we are assured that %
$\hat{v}_i(x)<\Xi$ holds for every position $i$ of $\hat\sigma_x$,
since clock $x$ (always the larger of the two) is incremented in $\hat \sigma_x$ by less than $\Xi-\cmax-\epsilon$ (since the duration of $\hat \sigma_x$ is the same as that of $\sigma'_x$).\\
If Condition (II) holds, instead, the  original maximal sequence $\sigma_x$ in $\tilde \rho$ was such that $x$ could be increased of less than $\Delta$, in order to have 
$v_i(x)<\bar{\param}$
for every position $i$ of $\sigma_x$;
since, by construction, the duration of $\sigma'_x$ is not greater than that of $\sigma_x$, the corresponding sequence $\hat\sigma_x$ must also increase $x$ of an amount less than $\Delta$---i.e., 
$\hat{v}_i(x)<\Xi$ holds for every position $i$ of $\hat\sigma_x$.
As in point 4.ii above, Condition (a) guarantees that the same constraints on $x$ and $y$ hold along $\sigma_x$ and $\hat{\sigma}_x$.
\end{enumerate}

\end{enumerate}

 \textbf{Let us now consider the case $\bar{\param} < \Xi$}.
 Again, we build $\hat \rho, \hat \tau$ based on $\rho$, $\tau$ by induction on the position $h\ge 0$.
 The induction hypothesis is the same as (*) before.

The base case is obvious (just let $\hat \cnf_0 = \cnf'_0=\cnf_0)$.
Let $\cnf_h=(q_h, v_h)$ be a configuration of $\tilde\rho$, with $h > 0$.
We define $\hat \cnf_h=(q_h,\hat v_h)$. The induction step considers the various cases of comparison with the parameter.   
\begin{enumerate}
	\item  If $\cnf_{h}$ verifies the constraint $x\geq{\param}\land y\geq{\param}$
	(hence, also $x>{\cmax}\land y>{\cmax}$), then we can simply define the delay $\hat \tau(h) - \hat \tau(h-1)$ such that $\hat v_{h}(x) > \Xi$ and $\hat v_{h}(y) > \Xi$ hold.
	
	\item If $\cnf_{h}$ verifies the constraint $x<{\param} \ \land \ y\ge{\param}$, then, as in the previous part of the proof, for simplicity we can assume that in $h-1$ constraint $y < \param$ holds.
	Hence, $h-1$ is the rightmost position of a complete $\param$-increasing sequence for $y$ $\sigma_y = \cnf_j\ldots \cnf_{h-1}$.
	As in point 2.ii of the first part of the proof, we separate the two cases $v_j(y) > \cmax$ and $v_j(y) \leq \cmax$.
	If $v_j(y) > \cmax$ holds, since $\A$ is an nrtTA, we increase the timestamp $\hat \tau(j)$ by quantity $\Xi - \bar{\param}$
	and we define $\hat \tau(i) = \hat \tau(i-1) + \tau'(i) - \tau'(i-1)$, for all $j < i \leq h$,
	thus obtaining that $\hat v_{h}(y) > \Xi$ holds.
	If $v_j(y) \leq \cmax$ holds, then there must be a position $j < k \leq h$ where $y > \cmax$ and $x > \cmax$ both hold (since $\bar{\param} > 2\cmax$ and $v_j(y) = v_j(y) - v_j(x) \leq \cmax$
	both hold and $x$ and $y$ advance with the same rate, which entails that $v_h(x) > \cmax$ must hold).
	If we increase the timestamp $\hat \tau(h)$ by quantity $\Xi - \bar{\param}$,
	we obtain that $\hat v_{h}(y) \geq \Xi$ and $\hat v_{h}(x) < \Xi$ hold.
	
    \item The case where $\cnf_{h}$  verifies the constraint $y<{\param} \ \land \ x\ge{\param}$ is symmetrical to the previous one. 

	\item
	If $\cnf_{h}$ verifies the constraint $x<{\param}\land y<{\param}$, then if we simply define $\hat \tau(h) = \hat \tau(h-1) + \tau'(h) - \tau'(h-1)$, we obtain that $\hat v_{h}(x) < \Xi$ and $\hat v_{h}(y) < \Xi$ both hold as $\Xi$ is greater than $\bar{\param}$ (also, the same constraints that hold in $v_h$ hold in $\hat{v}_h$).
\end{enumerate}
\end{proof}

\begin{remark}
\label{rem:over2cmax}

Thanks to Theorem \ref{thm:over2cmax}, we can separately deal with values $\mathcal{I}(\param) = \bar{\param}$ of the parameter such that $\bar{\param} > 2\cmax$ holds.
Indeed, to determine whether there is a run of automation $\A$ for some $\bar{\param} > 2\cmax$ it is enough to instantiate parameter $\param$ with value $\Xi$ and check the emptiness problem for that specific, non-parametric automaton.
Similarly, we can test, one by one, all cases in which the parameter $\param$ is a multiple of $\frac{1}{2}$ and it is less than ot equal to $2\cmax$, simply by instantiating the automaton with those values of the parameter.
If the language of the automaton is not empty for any of those values, the decision procedure stops.
Hence, in the rest of this paper we will consider that $\param < 2\cmax$ holds and it is not a multiple of $\frac{1}{2}$.
\end{remark}

Let ${X}$ be a set of clocks, and $\cmax \in \Natgez$ a constant.
A {\em clock region} \cite{Alur&Dill94} is a set of clock valuations that satisfies a maximal consistent set of constraints on clocks of the form $x \sim \const$, $x \sim y + \const$, and their negations,
with $\sim \in \set{<,=}$, $\const \in \Natgez$, $\const \leq 2\cmax$, $x,y \in {X}$ (notice that, unlike \cite{Alur&Dill94}, we need to define clock regions up to $2\cmax$, rather than $\cmax$).
We can define the \emph{time-successor} relationship among clock regions as in \cite{Alur&Dill94}.

Given the statement of Theorem \ref{thm:nrtTAdec}, we consider a set of clocks $X$ such that $|X| = 2$ holds and we use $x,y$ as names of the two clocks.
By symmetry, every statement about a 2-clock nrtTA can be given by exchanging $x,y$.
In the rest of this paper, when we need to indicate a generic clock in $X$, we use symbols $z, z_1, z_2$.

Let $\param$ be a \emph{parameter}, and let $\mathcal{I}(\param)$ be its value in $\Realgez$.
Given an interpretation $\mathcal{I}(\param) = \bar{\param}$ for parameter $\param$, a \emph{$\param$-parametric (clock) region} $R_{\bar{\param}}$ is the intersection of a clock region with clock valuations satisfying a maximally consistent set of clock constraints of the form $z \sim \param$, with $z \in X$ and $\sim \in \set{<,=}$, and their negations. 

Let $v$ be a clock valuation over set of clocks $X$ and $\delta \in \Real_{>0}$ a delay.
We define $v + \delta$ the clock valuation $v'$ such that, for all $z \in X$
it holds that $v'(z) = v(z) + \delta$.
We also define $v \oplus \delta$ as the set of clock valuations $v'$ such that, for all $z \in X$,
either $v'(z) = v(z) + \delta$, or $v'(z) = 0$.
Notice that $v \oplus \delta$ is the set of clock valuations that can be obtained from $v$ with delay $\delta$ considering all possible resets of clocks in $X$.

\begin{figure}[!tp]
\centering
\includegraphics[scale=0.55]{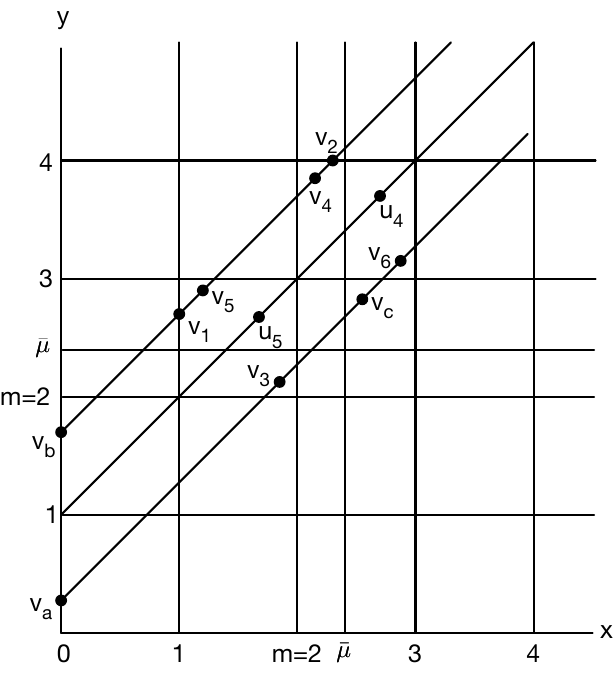}
\caption{Graphical depiction of examples of the cases of Proposition \ref{prop:critval}.
In all cases, we have $\intprm = 2$, $z_1 = x$ and $z_2 = y$.
Valuations $v_i$ and $u_i$ are examples of case $i$.
For instance, for valuation $v_3$ we have that $v = v_a$, $v' = v_3$, $\const = 2$, $z_{2,0} = 0$ hold.
For valuation $v_5$, instead, it holds that $v = v_b$, $v' = v_5$, $\const = 1$, $z_{2,0} = 1$.
Notice that multiple cases can hold for a critical valuation; for example, valuation $v_c$ corresponds to both cases 4 (with $\const = 3$) and 5 (with $\const = 2$).}
\label{fig:Prop1}
\end{figure}

Let $\intprm=\lfloor \bar\param \rfloor$. In the following, we call {\em critical} a clock valuation $v$ such that $\intprm<v(x)<\intprm+1$ or $\intprm<v(y)<\intprm+1$ hold.
The following proposition, which is illustrated by Figure \ref{fig:Prop1}, lists some properties that hold for critical valuations, which will be useful in the proof of Lemma \ref{lm:agreement}.

\begin{proposition}
\label{prop:critval}
Let $v$ be a clock valuation such that $v(z_1) = 0$ and $z_{2,0} < v(z_2) < z_{2,0}+1$ holds (with $z_1, z_2 \in X$), for some $z_{2,0} \in \Natgez$, with $z_{2,0} < 2\cmax$.
For all $\delta \in \Real_{>0}$, if $v' = v + \delta$ is a critical valuation, then $v'$ satisfies at least one of the following combinations of constraints (for some $\const \in \Natgez$):

\begin{enumerate}
\item 
\label{en:critval1}
$v'(z_1) = \const$ and $\intprm = \const + z_{2,0}$

\item 
\label{en:critval2}
$v'(z_2) = \const$ and $\intprm = \const - z_{2,0} -1$

\item 
\label{en:critval3}
$\const -1 < v'(z_1) < \const$ and $\intprm = \const + z_{2,0}$

\item 
\label{en:critval4}
$\const -1 < v'(z_2) < \const$ and $\intprm = \const - z_{2,0} -1$

\item 
\label{en:critval5}
$\const < v'(z_1) < \const+1$ and $\intprm = \const + z_{2,0}$

\item 
\label{en:critval6}
$\const < v'(z_2) < \const+1$ and $\intprm = \const - z_{2,0} -1$

\end{enumerate}
\end{proposition}

\begin{figure}[!tp]
\centering
\includegraphics[scale=0.55]{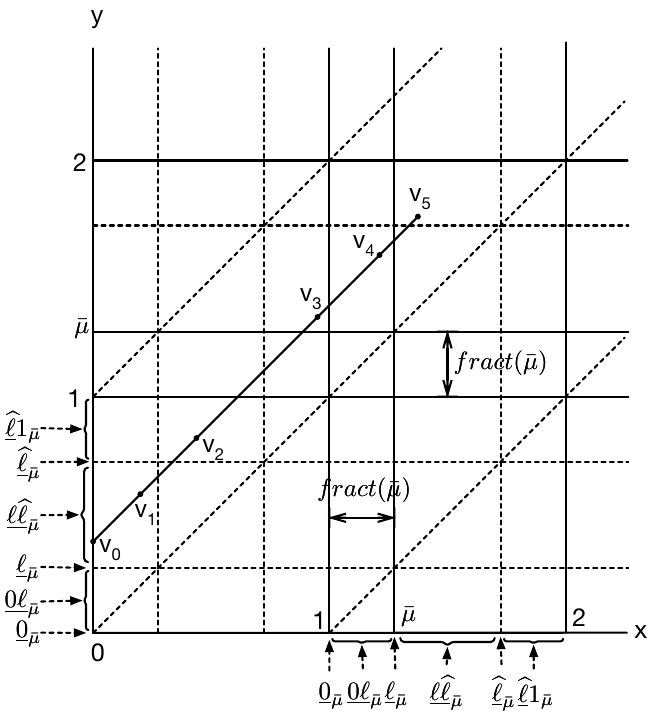}
\caption{Graphical depiction of intervals $\ZR_{\bar{\param}}$, $\ZA_{\bar{\param}}$, $\ZB_{\bar{\param}}$, $\ZC{\bar{\param}}$, $\ZD_{\bar{\param}}$, and $\ZE_{\bar{\param}}$ and example of \onerst{} sequence $v_0 v_1 v_2 v_3 v_4 v_5$ (notice that in this case the polarity is negative, as $\fract{\bar{\param}} < \frac{1}{2}$ holds).}
\label{fig:resetregions}
\end{figure}

Let $\ell_{\bar{\param}} = \min({\fract{\bar{\param}}, 1-\fract{\bar{\param}}})$.
Given a clock valuation $v$ and a clock $z \in X$, we identify the following possible intervals $\iota_{v(z)}$ for the fractional part of $v(z)$ (see Figure \ref{fig:resetregions} for a graphical depiction):
\begin{itemize}
  \item[$\ZR_{\bar{\param}}$:] if $\fract{v(z)}=0$;
  \item[$\ZA_{\bar{\param}}$:] if $0<\fract{v(z)}<\ell_{\bar{\param}}$; 
  \item[$\ZB_{\bar{\param}}$:] if $\fract{v(z)}=\ell_{\bar{\param}}$;
  \item[$\ZC_{\bar{\param}}$:] if $\ell_{\bar{\param}}<\fract{v(z)}<1-\ell_{\bar{\param}}$;
  \item[$\ZD_{\bar{\param}}$:] if $\fract{v(z)}=1-\ell_{\bar{\param}}$;
  \item[$\ZE_{\bar{\param}}$:] if $1-\ell_{\bar{\param}}<\fract{v(z)}<1$.
\end{itemize}

We straightforwardly introduce the $\prec$ order relation between intervals $\iota$ in the following way: $\ZR_{*} \prec \ZA_{*} \prec \ZB_{*} \prec \ZC_{*} \prec \ZD_{*} \prec \ZE_{*}$, where the '*' stands for any value of the parameter.
Notice that the same order relation holds between the intervals defined for two different values 
$\bar{\param}$ and $\hat{\param}$ of the parameter if the values are such that $\fract{\bar{\param}} < \fract{\hat{\param}} < \frac{1}{2}$ holds, e.g., in this case $\ZA_{\bar{\param}} \prec \ZB_{\hat{\param}}$ holds.
\todoPL{}{ho provato a spiegare questo pezzo, ma ho dovuto invertire $\fract{\hat{\param}} < \fract{\bar{\param}} $ in $\fract{\bar{\param}} < \fract{\hat{\param}}  $}
We also straightforwardly define relation $\preceq$ as the reflexive closure of $\prec$. 
We will sometime write $\iota_\alpha$, for $\alpha \in \Realgez$, to indicate the interval of set $\set{\ZR_{\bar{\param}}, \ZA_{\bar{\param}}, \ZB_{\bar{\param}}, \ZC_{\bar{\param}}, \ZD_{\bar{\param}}, \ZE_{\bar{\param}}}$ to which $\fract{\alpha}$ belongs.

We say that $\bar{\param}$ has \emph{positive polarity} (resp., \emph{negative polarity}) if $\fract{\bar{\param}} > \frac{1}{2}$ (resp., $\fract{\bar{\param}} < \frac{1}{2}$) holds.

We say that two clock evaluations $v,\hat{v}$ are {\em in agreement for interpretations $\mathcal{I}(\param) = \bar{\param}$, $\hat{\mathcal{I}}(\param)=\hat{\param}$} if
they satisfy the same constraints of the form $z \sim \const$, $z \sim \param$, and their negations,
with $\sim \in \set{<,=}$, $\const \in \Natgez$, $\const \leq 2\cmax$, $z \in {X}$.
Notice that
\todoPL{}{posticipare a dopo definizione complete agreement; inoltre, potrebbe essere utile fare un esempio, prendendo 2 valori specifici di $\bar{\param}$ e $\hat{param}$ e mostrare che in effetti i vincoli di confronto con il parametro danno luogo a valutazioni diverse che li soddisfano.}
the $\param$-parametric clock regions $R_{\bar{\param}}$ and $R_{\hat{\param}}$ to which $v$ and $\hat{v}$ belong, respectively, are not necessarily the same (i.e., they might not include exactly the same clock valuations), even if $v$ and $\hat{v}$ are in agreement for interpretations $\mathcal{I}$, $\hat{ \mathcal{I}}$, because the set of valuations that belong to a clock region depends on the value of parameter $\param$ if the valuations are critical.
Valuations $v$ and $\hat{v}$ are in {\em complete agreement for $z \in X$} if $v(z) \leq 2\cmax$ and $\hat{v}(z) \leq 2\cmax$ hold, they are in agreement, and $\iota_{v(z)} = \ZA_{\bar{\param}}$ if, and only if, $\iota_{\hat{v}(z)} = \ZA_{\hat{\param}}$, and similarly for $\ZR_{\bar{\param}}, \ZB_{\bar{\param}}$, $\ZC_{\bar{\param}}$, $\ZD_{\bar{\param}}$, $\ZE_{\bar{\param}}$.
We say that they are in {\em complete agreement} if they are in complete agreement for all clocks in $X$.
As a shorthand, we will sometime say that $v(z)$ and $\hat{v}(z)$ are in complete agreement, instead of saying that $v, \hat{v}$ are in complete agreement for $z$.
\todoMR{}{aggiunta considerazione, usata in prova lemma 4.}
Notice that both the agreement and the complete agreement relations are equivalence relations.
Also, we admit that $\bar{\param} = \hat{\param}$ holds.

Let $v_0 v_1 \ldots$ be a (possibly infinite) sequence of valuations such that $v_0(z) = 0$ for some $z \in X$ and for all position $i$ in the sequence such that $i \geq 1$ holds, $v_i = v_{i-1} + \delta_i$ holds for some $\delta_i \in \Real_{>0}$.
We call such a sequence (exemplified in Figure \ref{fig:resetregions}) a \emph{\onerst{} sequence} (we say that it is a \emph{\onerst{} sequence for $z$} when we want to single out the reset clock of interest).
With a slight abuse of notation, given a sequence of configurations $\runsym = C_{0} C_{1} \ldots $, where $C_{i} = (q_{i}, v_{i})$ holds for all $i \geq 0$, we also say that $\runsym$ is a \onerst{} sequence if the corresponding sequence of valuations $v_0 v_1 \ldots $ is.

\todoPL{}{messa frasetta di introduzione al Lemma 3}

The following lemma shows, given an interpretation  $\mathcal{I}$ and another interpretation $\hat{\mathcal{I}}$ with the same integer part and polarity, that for every  \onerst{} sequence $\orssym$ for $\mathcal{I}$ it is always possible to define another \onerst{} sequence $\hat\orssym$ for $\hat{\mathcal{I}}$ which is in agreement with $\orssym$. Therefore, a parametric TA cannot distinguish between the two sequences, since in each position their evaluations are in agreement. This allows to modify the value of the parameter in a given run.  

\begin{lemma}
\label{lm:agreement}
Let $\mathcal{I}(\param) = \bar{\param}$ be an interpretation for $\param$ and let $\orssym = v_0 v_1 \ldots$ be a (possibly infinite) \onerst{} sequence.
For all interpretation $\hat{\mathcal{I}}(\param) = \hat{\param}$ such that $\intprm = \lfloor \hat{\param} \rfloor$ holds and $\bar{\param}$ and $\hat{\param}$ have the same polarity, for all valuation $\hat{v}_0$ that is in complete agreement with $v_0$, we can build a new \onerst{} sequence of valuations $\hat{\orssym} = \hat{v}_0 \hat{v}_1 \ldots$ such that for all position $i \geq 1$ in the sequence, $v_i$, $\hat{v}_i$ are in agreement for interpretations $\mathcal{I}$, $\hat{\mathcal{I}}$.

\end{lemma}

\begin{proof}[Proof of Lemma \ref{lm:agreement}]
The proof is by induction on the length $i$ of prefix $\orssym_i = v_0 v_1 \ldots v_i$ of $\orssym$.

The induction hypothesis is that, if $\hat{v}_0 \hat{v}_1 \dots \hat{v}_{i-1}$ is a \onerst{} sequence where every $\hat{v}_j$ is in agreement with ${v}_j$ for all $j \leq i-1$, then
we can define $\hat{v}_i$ that is in agreement with ${v}_i$ such that $\hat{v}_0 \hat{v}_1 \dots \hat{v}_i$ is also a \onerst{} sequence.

The base case, $i=0$, is trivial, since by hypothesis $\hat{v}_0$ is in complete agreement with $v_0$.

\paragraph{Inductive step} 
Consider $i > 0$.
We assume that for all $j < i$, $\hat{v}_j$ is in agreement with $v_j$.

We separate two cases.
If $v_i$ is not a critical valuation, $\hat{v}_i = \hat{v}_0 + \delta_i$ can trivially be defined by exploiting the properties of time successors.

If, instead, $v_i$ is a critical valuation, then
it is enough to find the value, possibly depending on $\alpha$, for either of the 
clock assignments $\hat{v}_i(x)$ and $\hat{v}_i(y)$ (the other one being then determined),
such that
the same relation $v_i(y)\sim \bar \param$, $\hat{v}_i(y)\sim \hat{\param}$ holds (similarly for $v_i(x)\sim \bar \param$, $\hat{v}_i(x)\sim \hat{\param}$).
We separate two cases.
If $v_{i-1}$ and $v_i$ satisfy the same (parametric) clock constraints (i.e., they are in agreement), then they belong to the same parametric clock region, which must be open.
In this case, since by induction hypothesis $\hat{v}_{i-1}$ is in agreement with $v_{i-1}$ (hence also with $v_i$), there is $0 < \hat{\epsilon}$ such that $\hat{v}_i = \hat{v}_{i-1} + \hat{\epsilon}$ holds and $\hat{v}_i$ satisfies the same constraints as $\hat{v}_{i-1}$, hence it is also in agreement with $v_i$.
If, instead, $v_{i-1}$ and $v_i$ are not in agreement, then we need to consider the various cases of Proposition \ref{prop:critval} regarding the constraints that hold in $v_i$.

Without loss of generality, assume that $v_0(x) = 0$ holds (i..e, $\orssym$ is a \onerst{} sequence for $x$) and let $y_0 = \lfloor v_0(y) \rfloor$, $\beta = \fract{v_0(y)}$ (so $v_0(y) = y_0 + \beta$ holds), and $\hat{v}_0$ be a valuation that is in complete agreement with $v_0$ for $y$, where $\hat{v}_0(y) = y_0 + \alpha$ holds (with $0 < \alpha < 1$).
Notice that, since $\bar{\param}$ and $\hat{\param}$ have the same polarity and $v_0(y)$ and $\hat{v}_0(y)$ are in complete agreement, $\alpha$ and $\beta$ are such that, for any $\sim \in \set{<, =, >}$, the following holds:
\begin{equation}
\label{eq:betaalphaprop}
\beta \sim \fract{\bar{\param}} \text{ if, and only if, } \alpha \sim \fract{\hat{\param}}.
\end{equation}

Assume first that there exists an integer $\const < 2\cmax$ such that 
$v_i(x) = \const \land m=  \const+y_0$ hold---i.e., we consider case \ref{en:critval1} of Proposition \ref{prop:critval}.
Hence, $v_i(y) = y_0 + \const + \beta = \intprm + \beta$ and $\hat{v}_i(y)=y_0+\const+\alpha=\intprm+\alpha$ hold. 
From property \eqref{eq:betaalphaprop}, we have that $v_i(y) \sim \bar{\param}$ holds if, and only if, $\hat{v}_i(y) \sim \hat{\param}$ also holds.

Assume now that there exists an integer $\const < 2\cmax$ such that 
$v_i(y) = \const \land m=  \const - y_0 - 1$ hold,
i.e., we consider case \ref{en:critval2} of Proposition \ref{prop:critval}.
Hence $v_i(x) = \const - y_0 - \beta = \intprm + 1 - \beta$ and $\hat{v}_i(x)=\const - y_0 - \alpha=\intprm + 1 - \alpha$ hold. 
Again, from property \eqref{eq:betaalphaprop}, we have that $v_i(x) \sim \bar{\param}$ holds if, and only if, $\hat{v}_i(x) \sim \hat{\param}$ also holds.

Consider now the case where there is an integer $\const< 2\cmax$
such that 
$\const-1<v_i(x)< \const \land \intprm= \const+y_0$ holds, i.e., we are in case \ref{en:critval3} of Proposition \ref{prop:critval} (notice that, in this case, $\lfloor v_i(x) \rfloor = \const -1< \intprm$ holds, hence $v_i(x) < \bar{\param}$ also does).
Then, $v_i(x) = \const-1+\epsilon$ holds for some $0 < \epsilon < 1$ and $v_i(y) = y_0 + \beta + \const-1+\epsilon = \intprm -1 + \beta+ \epsilon$ also holds.
We need to show that there exists $0 < \hat{\epsilon} < 1$ such that, if $\hat{v}_i(x) = \const-1+\hat{\epsilon}$ holds (hence also $\hat{v}_i(x) < \hat{\param}$ holds), then $\hat{v}_i(y) = y_0 + \alpha + \const-1+\hat{\epsilon} = \intprm -1 + \alpha+ \hat{\epsilon}$ has the same relation with the parameter as ${v}_i(y)$.
Since $v_i$ is a critical valuation and, by hypothesis, $\const \leq \intprm$ holds, then it must be that $v_i(y) > \intprm$ holds, hence, $\beta + \epsilon > 1$ also holds.
Notice also that, since $\epsilon < 1$ holds, then $\fract{\beta + \epsilon} < \beta$ and $\fract{v_i(y)} = \beta + \epsilon -1 = \fract{\beta + \epsilon} < \beta$ hold.
We define $\hat{\epsilon}$ such that $\alpha + \hat{\epsilon} > 1$ holds---hence $\fract{\hat{v}_i(y)} = \alpha + \hat{\epsilon} -1$ holds---and such that $\fract{v_i(y)} \sim \fract{\bar{\param}}$ holds if, and only if, $\fract{\hat{v}_i(y)} \sim \fract{\hat{\param}}$ also holds.
If $\beta \leq \fract{\bar{\param}}$ holds (hence, by property \eqref{eq:betaalphaprop}, $\alpha \leq \fract{\hat{\param}}$ also holds), then it must be $\fract{v_i(y)} < \fract{\bar{\param}}$. In this case, any $0 < \hat{\epsilon} < 1$ such that $\alpha + \hat{\epsilon} > 1$ holds is such that $\fract{\hat{v}_i(y)} = \fract{\alpha + \hat{\epsilon}} < \alpha \leq \fract{\hat{\param}}$ holds.
If, instead, $\beta > \fract{\bar{\param}}$ holds (hence also $\alpha > \fract{\hat{\param}}$ holds), for all $\sim \in \set{<, =, >}$ there is $0 < \epsilon < 1$ such that $\fract{\beta + \epsilon} \sim \fract{\bar{\param}}$ holds.
In all cases, we can find $\hat{\epsilon}$ such that $\fract{\alpha + \hat{\epsilon}} \sim \fract{\hat{\param}}$ also holds.
For example, if $\fract{\beta + \epsilon} = \fract{\bar{\param}}$ holds, then it is enough to define $\hat{\epsilon} = 1 - \alpha + \fract{\hat{\param}}$ so that $\hat{v}_i(y) = m - 1 + \alpha + \hat{\epsilon} = m + \fract{\hat{\param}} = \hat{\param}$ holds.

Consider now the case where there is an integer $\const< 2\cmax$
such that 
$\const<v_i(y)< \const+1 \land \intprm= \const-y_0-1$ holds---i.e., we are in case \ref{en:critval6} of Proposition \ref{prop:critval} (notice that, in this case, $\lfloor v_i(y) \rfloor = \const > \intprm$ and $v_i(y) > \bar{\param}$ hold).
Hence, $v_i(y) = \const+\epsilon$ holds for some $0 < \epsilon < 1$  and $v_i(x) = \const +\epsilon - y_0 - \beta = \intprm + 1 - \beta + \epsilon$ also holds.
We need to show that there exists $0 < \hat{\epsilon} < 1$ such that, if $\hat{v}_i(y) = \const+\hat{\epsilon}$ holds (hence $\hat{v}_i(y) > \hat{\param}$ also holds), then $\hat{v}_i(x) = \const+\hat{\epsilon} - y_0 - \alpha = \intprm +1 - \alpha+ \hat{\epsilon}$ has the same relation with the parameter as ${v}_i(x)$.
Since $v_i$ is a critical valuation and, by hypothesis, $\const > \intprm$ holds,
then it must be that $\intprm < v_i(x) < \intprm + 1$ holds, hence, $\epsilon < \beta$ also holds.
Notice also that, since $\beta < 1$ holds, then ${1 - \beta + \epsilon} = \fract{v_i(x)}$ holds.
We define $\hat{\epsilon}$ such that $1 - \alpha + \hat{\epsilon} < 1$ holds---hence $\fract{\hat{v}_i(x)} = 1 - \alpha + \hat{\epsilon}$ holds---and such that $\fract{v_i(x)} \sim \fract{\bar{\param}}$ holds if, and only if, $\fract{\hat{v}_i(x)} \sim \fract{\hat{\param}}$ also holds.
If $1 - \beta \geq \fract{\bar{\param}}$ holds (hence, by property \eqref{eq:betaalphaprop}, $1- \alpha \geq \fract{\hat{\param}}$ also holds), then it must be $\fract{v_i(y)} > \fract{\bar{\param}}$. In this case, any $0 < \hat{\epsilon} < 1$ such that $1 - \alpha + \hat{\epsilon} < 1$ holds is such that $\fract{\hat{v}_i(y)} = 1 - \alpha + \hat{\epsilon} > 1 - \alpha \geq \fract{\hat{\param}}$ holds.
If, instead, $1 - \beta < \fract{\bar{\param}}$ (and $1 - \alpha < \fract{\hat{\param}}$) holds, for all $\sim \in \set{<, =, >}$ there is $0 < \epsilon < 1$ such that $1 - \beta + \epsilon \sim \fract{\bar{\param}}$ holds.
In all cases, we can find $\hat{\epsilon}$ such that $1 -\alpha + \hat{\epsilon} \sim \fract{\hat{\param}}$ also holds.
For example, if $1 - \beta + \epsilon = \fract{\bar{\param}}$ holds, then it is enough to define $\hat{\epsilon} = \fract{\hat{\param}} + \alpha - 1$ so that $\hat{v}_i(y) = m + 1 - \alpha + \hat{\epsilon} = m + \fract{\hat{\param}} = \hat{\param}$ holds.

Cases \ref{en:critval4} and \ref{en:critval5} of Proposition \ref{prop:critval} are similar.
\todoMR{}{i case 4 e 5 sono proprio quelli che possono valere in locale, i 3 e 6, da quel che dicevamo anche sopra, valgono da soli. Non so se questo può avere qualche impatto... 4 e 5 possono anche valere con 1 e 2...}
\todoMR{}{considerare se è il caso di dimostrare, invece che il caso 6, il caso 4, che è più diverso da 3.}
\end{proof}

\todoPL{}{aggiunta proposizione per spezzare il lemma. }
\todoMR{}{Si era detto che magari si poteva racchiudere tutto in una tabella, ma forse non serve, da ridiscutere.}
The following immediate proposition considers the case  of two clock valuations $v_1, v_2$ such that in $v_1$ one of the two clocks of $X$ (say, $z_1$) is reset and $v_2 = v_1 + \delta$ holds for some $\delta \in \Real_{>0}$ and lists the possible values of the integer parts of the clocks in $v_1,v_2$. 
 
\begin{proposition}\label{pr2}
	Let $\mathcal{I}(\param) = \bar{\param}$ be an interpretation for $\param$, with $\intprm$ the integer part of $\bar{\param}$.
	Let $v_1, v_2$ be two clock valuations over set of clocks $X = \set{x, y}$ such that $v_1(z_1) = 0$ holds for some $z_1 \in X$, $v_1(z_2) > 0$ holds for $z_2 \in X - \set{z_1}$ and $v_2 = v_1 + \delta$ holds for some $\delta \in \Real_{>0}$, where also $\fract{v_2(z_1)} > 0$ and $\fract{v_2(z_2)} > 0$ hold.
	Let 	$z_{2,0} = \lfloor v_1(z_2) \rfloor$, $\const_1 = \lfloor v_2(z_1) \rfloor$  and $\const_2 = \lfloor v_2(z_2) \rfloor$. 
	The values $\const_1,\const_2,z_{2,0}$ verify either one of the following conditions: 
	
\noindent	$\const_2 = z_{2,0} + \const_1$ or 
	$\const_2 = z_{2,0} + \const_1 +1$.
\end{proposition}
The following lemma establishes, given two clock valuations $v_1, v_2$ verifying the previous Proposition, how the fractional parts of the values of clocks in $v_2$ are related to those of the clocks in $v_1$ depending also on the values of the integer parts of the clocks.
In particular, cases 1-4 define the shape of the fractional part of clock $z_2$ (i.e., the one that is not reset in $v_1$) depending on $v_2(z_1)$ being critical or not and on $\const_2$ being equal to $ z_{2,0} + \const_1$ or to  $z_{2,0} + \const_1 +1$; cases 5-8 define that of clock $z_1$ (i.e., the one that is reset in $v_1$) depending on $v_2(z_2)$ being critical or not and either one of the above cases for  $\const_2$. 
\todoPL{}{enunciato nuovo}
\begin{lemma}\label{lm:fracvalue}
	Let $\bar{\param}, v_1, v_2,\delta,m, 	z_{2,0},\const_1,  \const_2$ as in Proposition~\ref{pr2}, and let $b = \fract{v_1(z_2)}$. The following properties hold:
\begin{enumerate}
	
	\item 
	\label{lm4case:1}
	$\const_1 \neq \intprm$ and $\const_2 = z_{2,0} + \const_1$ $\Longrightarrow$ $\exists 0 \leq \epsilon < 1-b \,\mid\, \fract{v_2(z_2)} = b + \epsilon$

	\item 
	\label{lm4case:2}
	$\const_1 \neq \intprm$ and $\const_2 = z_{2,0} + \const_1 +1$ $\Longrightarrow$ $\exists 0 \leq \epsilon < b \,\mid\, \fract{v_2(z_2)} = \epsilon$

	\item
	\label{lm4case:3}
	$\const_1 = \intprm$ and $\const_2 = z_{2,0} + \const_1$ $\Longrightarrow$
	\begin{enumerate}
		\item 
		\label{lm4case:3a}
		$v_2(z_1) \sim \bar{\param}$ and $\sim \in \set{=, >}$ $\Longrightarrow$ $\exists 1-(b + \fract{\bar{\param}})> \epsilon \sim 0 \,\mid\,  \fract{v_2(z_2)} = b + \fract{\bar{\param}} + \epsilon$
		
		\item
		\label{lm4case:3b}
		$v_2(z_1) < \bar{\param}$ $\Longrightarrow$  $\exists 0 \leq \epsilon < \fract{\bar{\param}} \,\mid\,  \fract{v_2(z_2)} = b + \epsilon < \min(1,b+\fract{\bar{\param}})$
		
	\end{enumerate}

	\item
	\label{lm4case:4}
	$\const_1 = \intprm$ and $\const_2 = z_{2,0} + \const_1 + 1$ $\Longrightarrow$
	\begin{enumerate}
		\item
		$1-b < \fract{\bar{\param}}$ $\Longrightarrow$
			
		\begin{enumerate}
			
			\item 
			\label{lm4case:4ai}
			$v_2(z_1) \sim \bar{\param}$ and $\sim \in \set{=, >}$ $\Longrightarrow$ $\exists  1 - \fract{\bar{\param}} > \epsilon \sim 0 \,\mid\, \fract{v_2(z_2)} = b - (1 - \fract{\bar{\param}}) + \epsilon < b$
			
			\item 
			\label{lm4case:4aii}
			$v_2(z_1) < \bar{\param}$ $\Longrightarrow$ $\exists 0 \leq \epsilon < b - (1 -\fract{\bar{\param}})  \,\mid\,   \fract{v_2(z_2)} = \epsilon$
			
		\end{enumerate}
		
		\item
		\label{lm4case:4b}
		$1-b \geq \fract{\bar{\param}}$ $\Longrightarrow$ $\exists 0 \leq \epsilon < b \,\mid\,  \fract{v_2(z_2)} = \epsilon$
		
	\end{enumerate}

	\item 
	\label{lm4case:5}
	$\const_2 \neq \intprm$ and $\const_1 = \const_2 - z_{2,0} -1$ $\Longrightarrow$ $\exists 0 \leq \epsilon < b \,\mid\, \fract{v_2(z_1)} = 1 - b + \epsilon$

	\item 
	\label{lm4case:6}
	$\const_2 \neq \intprm$ and $\const_1 = \const_2 - z_{2,0}$ $\Longrightarrow$ $\exists 0 \leq \epsilon < 1-b  \,\mid\,  \fract{v_2(z_1)} = \epsilon$

	\item 
	\label{lm4case:7}
	$\const_2 = \intprm$ and $\const_1 = \const_2 - z_{2,0} -1$ $\Longrightarrow$
	
	\begin{enumerate}
		
		\item 
		\label{lm4case:7a}
		$v_2(z_2) \sim \bar{\param}$ and $\sim \in \set{=, >}$ $\Longrightarrow$ $\exists b - \fract{\bar{\param}} > \epsilon \sim 0  \,\mid\,  \fract{v_2(z_1)} = 1 - b + \fract{\bar{\param}} + \epsilon$
		
		\item
		\label{lm4case:7b}
		$v_2(z_2) < \bar{\param}$ $\Longrightarrow$ $\exists 0 \leq \epsilon < \fract{\bar{\param}}  \,\mid\,  \fract{v_2(z_1)} = 1 - b + \epsilon < \min(1, 1-b+ \fract{\bar{\param}})$
		
	\end{enumerate}

	\item
	\label{lm4case:8}
	$\const_2 = \intprm$ and $\const_1 = \const_2 - z_{2,0}$ $\Longrightarrow$
	\begin{enumerate}
		\item
		$b < \fract{\bar{\param}}$ $\Longrightarrow$
		
		\begin{enumerate}
			
			\item 
			\label{lm4case:8ai}
			$v_2(z_2) \sim \bar{\param}$ and $\sim \in \set{=, >}$ $\Longrightarrow$ $\exists 1-\fract{\bar{\param}} > \epsilon \sim 0  \,\mid\,  \fract{v_2(z_1)} = 1- b - (1- \fract{\bar{\param}}) + \epsilon < 1-b$
			
			\item 
			\label{lm4case:8aii}
			$v_2(z_2) < \bar{\param}$ $\Longrightarrow$ $\exists 0 \leq \epsilon < 1-b-(1-\fract{\bar{\param}}) \,\mid\,  \fract{v_2(z_1)} = \epsilon$
			
		\end{enumerate}
		
		\item
		\label{lm4case:8b}
		$b \geq \fract{\bar{\param}}$ $\Longrightarrow$ $\exists 0 \leq \epsilon < 1-b  \,\mid\,  \fract{v_2(z_1)} = \epsilon$.
		
	\end{enumerate}

\end{enumerate}

\begin{figure}[!tp]
\centering
\includegraphics[scale=0.55]{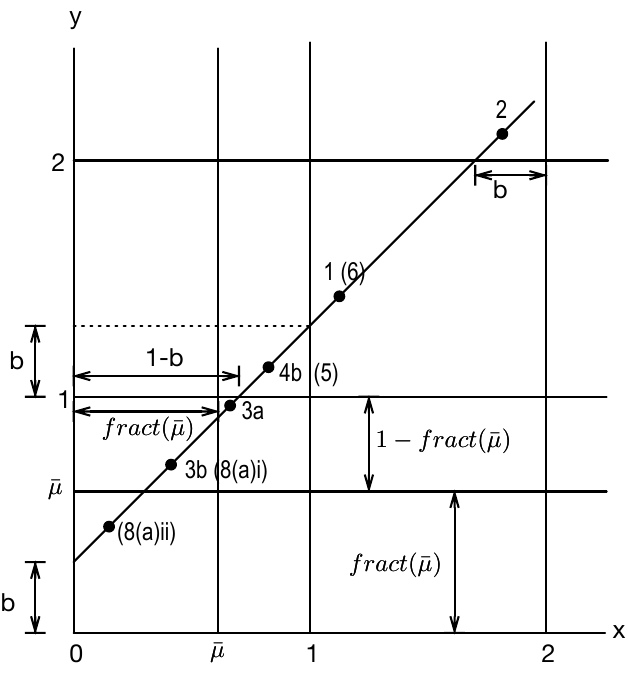}
\hspace{0.2cm}
\includegraphics[scale=0.55]{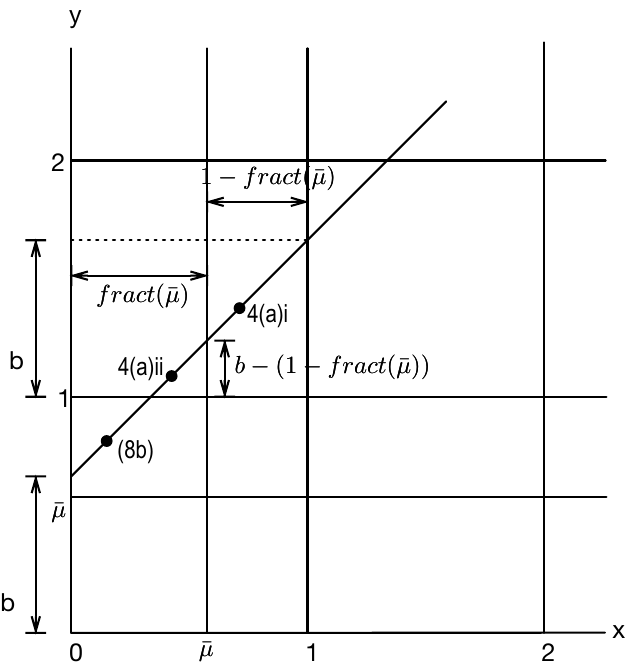}
\includegraphics[scale=0.55]{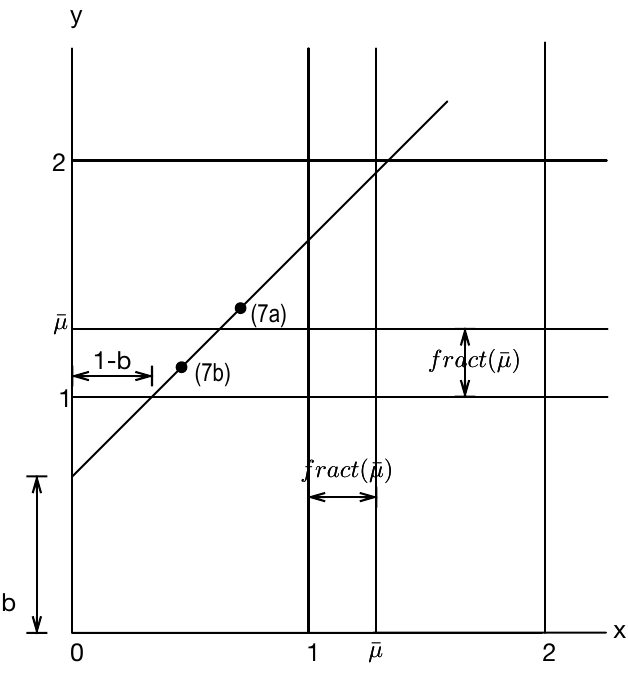}
\caption{Graphical depiction of cases of Lemma \ref{lm:fracvalue}, where $z_1 = x$ and $z_2 = y$. Cases in which the fractional value of $z_1$ is of interest are enclosed in brackets (e.g., (5) in the top left figure and (7a) in the bottom figure).}
\label{fig:Lemmafracvalue}
\end{figure}

\begin{proof}
Figure \ref{fig:Lemmafracvalue}
provides a graphical depiction of the cases enumerated by the statement of Lemma \ref{lm:fracvalue}.
It is easy to see that either $\const_2 = z_{2,0} + \const_1$ or $\const_2 = z_{2,0} + \const_1 + 1$ must hold. \todoPL{}{By Proposition ... it follows that ...}

In all cases we have that $\delta = \const_1 + \varepsilon_1 = v_2(z_1)$ and $v_2(z_2) = \const_2+\varepsilon_2 = z_{2,0} + b + \delta = z_{2,0} + b + \const_1 + \varepsilon_1$ hold for some $\varepsilon_1, \varepsilon_2 \geq 0$.

In cases \ref{lm4case:1}-\ref{lm4case:4} we study the value of $\fract{v_2(z_2)}$---i.e., of $\varepsilon_2$.

Let us consider case \ref{lm4case:1}.
 Since $c_1 \neq \intprm$ holds, then $\bar{\param} < \const_1$ or $\bar{\param} > \const_1+1$ hold (recall that we are assuming that $\bar{\param} < 2\cmax$ and $\bar{\param} \notin \Natgez$ hold).
Since $\const_2 = z_{2,0} + \const_1$ holds, then so does $\varepsilon_2 = b + \varepsilon_1 < 1$, so also $\varepsilon_1< 1-b$ holds.

In case \ref{lm4case:2}, instead, $\const_2 = z_{2,0} + \const_1 + 1$ holds, so $b + \varepsilon_1 = 1 + \epsilon \geq 1$
holds, and also $\varepsilon_2 = \epsilon = b +\varepsilon_1 - 1 < b$ holds, since $\varepsilon_1 < 1$ does.

In case \ref{lm4case:3}, since it holds that $\const_1 = \intprm$, then if $v_2(z_1) \geq \bar{\param}$ holds (case \ref{lm4case:3a}), then $\varepsilon_1 = \fract{\bar{\param}} + \epsilon$ holds, (with $\epsilon = 0$ if $v_2(z_1) = \bar{\param}$ and $\epsilon > 0$ if $v_2(z_1) > \bar{\param}$).
Hence, it holds that $\varepsilon_2 = b + \fract{\bar{\param}} + \epsilon < 1$ (and $\epsilon < 1- (b + \fract{\bar{\param}})$).
If, instead, $v_2(z_1) < \bar{\param}$ holds (case \ref{lm4case:3b}), the situation is similar to the one in case \ref{lm4case:1}, so $\fract{v_2(z_2)} = \varepsilon_2 = b + \varepsilon_1$, except that it must hold that $\varepsilon_1 < \fract{\bar{\param}}$, hence also $b + \varepsilon_1 < b + \fract{\bar{\param}}$, in addition to $b + \varepsilon_1 < 1$.

In case \ref{lm4case:4}, again $b + \varepsilon_1 \geq 1$ (hence $\varepsilon_1 \geq 1 - b$ and $\varepsilon_1 = 1 - b + \epsilon$ for some $\epsilon \geq 0$) holds and also $\varepsilon_2 = b + \varepsilon_1 - 1$.
If $1-b < \fract{\bar{\param}}$ holds, then both $\varepsilon_1 < \fract{\bar{\param}}$ (and $v_2(z_1) < \bar{\param}$) and $\varepsilon_1 \geq \fract{\bar{\param}}$ (and $v_2(z_1) \geq \bar{\param}$) can hold.
If $\varepsilon_1 \geq \fract{\bar{\param}}$ holds (case \ref{lm4case:4ai}), then $\varepsilon_2 = b + \fract{\bar{\param}} - 1 + \epsilon = b - (1 - \fract{\bar{\param}}) + \epsilon < b$ holds since $\varepsilon_1 < 1$ does.
If, instead, $\varepsilon_1 < \fract{\bar{\param}}$ holds (case \ref{lm4case:4aii}), then $\varepsilon_2 = b + \varepsilon_1 - 1 = \epsilon < b + \fract{\bar{\param}} - 1 = b - (1 - \fract{\bar{\param}})$ holds.
If $1-b \geq \fract{\bar{\param}}$ holds (case \ref{lm4case:4b}), then also $\varepsilon_1 \geq \fract{\bar{\param}}$ and $v_2(z_1) \geq \bar{\param}$ hold, and so does $\varepsilon_2 = b + \varepsilon_1 - 1 = \epsilon < b$, since $\varepsilon_1 < 1$ holds.

In cases \ref{lm4case:5}-\ref{lm4case:8} we study the value of $\fract{v_2(z_1)}$---i.e., of $\varepsilon_1$.
Notice that $v_2(z_1) = \const_1 + \varepsilon_1 = \const_2 +\varepsilon_2 - z_{2,0} -b$ holds.

Consider case \ref{lm4case:5}.
Since $\const_1 = \const_2 - z_{2,0} - 1$ holds, then $\varepsilon_1 = 1 - b + \varepsilon_2$ holds, for $\varepsilon_2 = \epsilon < b$, since $\varepsilon_1 < 1$ holds.

In case \ref{lm4case:6}, $\varepsilon_1 = \varepsilon_2 - b$ holds, so it must hold that $\varepsilon_2 \geq b$ and $\varepsilon_1 = \epsilon < 1- b$, since $\varepsilon_2 < 1$ holds.

In case \ref{lm4case:7}, since $v_2(z_2) = \intprm + \varepsilon_2$ holds, we need to separate the cases $\varepsilon_2 \geq \fract{\bar{\param}}$ and $\varepsilon_2 < \fract{\bar{\param}}$.
If $\varepsilon_2 \geq \fract{\bar{\param}}$ holds (case \ref{lm4case:7a}), then $\varepsilon_2 = \fract{\bar{\param}} + \epsilon$ holds, (with $\epsilon = 0$ if $\varepsilon_2 = \fract{\bar{\param}}$ and $\epsilon > 0$ if $\varepsilon_2 > \fract{\bar{\param}}$).
Hence, it holds that $\varepsilon_1 = 1 -b + \fract{\bar{\param}} + \epsilon < 1$ (and $\epsilon < 1- (1-b + \fract{\bar{\param}})$).
If $\varepsilon_2 < \fract{\bar{\param}}$ holds (case \ref{lm4case:7b}), then this is similar to case \ref{lm4case:5}, and $\varepsilon_1 = 1 - b + \varepsilon_2$ holds, except that it must also hold $1 - b + \varepsilon_2 < 1 - b + \fract{\bar{\param}}$, in addition to $1 - b + \varepsilon_2 < 1$.

In case \ref{lm4case:8}, $\varepsilon_1 = \varepsilon_2 - b \geq 0$ holds, hence also $\varepsilon_2 \geq b$.
If $b < \fract{\bar{\param}}$ holds, then both $\varepsilon_2 < \fract{\bar{\param}}$ (and $v_2(z_2) < \bar{\param}$) and $\varepsilon_2 \geq \fract{\bar{\param}}$ (and $v_2(z_2) \geq \bar{\param}$) can hold.
If $\varepsilon_2 \geq \fract{\bar{\param}}$ holds (case \ref{lm4case:8ai}), then $\varepsilon_1 = \fract{\bar{\param}} + \epsilon - b =  1- b - (1 - \fract{\bar{\param}}) + \epsilon < 1-b$ since $\varepsilon_2 < 1$ holds.
If, instead, $\varepsilon_2 < \fract{\bar{\param}}$ holds (case \ref{lm4case:8aii}), then $\varepsilon_1 = \varepsilon_2 - b = \epsilon < \fract{\bar{\param}} - b = 1 - b - (1 - \fract{\bar{\param}})$ hold.
If $b \geq \fract{\bar{\param}}$ holds (case \ref{lm4case:8b}), then also $\varepsilon_2 \geq \fract{\bar{\param}}$ and $v_2(z_2) \geq \bar{\param}$ hold, and so does $\varepsilon_1 = \varepsilon_2 - b = \epsilon < 1-b$, since $\varepsilon_2 < 1$ holds.
\end{proof}

\end{lemma}

Given an interval $I = (e_1, e_2)$ (with $e_1 \leq e_2$), we indicate with $\leftend{I}$ (resp., $\rightend{I}$) the left (resp., right) endpoint of $I$, that is $e_1$ (resp., $e_2$).
Notice that $\ZA_{\bar{\param}}, \ZB_{\bar{\param}}, \ZC_{\bar{\param}}, \ZD_{\bar{\param}}, \ZE_{\bar{\param}}$ are all intervals in $(0,1)$, so, for example, we have that $\leftend{\ZC_{\bar{\param}}} = \ell_{\bar{\param}}$ and $\rightend{\ZC_{\bar{\param}}} = 1-\ell_{\bar{\param}}$.

\begin{figure}[!tp]
\centering
\includegraphics[scale=0.55]{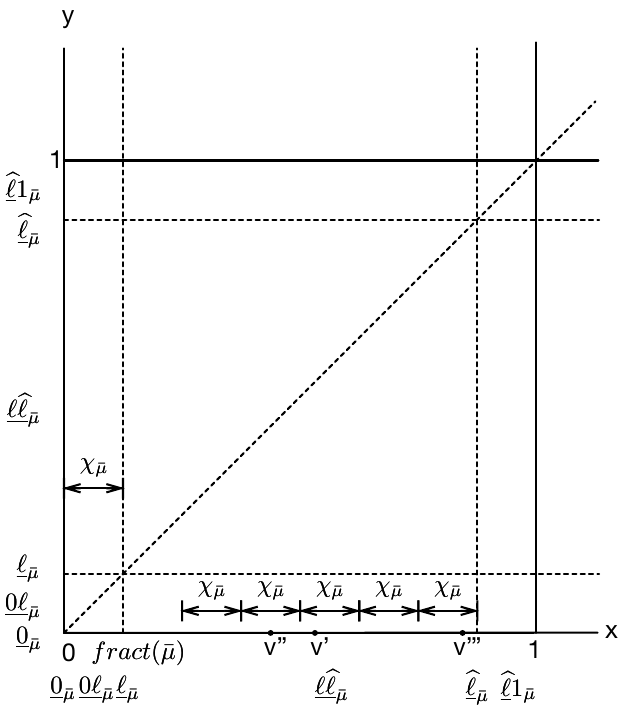}
\hspace{0.2cm}
\includegraphics[scale=0.55]{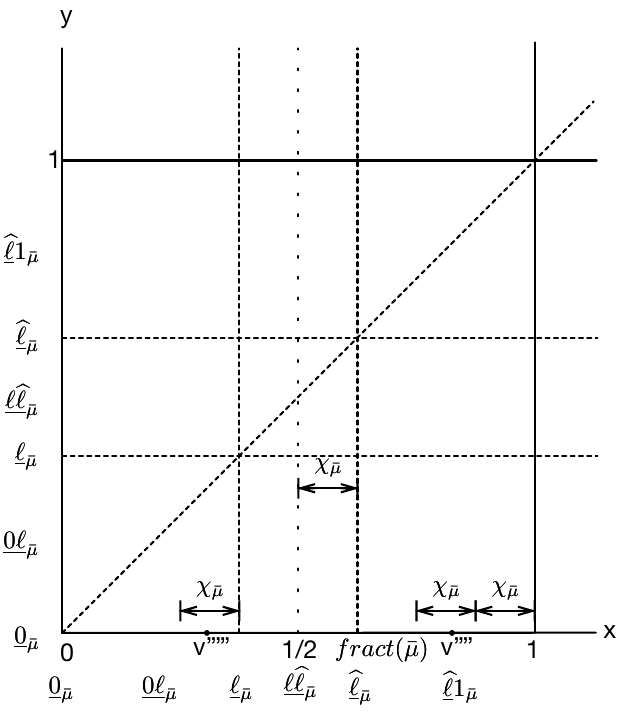}
\caption{Examples of values for $\LowK{v(x),\chi_{\bar{\param}}}$. In the figure on the left, the polarity is negative, and it holds that $\LowK{v'(x),\chi_{\bar{\param}}} = 3$, $\LowK{v''(x),\chi_{\bar{\param}}} = 4$, $\LowK{v'''(x),\chi_{\bar{\param}}} = 1$. In the figure on the right, instead, the polarity is positive (hence value $\frac{1}{2}$ has been highlighted, to show the definition of $\chi_{\bar{\param}}$ in this case) and it holds that $\LowK{v''''(x),\chi_{\bar{\param}}} = 2$ and $\LowK{v'''''(x),\chi_{\bar{\param}}} = 1$ (notice that the value of $\LowK{v(x),\chi_{\bar{\param}}}$ always refers to interval $\iota_{v(x)}$).}
\label{fig:Dnk}
\end{figure}

Let $v(z)$ be the value of some clock $z$ such that $\iota_{v(z)} \in \set{\ZA_{\bar{\param}}, \ZC_{\bar{\param}}, \ZE_{\bar{\param}}}$ holds.
Let $\chi \in \Real_{>0}$ be such that $\chi < \rightend{\iota_{v(z)}} - \leftend{\iota_{v(z)}}$ holds.
We indicate with $\LowK{v(z),\chi}$ the value $k$ (with $k \geq 1$) such that $\rightend{\iota_{v(z)}} - k\chi \leq \fract{v(z)} < \rightend{\iota_{v(z)}} - (k -1) \chi$ (see Figure \ref{fig:Dnk} for some examples).
\todoMR{}{ho cambiato la definizione, ho dovuto incrementare $k$ di 1, e ho specificato che $k \ge 1$}
Essentially, $\LowK{v(z),\chi}$ counts how many intervals of length $\chi$ there are between $v(z)$ and the right endpoint of $\iota_{v(z)}$ (including the one in which $v(z)$ resides).
For example, in Figure \ref{fig:Dnk}, $\LowK{v'(x),\chi_{\bar{\param}}} = 3$ holds because $v'(x)$ is in the third interval of length $\chi_{\bar{\param}}$ moving away (i.e., ``down'') from $\rightend{\iota_{v'(x)}}$.
\todoMR{}{la modific impattava sul resto; il problema quindi era la definizione, che è stata cambiata, insieme alle spiegazioni e gli esempi}

Given an interpretation $\mathcal{I}(\param) = \bar{\param}$, we indicate with ${\ellgenneg_{\bar{\param}}}$ the value $\ell_{\bar{\param}}$ and with ${\ellgenpos_{\bar{\param}}}$ the value $\frac{1}{2} - \ell_{\bar{\param}}$.
We also define $\SetZZ^- = \set{\ZC_{\bar{\param}}}$ and $\SetZZ^+ = \set{\ZA_{\bar{\param}}, \ZE_{\bar{\param}}}$; since, by definition, all intervals of $\SetZZ^-$ (resp., $\SetZZ^+$) have the same size (i.e., for all $\ZZ^1, \ZZ^2$ in the set, $\rightend{\ZZ^1} - \leftend{\ZZ^1} = \rightend{\ZZ^2} - \leftend{\ZZ^2}$ holds), we indicate it with $\widiota^-$ (resp., $\widiota^+$).
Notice that the following relations hold: $\widiota^- = 1 -2\ellgenneg_{\bar{\param}} $ and $\widiota^+ = \frac{1 -2\ellgenneg_{\bar{\param}}}{2} $.
If the polarity is negative (resp., positive) let ${\chi}_{\bar{\param}} = \ellgenneg_{\bar{\param}}$,
$\SetZZ = \SetZZ^-$, and $\widiota = \widiota^-$ (resp., ${\chi}_{\bar{\param}} = \ellgenpos_{\bar{\param}}$,
$\SetZZ = \SetZZ^+$, and $\widiota = \widiota^+$).
Figure \ref{fig:Dnk} shows examples of definition of $\chi_{\bar{\param}}$ in cases of positive and negative polarity.

The next lemma shows that, given a run $\runsym$ for a parametric nrtTA $\A$ with one parameter and two clocks, we can always build a run $\runsym'$ of $\A$ that does not include more than $|Q|$ consecutive \onerst{} sequences whose initial configurations are such that the clock valuations are in agreement.

\begin{lemma}
\label{lm:shortening}
Let $\A=(\Sigma, Q, T, q_0, B)$ be a parametric nrtTA with one parameter, whose set of clocks $X$ is such that $|X| = 2$.
Let $\mathcal{I}(\param) = \bar{\param}$ be a parameter evaluation such that there is a {parametric run} $\runsym$ for $\A$ over a timed word $\pair{\pi}{\tau}$.
Let $\runsym$ be of the form $\runsym_{\textit{pref}} \runsym_0 \runsym_1 \ldots \runsym_n \runsym_{\textit{suff}}$ where, for all $0 \leq i \leq n$,
$\runsym_i$ is a \onerst{} sequence for $z_1$.
If all $\runsym_i$ are such that their initial valuations are in agreement,
then there are sequences of configurations $\runsym'_0, \runsym'_1, \ldots, \runsym'_{n'}$ such that $\runsym_{\textit{pref}} \runsym'_0 \runsym'_1 \ldots \runsym'_{n'} \runsym_{\textit{suff}}$ is also a run of $\A$, with $n' \leq |Q|$.

\end{lemma}

\begin{proof}
Let $\runsym_i = C_{i,0}, C_{i, 1}, \ldots, C_{i, n_i}$, with $C_{i,j} = (q_{i,j}, v_{i,j})$.
Assume, by contradiction, that $n > |Q|$ holds.
Then, there must be $0 \leq i'_1 < i'_2 \leq n$ such that $q_{i'_1,0} = q_{i'_2,0}$ (recall that, by hypothesis, also $v_{i'_1,0}(z_1) = v_{i'_2,0}(z_1) = 0$ holds and $v_{i'_1,0}(z_2)$ and $v_{i'_2,0}(z_2)$ are in agreement).
Hence, we can eliminate
the subsequence $\runsym_{i_1} \ldots \runsym_{i_2-1}$ from $\runsym$, and $\runsym_{\textit{pref}} \runsym_0 \ldots \runsym_{i_2} \ldots \runsym_{n} \runsym_{\textit{suff}}$ is still a run for $\A$.
We iterate the procedure until the middle sequence is not longer than $|Q|$. 
\end{proof}

The next theorem shows that, given a parametric nrtTA $\A$ with one parameter and two clocks, there is a value $\alpha$ such that, if there a run $\runsym$ for $\A$ for an interpretation of the parameter that is less than $2\cmax$, there is also a run $\hat{\runsym}$ for $\A$ for an interpretation of the parameter that is of the form $\halffrac + \alpha$, for some $\floorprm < 4\cmax$.
This allows us to handle the case in which the value of the parameter is less than $2\cmax$.
More precisely, we can determine if there is a parametric run for $\A$ with parameter evaluation less than $2\cmax$ simply by checking all values of the parameter of the form $\halffrac + \alpha$.
This, combined with Theorem \ref{thm:over2cmax}, allows us to prove Theorem \ref{thm:nrtTAdec}.

\begin{theorem}
\label{thm:below2cmax}
Let $\A=(\Sigma, Q, T, q_0, B)$ be a parametric nrtTA with one parameter, whose set of clocks $X$ is such that $|X| = 2$.
There exists a value $0 < \alpha < \frac{1}{2}$ such that, for all $\floorprm < 4\cmax$, with $\floorprm \in \Natgez$, if there is a {parametric run} $\runsym$ for $\A$ over a timed word $\pair{\pi}{\tau}$
with  parameter evaluation $\mathcal{I}(\param) = \bar{\param}$ with $\halffrac < \bar{\param} < \frac{\floorprm+1}{2}$, 
then there is also a {parametric run} $\hat \runsym$ for $\A$ over a timed word $\pair{\pi}{\hat \tau}$ such that $\hat{ \mathcal{I}}(\param) = \hat{\param} = \halffrac+\alpha$ holds.
\end{theorem}

\begin{proof}

Let $A$ be $\max\{|Q|, 4\cmax\}$.
Let $\alpha$ be any value less than %
$\frac{1}{4(1+\cmax A)}$.
\todoMR{}{valore aggiustato, controllare che nel prosieguo funzioni bene}
Notice that it holds that $\alpha < \frac{1}{20}$,
since $\cmax$ is at least 1, or the case is trivial.
Assume that there is a {parametric run} $\runsym$ for $\A$ over a timed word $\pair{\pi}{\tau}$ with  parameter evaluation $\mathcal{I}(\param) = 
\bar{\param}$. 
Let $\beta$ be the real number, with $0<\beta<\frac{1}{2}$, such that $\bar\param=\halffrac+\beta$ holds. 
Notice that, if the polarity of $\bar{\param}$ is negative, then $\ell_{\bar{\param}} = \beta$ (and $\ell_{\hat{\param}} = \alpha$) holds; otherwise, $\ell_{\bar{\param}} = 1 - \frac{1}{2} - \beta = \frac{1}{2} - \beta$ (and $\ell_{\hat{\param}} = \frac{1}{2} - \alpha$) holds.
In addition, we have that $\beta = \chi_{\bar{\param}}$ and $\alpha = \chi_{\hat{\param}}$ hold, no matter the polarity.
\todoMR{}{aggiunto questo commento; forse possiamo direttamente usare $\chi_{\bar{\param}}$ e $\chi_{\hat{\param}}$, e dire ``let $\chi_{\hat{\param}}$ be less than...''.}
For simplicity, in the following we ignore the input alphabet---i.e., $\Sigma$ can be assumed to be a singleton. The timed word can thus be represented just by the mapping $\tau$.

Let run $\runsym$ be such that $\runsym = C_0 C_1 C_2\dots$, where every configuration $C_i$ is $(q_i,v_i)$.
Let $\Xi = v_0 v_1 v_2 \dots$ be the corresponding sequence of clock valuations.
We can see sequence $\Xi$ as a (possibly finite) sequence of \onerst{} sequences $\orssym_1 \orssym_2 \orssym_3 \ldots$ where each \onerst{}  sequence $\orssym_i$ contains a finite number $n_i+1$ of valuations $v_{0,i} v_{1,i} \ldots v_{n_i, i}$, except possibly the last one (which could be infinite).
We need to show that we can build a new parametric run $\hat{\runsym} = \hat{C}_0 \hat{C}_1 \hat{C}_2\dots $ for $\A$ such that $\hat{ \mathcal{I}}(\param) = \hat{\param} = \halffrac+\alpha$ holds.
Similarly to $\runsym$, we call $\hat{\Xi} = \hat{v}_0 \hat{v}_1 \hat{v}_2 \dots$ the corresponding sequence of clock valuations which we see as a (possibly finite) sequence of \onerst{} sequences $\hat{\orssym}_1 \hat{\orssym}_2 \hat{\orssym}_3 \ldots$.
To obtain the desired result, it is enough to show that we can build $\hat{\Xi}$ such that each clock valuation $\hat{v}_i$ satisfies the same clock constraints as $v_i$, for all $i \in \Natgez$.

First of all, thanks to Lemma \ref{lm:shortening}, we can assume, without loss of generality, that $\runsym$ does not include a sequence of \onerst{} sequences $\orssym_i \orssym_{i+1} \ldots \orssym_{i+n-1}$ such that all $v_{0,j}$ (with $i \leq j \leq i+n-1$) are in agreement and $n > |Q|$ holds.

\begin{figure}[tb]
\centering
\includegraphics[width=\columnwidth]{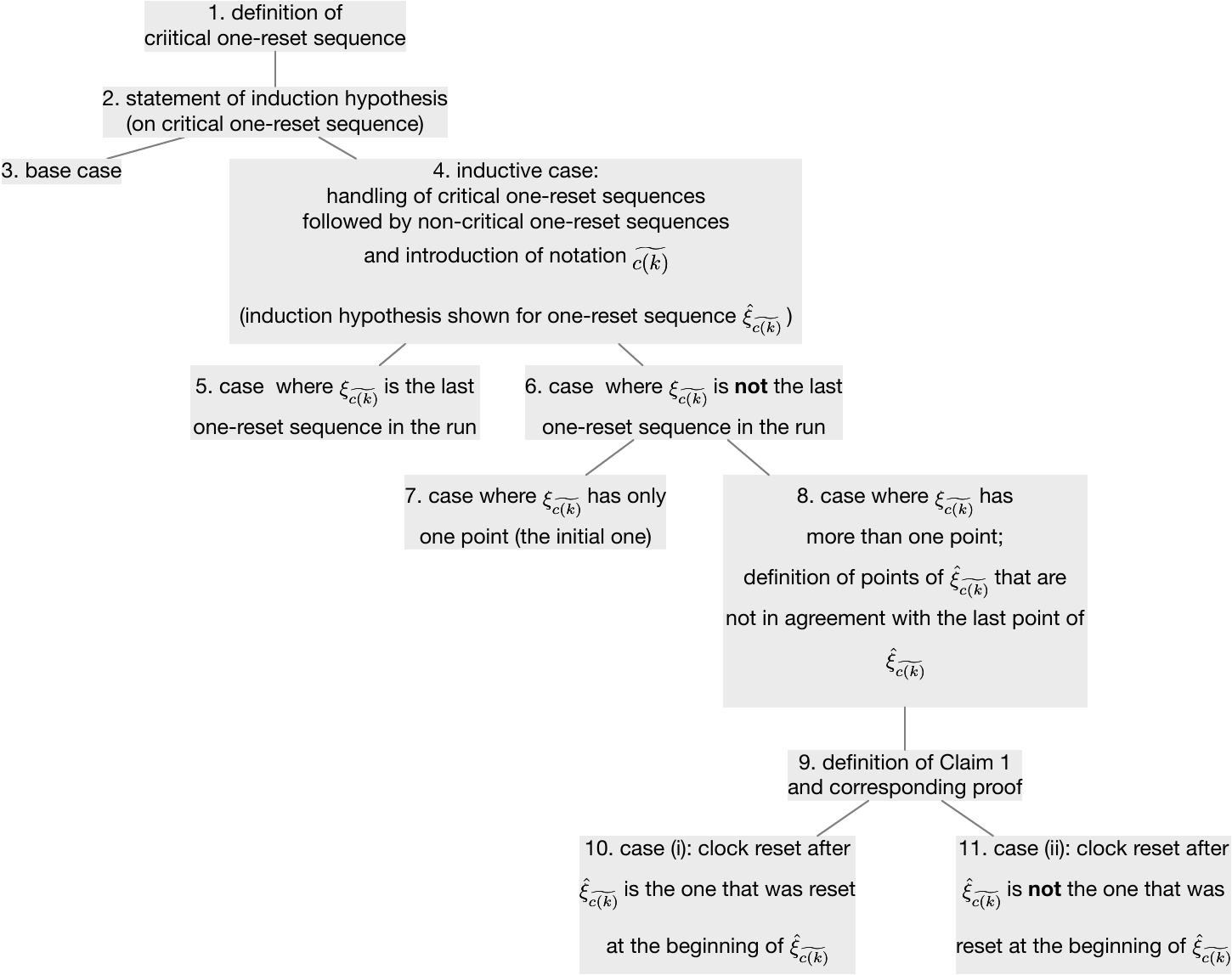}
\caption{Overall structure of the proof.}
\label{fig:proofStructureGeneral}
\end{figure}

The proof is by induction on the number of \onerst{} sequences.
Since the proof is rather articulated, and it deals with many cases and sub-cases, Figure~\ref{fig:proofStructureGeneral} provides a graphical depiction of its structure to help the reader follow the various steps.
To this end, Figure~\ref{fig:proofStructureGeneral} numbers the different points in the proof, which are used in the text to indicate when the discussion of that point begins in the proof.
As mentioned above, the proof is by induction on \onerst{} sequences; it starts (steps 1 and 2) by introducing some definitions, and the induction hypothesis.
The base case (step 3) is rather simple, but the inductive case (which starts with step 4) requires handling various situations and sub-cases.
In particular, step 4 identifies a specific \onerst{} sequence, which we indicate as $\hat{\orssym}_{\widetilde{\crtind{k}}}$, which will be the object of the analysis in the subsequent steps.
The rest of the proof deals with various cases depending on the form of $\hat{\orssym}_{\widetilde{\crtind{k}}}$.
Notice that steps 11 and 12 are themselves structured in various sub-cases, which are depicted in Figure~\ref{fig:proofStructureCasei} and Figure~\ref{fig:proofStructureCaseii}, respectively.
\todoMR{}{se mettiamo delle parti in appendice, occorre menzionare la cosa qui, magari dicendo quali parti (per esempio il punto 13) sono in appendice.}

\textbf{Point 1} of Figure \ref{fig:proofStructureGeneral}.
The proof focuses on \onerst{} sequences $\orssym_i$, which we call \emph{\crtonerst}, that have the following characteristics (and which are exemplified in Figure \ref{fig:CrtOneResetSeq}):
\begin{itemize}
\item
for all $z \in X$ it holds that $v_{0,i}(z) = 0$, or

\item
for some $z_1, z_2 \in X$ it holds that $v_{0,i}(z_1) = 0$, $v_{0,i}(z_2) \neq 0$, and %
$\iota_{v_{0,i}(z_2)} \notin \SetZZ$, or

\item
for some $z_1, z_2 \in X$ it holds that $v_{0,i}(z_1) = 0$, $v_{0,i}(z_2) \neq 0$, %
$\iota_{v_{0,i}(z_2)} \in \SetZZ$ and:

   \begin{itemize}
   \item 
   for all $z \in X$ it holds that $v_{0,i-1}(z) = 0$, or

   \item
   $v_{0,i-1}(z_2) = 0$ and $v_{0,i-1}(z_1) \neq 0$, or

   \item
   $v_{0,i-1}(z_1) = 0$ and $v_{0,i-1}(z_2) \neq 0$ and $\iota_{v_{0,i-1}(z_2)} \notin \SetZZ$ or $\iota_{v_{0,i-1}(z_2)} \neq \iota_{v_{0,i}(z_2)}$, or

   \item
   $v_{0,i-1}(z_1) = 0$ and $v_{0,i-1}(z_2) \neq 0$ and $\iota_{v_{0,i-1}(z_2)} \in \SetZZ$ and $\iota_{v_{0,i-1}(z_2)} = \iota_{v_{0,i}(z_2)}$ and $\lfloor v_{0,i-1}(z_2) \rfloor < \lfloor v_{0,i}(z_2) \rfloor$ hold.
   \end{itemize}

\end{itemize}

Notice that the \onerst{} sequences that are \emph{not \crtonerst{}} are those where $v_{0,i-1}(z_1) = 0$ and $v_{0,i-1}(z_2) \neq 0$, $\iota_{v_{0,i-1}(z_2)} \in \SetZZ$ and $\iota_{v_{0,i-1}(z_2)} = \iota_{v_{0,i}(z_2)}$ and $\lfloor v_{0,i-1}(z_2) \rfloor = \lfloor v_{0,i}(z_2) \rfloor$ hold (see Figure \ref{fig:CrtOneResetSeq} for some examples of non-\crtonerst{} \onerst{} sequences).

\begin{figure}[!tp]
\centering
\includegraphics[scale=0.55]{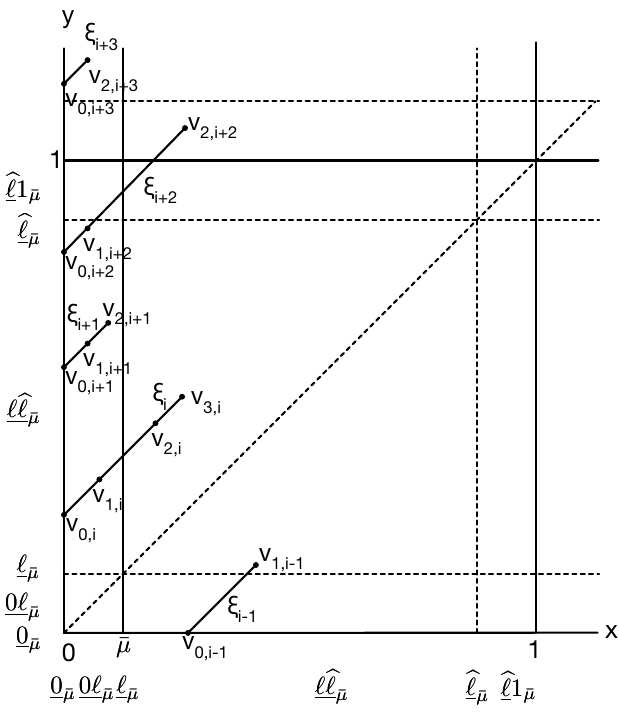}
\caption{Examples of \crtonerst{} (and non-\crtonerst) \onerst{} sequences (in this case, since the polarity is negative, it holds that $\SetZZ = \set{\ZC_{\bar{\param}}}$). Sequences $\orssym_i$ and $\orssym_{i+3}$ are \crtonerst{}, whereas sequences $\orssym_{i+1}$ and $\orssym_{i+2}$ are not. Notice that sequence $\orssym_{i-1}$ might or might not be \crtonerst{}, depending on the shape of $\orssym_{i-2}$.}
\label{fig:CrtOneResetSeq}
\end{figure}

\textbf{Point \hypertarget{bmk:p2}{2}}
of Figure \ref{fig:proofStructureGeneral}.
Consider the $k$-th \crtonerst{} \onerst{} sequence.
Let $\crtind{k} = i$ be the index in $\Xi$ of the $k$-th \crtonerst{} \onerst{} sequence (where $k \leq \crtind{k}$ holds by definition).
We introduce the following induction hypothesis.
For all \crtonerst{} \onerst{} sequence $\orssym_{\crtind{k'}}$ such that $k' \leq k$ holds we have that:
\begin{enumerate}[label=(I\arabic*)]
\item
\label{I1}
$\hat{\orssym}_{\crtind{k'}}$ is also \crtonerst{};

\item
\label{I2}
$\hat{v}_{0,\crtind{k'}}$ is in complete agreement with ${v}_{0,\crtind{k'}}$;

\item
\label{I3}
if $\iota_{{v}_{0,\crtind{k'}}(z_2)} \in \SetZZ$ holds, there is \onerst{} sequence $\crtind{k'}+1$, and ${v}_{0,\crtind{k'}+1}(z_1) = 0$ and ${v}_{0,\crtind{k'}+1}(z_2) > 0$ hold, then
  \begin{enumerate}
  \item 
  \label{I3a}
  if the polarity is negative, then
  \linebreak
  $\LowK{\hat{v}_{0,\crtind{k'}}(z_2), \chi_{\hat{\param}}} = (2\cmax - \lfloor \hat{v}_{0,\crtind{k'}}(z_2) \rfloor)\cdot|Q|+1$ 
  holds;
  \todoMR{}{aggiunto +1, ci saranno da ricontrollare i conti.}
  
  \item 
  \label{I3b}
  if the polarity is positive, then
  \linebreak
  $\LowK{\hat{v}_{0,\crtind{k'}}(z_2), \chi_{\hat{\param}}} = 2(2\cmax - \lfloor \hat{v}_{0,\crtind{k'}}(z_2) \rfloor)$ holds;

  \end{enumerate}

\item
\label{I4}
if $\iota_{{v}_{0,\crtind{k'}}(z_2)} \in \SetZZ$, there is \onerst{} sequence $\crtind{k'}+1$, and ${v}_{0,\crtind{k'}+1}(z_2) = 0$ and ${v}_{0,\crtind{k'}+1}(z_1) > 0$ hold, then $\LowK{\hat{v}_{0,\crtind{k'}}(z_2), \chi_{\hat{\param}}} = 1$ holds;

\item 
\label{I5}
for all clock valuation $\hat{v}_{i,k''}$ with $k'' < k'$ and $0 \leq i \leq n_{k''}$, $\hat{v}_{i,k''}$ and ${v}_{i,k''}$ are in agreement.

\end{enumerate}

\hypertarget{bmk:enoughintervals}{Notice} that, in the induction hypothesis above, the greatest value that $\LowK{\hat{v}_{0,\crtind{k'}}(z_2), \chi_{\hat{\param}}}$ can take is $2\cmax\cdot |Q|+1$ if the polarity is negative, and $4\cmax$ if the polarity is positive.
We remark that in both cases, the number of intervals of length $\alpha = \chi_{\hat{\param}}$ (where $\alpha < \frac{1}{4(1+\cmax A)}$
and $A = \max\{|Q|, 4\cmax\}$ hold) that can fit in the interval(s) of $\SetZZ$ is greater than or equal to the maximum value of $\LowK{\hat{v}_{0,\crtind{k'}}(z_2), \chi_{\hat{\param}}}$ (hence, the induction hypothesis is well-defined).
\todoMR{}{valutare se è il caso di tenere tutti i conti così dettagliati. Comunque controlare che anche nel seguito la modifica al valore di $\alpha$ funzioni}
In particular, if the polarity is negative, then $\alpha = \ell_{\hat{\param}}$ holds, so it also holds that $\rightend{\ZC_{\hat{\param}}} - \leftend{\ZC_{\hat{\param}}} = 1 - 2\alpha$ and $\frac{1 - 2\alpha}{\alpha} = \frac{1}{\alpha} - 2 > 4(1+\cmax A) - 2 = 2 + 4\cmax A > 2\cmax |Q|+1$ (since $A \geq |Q|$ holds).
If, instead, the polarity is positive, then $\alpha = \frac{1}{2} - \ell_{\hat{\param}}$ holds, so it also holds that $\rightend{\ZA_{\hat{\param}}} - \leftend{\ZA_{\hat{\param}}} = \frac{1}{2} - \alpha$ and $\frac{\frac{1}{2} - \alpha}{\alpha} = \frac{1}{2\alpha} - 1 > 2(1+A) - 1 = 1 + 2A > 4\cmax$ (since $A \geq 4\cmax$ holds).

\textbf{Point 3} of Figure \ref{fig:proofStructureGeneral}.
In the base case $k=0$ holds, which, by definition, is such that $\crtind{0} = 0$ (i.e., the first \onerst{} sequence is critical).
If no clock is reset (i.e., $\Xi$ contains a single \onerst{} sequence, so it holds that $\Xi = \orssym_0$) then, by Lemma \ref{lm:agreement}, $\hat{\Xi} = \hat{\orssym}_0$ can be defined so that each valuation $\hat{v}_i$ is in agreement with $v_i$.
Otherwise, if some clock $z_1 \in X$ is eventually reset, by definition of \crtonerst{} \onerst{} sequence, it also holds that $\crtind{1} = 1$ and $\orssym_0 = v_{0,0} v_{1,0} \ldots v_{n_0,0}$.
In this case, for all $0 < i \leq n_0$, for all $\tau > 0$ we can define $\hat{\delta}_0, \hat{\delta}_1, \ldots, \hat{\delta}_{i-1}$ such that $\hat{v}_{i,0}(z_1) = \hat{v}_{i,0}(z_2) = 0 + \sum_{j=0}^{i-1}\hat{\delta}_{j} = \tau$. 
Then, we can define $\hat{\delta}_0, \hat{\delta}_1, \ldots, \hat{\delta}_{n_0} > 0$ such that, for all $0 < i \leq n_0$, $\hat{v}_{i,0}$ is in agreement with ${v}_{i,0}$ and $\hat{v}_{0,1}(z_2) = \hat{v}_{n_0,0}(z_2) + \hat{\delta}_{n_0}$ satisfies the induction hypothesis.
In fact, if it holds that $\hat{v}_{0,1}(z_2) \neq 0$, since, given that $\A$ is an nrtTA, there is no constraint on $\hat{v}_{n_0,0}(z_1) + \hat{\delta}_{n_0}$ (i.e., on the value of clock $z_1$ before the reset), $\hat{\delta}_{n_0}$ can be suitably chosen to satisfy the induction hypothesis (the case where it must hold that $\hat{v}_{0,1}(z_1) = \hat{v}_{0,1}(z_2) = 0$ is trivial).
\todoMR{}{ho leggermente rifrasato questo pezzo, vedere se funziona}

\textbf{Point \hypertarget{bmk:p4}{4}}
of Figure \ref{fig:proofStructureGeneral}.
Let us now consider the $k$-th \crtonerst{} \onerst{} sequence (with $k>0$) of $\Xi$ (i.e., of run $\runsym$).
Let ${\Xi}_k = {\orssym}_0 {\orssym}_1 \ldots {\orssym}_{\crtind{k}-1}{v}_{0,\crtind{k}}$ be the prefix of $\Xi$ that ends in the initial point of the $k$-th \crtonerst{} \onerst{} sequence.
By construction, the corresponding sequence $\hat{\Xi}_k = \hat{\orssym}_0 \hat{\orssym}_1 \ldots \hat{\orssym}_{\crtind{k}-1}\hat{v}_{0,\crtind{k}}$ satisfies the induction hypothesis (notice that the induction hypothesis constrains \onerst{} sequences up to the first point of the $k$-th one).
We need to show that we can extend $\hat{\Xi}_k$ to $\hat{\Xi}_{k+1}$ in a way that preserves the induction hypothesis.

We identify two cases:
\begin{enumerate}
\item 
\label{case:1}
$v_{0,\crtind{k}}(z_2)$ is \emph{not} a point in $\SetZZ$;

\item 
$v_{0,\crtind{k}}(z_2)$ is a point in $\SetZZ$.
\label{case:2}

\end{enumerate}

Notice that, if \onerst{} sequence $\orssym_{\crtind{k}}$ is then followed by at least another \onerst{} sequence $\orssym_{\crtind{k}+1}$, the nature of the latter can be different in cases \ref{case:1} and \ref{case:2}.
More precisely, in case \ref{case:1}, by definition $\orssym_{\crtind{k}+1}$ is also \crtonerst{}, that is, it holds that $\crtind{k}+1 = \crtind{k+1}$.
In case \ref{case:2}, instead, while there is still the possibility that $\orssym_{\crtind{k}+1}$ is also \crtonerst{}, it is also possible that $\orssym_{\crtind{k}}$ is followed by one or more non-\crtonerst{} \onerst{} sequences.
Let us consider the case where $\orssym_{\crtind{k}+1}$ is non-\crtonerst{} (see the left-hand side of Figure \ref{fig:Compacting} for an example).
By definition, ${v}_{0,\crtind{k}+1}$ is in complete agreement with ${v}_{0,\crtind{k}}$.
If there is another \onerst{} sequence $\orssym_{\crtind{k}+2}$ that is non-\crtonerst{}, then again, by definition, ${v}_{0,\crtind{k}+2}$ is in complete agreement with ${v}_{0,\crtind{k}+1}$ (hence also with ${v}_{0,\crtind{k}}$).
By Lemma \ref{lm:shortening}, the number $R$ of non-\crtonerst{} \onerst{} sequences $\orssym_{\crtind{k}+1}, \orssym_{\crtind{k}+2}, \ldots, \orssym_{\crtind{k}+R}$ that follow $\orssym_{\crtind{k}}$ (which are all such that ${v}_{0,\crtind{k}+i}$, with $1 \leq i \leq R$, is in complete agreement with ${v}_{0,\crtind{k}}$) is less than $|Q|$ (i.e., $R < |Q|$ holds).
Since by hypothesis $z_2$ is never reset along $\orssym_{\crtind{k}}, \orssym_{\crtind{k}+1}, \ldots, \orssym_{\crtind{k}+R}$, for all ${v}_{j,\crtind{k}+i}$, with $0 \leq i < R$ and $0 \leq j \leq n_{\crtind{k}+i}$, the same constraints $\const < {v}_{j,\crtind{k}+i}(z_2) < \const+1$ (with $\const < 2\cmax$) and ${v}_{j,\crtind{k}+i}(z_2) \sim \bar{\param}$ hold; in addition, it holds that $0 < {v}_{j,\crtind{k}+i}(z_1) < 1$ if $j > 0$ holds (and obviously ${v}_{0,\crtind{k}+i}(z_1) = 0$ holds).
On the other hand, there can be some $j, i$ such that ${v}_{j-1,\crtind{k}+i}(z_1) < \bar{\param}$ and ${v}_{j,\crtind{k}+i}(z_1) \geq \bar{\param}$ hold (i.e., clock $z_1$ surpasses parameter $\param$ along the \onerst{} sequence); notice that, in this case, it must hold that $\intprm = 0$, as ${v}_{j,\crtind{k}+i}(z_1) < 1$ holds, hence also $\fract{\bar{\param}} = \bar{\param}$ and $\fract{\hat{\param}} = \hat{\param}$ hold.
The same constraints must also hold for valuations $\hat{v}_{j,\crtind{k}+i}$.
By Lemma \ref{lm:fracvalue}, case \ref{lm4case:3a}
(and the fact that $\fract{\bar{\param}} = \bar{\param}$ and $\fract{\hat{\param}} = \hat{\param}$ hold), for all ${v}_{j,\crtind{k}+i}(z_1)$ such that ${v}_{j,\crtind{k}+i}(z_1) \geq \bar{\param}$ holds, the value of $\fract{{v}_{j,\crtind{k}+i}(z_2)}$ has the form $\fract{{v}_{0,\crtind{k}+i}(z_2)} + \bar{\param} + \epsilon$, for some $\epsilon \geq 0$; similarly, $\fract{\hat{v}_{j,\crtind{k}+i}(z_2)}$ has the form $\fract{\hat{v}_{0,\crtind{k}+i}(z_2)} + \hat{\param} + \hat{\epsilon}$.
Consider now \onerst{} sequence $\orssym_{\crtind{k}}$.
If, for all $0 \leq j \leq n_{\crtind{k}}$, it holds that ${v}_{j,\crtind{k}}(z_1) < \bar{\param}$, then for any $0 < \hat{\varepsilon} < \hat{\param}$ we can define positive delays $\hat{\delta}_{0,\crtind{k}}, \hat{\delta}_{1,\crtind{k}}, \ldots,  \hat{\delta}_{n_{\crtind{k}}-1,\crtind{k}}$ such that $\hat{v}_{j,\crtind{k}} = \hat{v}_{0,\crtind{k}} + \sum_{h=0}^{j-1} \hat{\delta}_{h,\crtind{k}}$ are in complete agreement with ${v}_{j,\crtind{k}}$ and $\hat{v}_{n_{\crtind{k}},\crtind{k}} = \hat{v}_{0,\crtind{k}} + \hat{\varepsilon}$ holds (i.e., we can make \onerst{} sequence $\hat{\orssym}_{\crtind{k}}$ as short as necessary).
If, instead, it holds that ${v}_{n_{\crtind{k}},\crtind{k}}(z_1) \geq \bar{\param}$, then for any $0 \leq \hat{\varepsilon} < 1-\hat{\param}$, we can define positive delays $\hat{\delta}_{0,\crtind{k}}, \hat{\delta}_{1,\crtind{k}}, \ldots,  \hat{\delta}_{n_{\crtind{k}}-1,\crtind{k}}$ such that for all $0 < j < n_{\crtind{k}}$, $\hat{v}_{j,\crtind{k}} = \hat{v}_{0,\crtind{k}} + \sum_{h=0}^{j-1} \hat{\delta}_{h,\crtind{k}}$ is in complete agreement with ${v}_{j,\crtind{k}}$ and $\hat{v}_{n_{\crtind{k}},\crtind{k}} = \hat{v}_{0,\crtind{k}} + \hat{\param} + \hat{\varepsilon}$ holds, where $\hat{\varepsilon} = 0$ holds if ${v}_{n_{\crtind{k}},\crtind{k}}(z_1) = \bar{\param}$ holds.
In other words, we can again make the length of \onerst{} sequence $\hat{\orssym}_{\crtind{k}}$ as close to $\hat{\param}$ as necessary (see the right-hand side of Figure \ref{fig:Compacting} for an example).
In addition, since ${v}_{n_{\crtind{k}},\crtind{k}}(z_2)$ and ${v}_{0,\crtind{k}+1}(z_2)$ are in complete agreement and ${v}_{0,\crtind{k}+1}(z_1) = 0$ holds, then the delay $\hat{\delta}_{n_{\crtind{k}},\crtind{k}} = \hat{v}_{0,\crtind{k}+1}(z_2) - \hat{v}_{n_{\crtind{k}},\crtind{k}}(z_2)$ between $\hat{v}_{n_{\crtind{k}},\crtind{k}}$ and $\hat{v}_{0,\crtind{k}+1}$ can also be as small as necessary.
As a consequence, delays $\hat{\delta}_{0,\crtind{k}}, \hat{\delta}_{1,\crtind{k}}, \ldots,  \hat{\delta}_{n_{\crtind{k}},\crtind{k}}$ can be defined such that, in one case (i.e., ${v}_{n_{\crtind{k}},\crtind{k}}(z_1) < \bar{\param}$),
it holds that $\hat{v}_{0,\crtind{k}+1}(z_2) = \hat{v}_{0,\crtind{k}}(z_2) + \hat{\varepsilon}'$, and in the other case (i.e., ${v}_{n_{\crtind{k}},\crtind{k}}(z_1) \geq \bar{\param}$) it holds that $\hat{v}_{0,\crtind{k}+1}(z_2) = \hat{v}_{0,\crtind{k}}(z_2) + \hat{\param} + \hat{\varepsilon}'$, for any $\hat{\varepsilon}'$ as small as necessary.
The same reasoning can be repeated for non-\crtonerst{} \onerst{} sequences $\orssym_{\crtind{k}+1}, \ldots, \orssym_{\crtind{k}+R-1}$.
Let $P \le R$ be the number of \onerst{} sequences $i$ among $\orssym_{\crtind{k}}, \ldots, \orssym_{\crtind{k}+R-1}$ such that ${v}_{n_{\crtind{k}+i},\crtind{k}+i}(z_1) \geq \bar{\param}$ holds.
For each of them we can define delays $\hat{\delta}_{j,\crtind{k}+i}$, for $0 \leq j \leq n_{\crtind{k}+i}$ such that each length is $\hat{\param} + \hat{\varepsilon}_i'$, for any $\hat{\varepsilon}_i'$ as small as necessary; hence, we can define the delays such that $\hat{v}_{0,\crtind{k}+R}(z_2) = \hat{v}_{0,\crtind{k}}(z_2) + \hat{\param}P + \hat{\varepsilon}'$ holds, for any $\hat{\varepsilon}'$ as small as necessary (see the right-hand side of Figure \ref{fig:Compacting} for an example in which $R = 3$ and $P = 2$ hold).

\begin{figure}[!tp]
\centering
\includegraphics[scale=0.55]{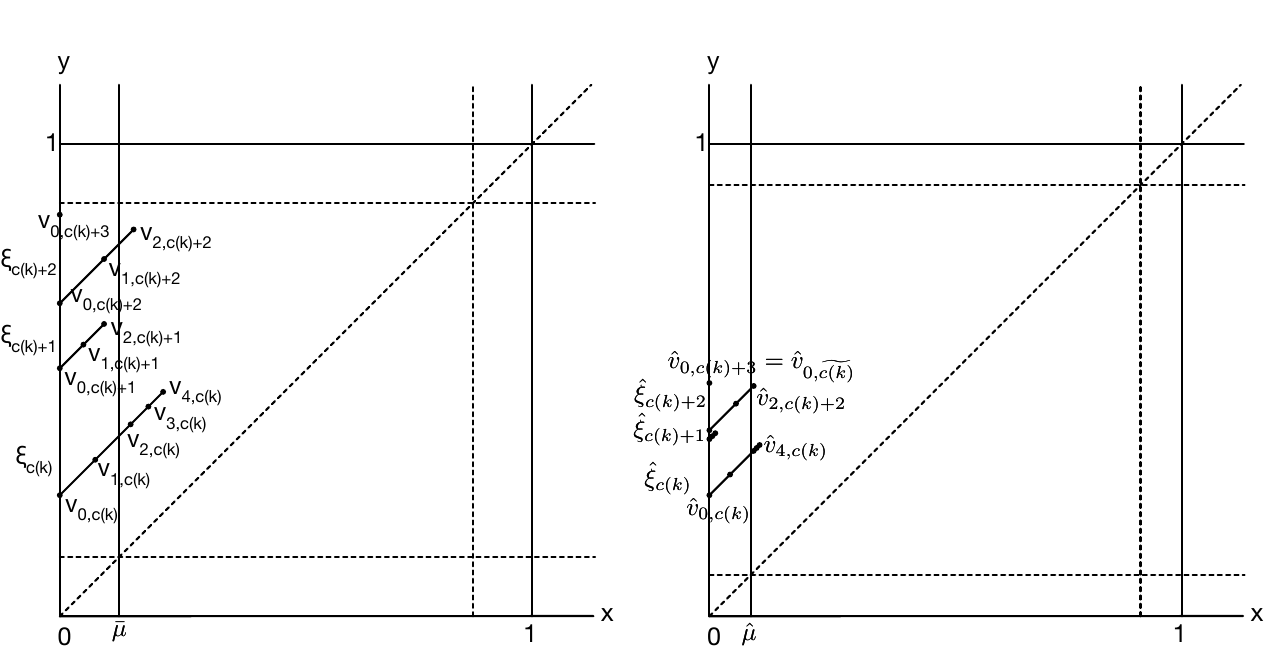}
\caption{Left-hand side: example of \crtonerst{} \onerst{} sequence $\orssym_{\crtind{k}}$ followed by three non-\crtonerst{} sequences (only the first valuation of sequence $\orssym_{\crtind{k}+3}$, $v_{0,\crtind{k}+3}$, is shown). Right-hand side: example of corresponding, ``compressed'' \onerst{} sequences $\hat{\orssym}_{\crtind{k}}$, $\hat{\orssym}_{\crtind{k}+1}$, $\hat{\orssym}_{\crtind{k}+2}$ and first valuation $\hat{v}_{0,\crtind{k}+3}$ of sequence $\hat{\orssym}_{\crtind{k}+3}$; in this case it holds that $\widetilde{\crtind{k}} = \crtind{k}+3$.}
\label{fig:Compacting}
\end{figure}

Notice that, if the polarity is positive and $\orssym_{\crtind{k}}$ starts from a point in an interval of $\SetZZ$ (i.e., $\ZA_{\bar{\param}}$ or $\ZE_{\bar{\param}}$), and it is followed by $R$ non-\crtonerst{} \onerst{} sequences, none of the $R$ sequences $\orssym_{\crtind{k}}, \ldots, \orssym_{\crtind{k}+R-1}$ can be such that clock $z_1$ surpasses $\param$ (i.e., $P = 0$ must hold); in fact, if such a sequence $\orssym_{\crtind{k}+i}$ existed, then ${v}_{n_{\crtind{k}+i},\crtind{k}+i}(z_2)$ would not be in complete agreement with ${v}_{0,\crtind{k}+i}(z_2)$.
Indeed, the only case where $P > 0$ can hold is if $\intprm = 0$ and the polarity is negative, hence $\SetZZ = \set{\ZC_{\bar{\param}}}$, $\bar{\param} = \chi_{\bar{\param}}$ and $\hat{\param} = \chi_{\hat{\param}}$ hold.
\hypertarget{bmk:condDn}{Then},
since we are assuming that $\hat{v}_{0,\crtind{k}}$ satisfies the induction hypothesis, we have that $\LowK{\hat{v}_{0,\crtind{k}}(z_2), \chi_{\hat{\param}}} = (2\cmax - \lfloor \hat{v}_{0,\crtind{k}}(z_2) \rfloor)\cdot|Q|+1$ holds; in other words, it holds that
$1-((2\cmax - \lfloor \hat{v}_{0,\crtind{k}}(z_2) \rfloor)\cdot|Q| + 2)\hat{\param} < \fract{\hat{v}_{0,\crtind{k}}(z_2)} < 1-((2\cmax - \lfloor \hat{v}_{0,\crtind{k}}(z_2) \rfloor)|Q|+1)\hat{\param}$.
Hence, since $P \leq |Q|-1$ is true, it also holds that $\fract{\hat{v}_{0,\crtind{k}}(z_2)} + \hat{\param}P < 1-(2\cmax - \lfloor \hat{v}_{0,\crtind{k}}(z_2) \rfloor -1)|Q|\hat{\param} - 2\hat{\param} \leq 1 - \hat{\param}$ holds, as $\lfloor \hat{v}_{0,\crtind{k}}(z_2) \rfloor \leq 2\cmax - 1$ holds; then, since we have that $\fract{\hat{v}_{0,\crtind{k}+R}(z_2)} = \fract{\hat{v}_{0,\crtind{k}}(z_2)} + \hat{\param}P + \hat{\varepsilon}$, we can choose a suitably small $\hat{\varepsilon}$ so that $\hat{v}_{0,\crtind{k}+R}(z_2)$ is still in $\ZC_{\hat{\param}}$.
\todoMR{}{cambiato i conti per adattarli alle nuove definizioni, dovrebbe ancora tutto tornare}

Consider now cases \ref{case:1} and \ref{case:2} again.
Let us define the following: $\widetilde{\crtind{k}} = \crtind{k}$ in case \ref{case:1}, and $\widetilde{\crtind{k}} = \crtind{k}+R$ in case \ref{case:2} (for example, in Figure \ref{fig:Compacting} it holds that $\widetilde{\crtind{k}} = \crtind{k}+3$).
As mentioned above, in the latter case we can assume that $\hat{v}_{0,\crtind{k}+R}(z_2) = \hat{v}_{0,\crtind{k}}(z_2) + \hat{\param}P + \hat{\varepsilon}'$ holds for an infinitesimal $\hat{\varepsilon}'$, and $P \leq R < |Q|$.
In addition, by construction we have that, if $\orssym_{\widetilde{\crtind{k}}}$ is not the last \onerst{} sequence of $\Xi$, then $\orssym_{\widetilde{\crtind{k}}+1}$ is \crtonerst{}, and it holds that $\crtind{k+1} = \widetilde{\crtind{k}}+1$.

In the rest of the proof we show that we can extend sequence $\widetilde{\hat{\Xi}}_{k} = \hat{\orssym}_0 \hat{\orssym}_1 \ldots \hat{\orssym}_{\widetilde{\crtind{k}}-1}\hat{v}_{0,\widetilde{\crtind{k}}}$
to $\widetilde{\hat{\Xi}}_{k+1}$
\todoMR{}{direi che dovrebbe essere $\widetilde{\hat{\Xi}}_{k+1} = {\hat{\Xi}}_{k+1}$, quindi a quel punto meglio usare direttamente ${\hat{\Xi}}_{k+1}$}
in a way that preserves the induction hypothesis.

The extended sequence $\widetilde{\hat{\Xi}}_{k+1}$ has different forms depending on the form of $\Xi$.
We identify the following two cases:
\begin{enumerate}[label=\alph*.,ref=\alph*]
\item 
\label{case:a}
${\orssym}_{\widetilde{\crtind{k}}}$ is the last \onerst{} sequence of $\Xi$ (which hence contains only a finite number of \onerst{} sequences);

\item
\label{case:b}
there exists in $\Xi$ \crtonerst{} \onerst{} sequence ${\orssym}_{\crtind{k+1}}$.

\end{enumerate}

\textbf{Point 5} of Figure \ref{fig:proofStructureGeneral}.
In {\em case \ref{case:a}}, it holds that ${\Xi} = {\orssym}_0 {\orssym}_1 \ldots {\orssym}_{\crtind{k}-1}{\orssym}_{\widetilde{\crtind{k}}}$,
hence it must also hold that $\hat{\Xi} = \widetilde{\hat{\Xi}}_{k+1} = \hat{\orssym}_0 \hat{\orssym}_1 \ldots \hat{\orssym}_{\crtind{k}-1}\hat{\orssym}_{\widetilde{\crtind{k}}}$.
\todoMR{}{usare ${\hat{\Xi}}_{k+1}$}
Then, by Lemma \ref{lm:agreement}, $\hat{\orssym}_{\widetilde{\crtind{k}}}$ can be defined so that each valuation $\hat{v}_{i,\widetilde{\crtind{k}}}$ is in agreement with $v_{i,\widetilde{\crtind{n}}}$.

\textbf{Point 6} of Figure \ref{fig:proofStructureGeneral}.
In {\em case \ref{case:b}}, by construction \onerst{} sequence $\orssym_{\widetilde{\crtind{k}}+1}$ is also the next \crtonerst{} one, that is, $\widetilde{\crtind{k}}+1 = \crtind{k+1}$ holds, and it also holds that $\widetilde{\hat{\Xi}}_{k+1} = \hat{\orssym}_0 \hat{\orssym}_1 \ldots \hat{\orssym}_{\widetilde{\crtind{k}}-1}\hat{\orssym}_{\widetilde{\crtind{k}}}\hat{v}_{0,\widetilde{\crtind{k}}+1}$.
\todoMR{}{usare ${\hat{\Xi}}_{k+1}$}
We further separate two cases:
\begin{enumerate}[label=(\roman*),ref=(\roman*)]
\item
\label{case:i}
$v_{0,\widetilde{\crtind{k}}+1}(z_1) = 0$ (and $v_{0,\widetilde{\crtind{k}}+1}(z_2) \neq 0$) and
\item
\label{case:ii}
$v_{0,\widetilde{\crtind{k}}+1}(z_2) = 0$ (and $v_{0,\widetilde{\crtind{k}}+1}(z_1) \neq 0$).
\end{enumerate}
Notice that the case where $v_{0,\widetilde{\crtind{k}}+1}(z_1) = v_{0,\widetilde{\crtind{k}}+1}(z_2) = 0$ holds is trivial since by Lemma \ref{lm:agreement} we can define each $\hat{v}_{i,\widetilde{\crtind{k}}}$ such that it is in agreement with $v_{i,\widetilde{\crtind{n}}}$, and then $\hat{v}_{0,\widetilde{\crtind{k}}+1}(z_1) = \hat{v}_{0,\widetilde{\crtind{k}}+1}(z_2) = 0$ trivially satisfies the induction hypothesis.

We separate two cases: $n_{\widetilde{\crtind{k}}} = 0$ (i.e., \onerst{} sequence $\orssym_{\widetilde{\crtind{k}}}$ contains only the reset, there are no other valuations) and  $n_{\widetilde{\crtind{k}}} > 0$.

\textbf{Point 7} of Figure \ref{fig:proofStructureGeneral}.
The case where $n_{\widetilde{\crtind{k}}} = 0$ holds (that is, $\hat{\xi}_{\widetilde{\crtind{k}}}$ has just one point) is straightforward since there are no other valuations between $v_{0,\widetilde{\crtind{k}}}$ and $v_{0,\widetilde{\crtind{k}}+1}$ (i.e., if $v_{0,\widetilde{\crtind{k}}}$ is the $p$-th valuation in $\Xi$, then $v_{0,\widetilde{\crtind{k}}+1}$ is the $p+1$-th).
Indeed, given that $\A$ is an nrtTA, in case \ref{case:i} there is no constraint on ${v}_{0,\widetilde{\crtind{k}}}(z_1) + {\delta}_{0,\widetilde{\crtind{k}}}$ (where ${\delta}_{0,\widetilde{\crtind{k}}}$ is the delay between valuations $v_{0,\widetilde{\crtind{k}}}$ and $v_{0,\widetilde{\crtind{k}}+1}$, i.e., ${v}_{0,\widetilde{\crtind{k}}+1}(z_2) = {v}_{0,\widetilde{\crtind{k}}}(z_2) + {\delta}_{0,\widetilde{\crtind{k}}}$ holds)---i.e., on the value of $z_1$ when the reset occurs---hence neither on $\hat{v}_{0,\widetilde{\crtind{k}}}(z_1) + \hat{\delta}_{0,\widetilde{\crtind{k}}}$; then, $\hat{\delta}_{0,\widetilde{\crtind{k}}}$ can be independently chosen to have $\hat{v}_{0,\widetilde{\crtind{k}}+1}(z_2)$ satisfy the desired constraint (case \ref{case:ii} is similar).

\textbf{Point 8} of Figure \ref{fig:proofStructureGeneral}.
Let us now consider the case in which $n_{\widetilde{\crtind{k}}} > 0$ holds (that is, $\hat{\xi}_{\widetilde{\crtind{k}}}$ has more than one point).
Let $z_r$ be the clock that is reset in valuation ${v}_{0,\widetilde{\crtind{k}}+1}$, where $z_r = z_1$ in case \ref{case:i} and $z_r = z_2$ in case \ref{case:ii}, and let $z_{nr}$ be the other clock.

Let $0 < i' \leq n_{\widetilde{\crtind{k}}}$ be such that $v_{i'-1,\widetilde{\crtind{k}}}$ is not in agreement with $v_{n_{\widetilde{\crtind{k}}},\widetilde{\crtind{k}}}$, whereas for all $i' \leq i'' \leq n_{\widetilde{\crtind{k}}}$, $v_{i'',\widetilde{\crtind{k}}}$ is in agreement with $v_{n_{\widetilde{\crtind{k}}},\widetilde{\crtind{k}}}$ (notice that such $i'$ must exist since $n_{\widetilde{\crtind{k}}} > 0$ and $v_{0,\widetilde{\crtind{k}}}(z_1) = 0$ hold).
By Lemma \ref{lm:agreement}, for all $0 \leq u \leq i'-1$ we can suitably define $\hat{v}_{u,\widetilde{\crtind{k}}}$ so that it is in agreement with ${v}_{u,\widetilde{\crtind{k}}}$.
In addition, if $i' < n_{\widetilde{\crtind{k}}}$ holds, once $\hat{v}_{n_{\widetilde{\crtind{k}}},\widetilde{\crtind{k}}}$ is defined, valuations $\hat{v}_{i',\widetilde{\crtind{k}}}, \hat{v}_{i'+1,\widetilde{\crtind{k}}}, \ldots, \hat{v}_{n_{\widetilde{\crtind{k}}}-1,\widetilde{\crtind{k}}}$ can be suitably distributed between $\hat{v}_{i'-1,\widetilde{\crtind{k}}}$ and $\hat{v}_{n_{\widetilde{\crtind{k}}},\widetilde{\crtind{k}}}$ so that they are all in agreement with $\hat{v}_{n_{\widetilde{\crtind{k}}},\widetilde{\crtind{k}}}$, since for each clock $z \in X$, $\hat{v}_{n_{\widetilde{\crtind{k}}},\widetilde{\crtind{k}}}(z)$ belongs to an open interval of the form $(\const, \const+1)$, $(\intprm, \hat{\param})$ or $(\hat{\param}, \intprm+1)$.
\todoMR{}{modificato, ricontrollare.}

Lemma \ref{lm:agreement} also allows us to define $\hat{v}_{n_{\widetilde{\crtind{k}}},\crtind{k}}$ so that it is in agreement with ${v}_{n_{\widetilde{\crtind{k}}},\widetilde{\crtind{k}}}$.
However, we need $\fract{\hat{v}_{n_{\widetilde{\crtind{k}}},\widetilde{\crtind{k}}}}$ to obey finer constraints, in addition to being in agreement with $\fract{{v}_{n_{\widetilde{\crtind{k}}},\widetilde{\crtind{k}}}}$.
Indeed, we need to show that the induction hypothesis holds in $\hat{v}_{0,\widetilde{\crtind{k}}+1}$, which is the clock valuation that immediately follows $\hat{v}_{n_{\widetilde{\crtind{k}}},\widetilde{\crtind{k}}}$ (i.e., there is $j \in \Natgez$ such that $\hat{v}_j = \hat{v}_{n_{\widetilde{\crtind{k}}},\widetilde{\crtind{k}}}$ and $\hat{v}_{j+1} = \hat{v}_{0,\widetilde{\crtind{k}}+1}$ hold), and in which we assume that clock $z_r$ is reset.
Since $\A$ is an nrtTA, the delay $\hat{\delta}_{n_{\widetilde{\crtind{k}}},\widetilde{\crtind{k}}} = \hat{v}_{0,\widetilde{\crtind{k}}+1}(z_{nr})-\hat{v}_{n_{\widetilde{\crtind{k}}},\widetilde{\crtind{k}}}(z_{nr})$ between the last valuation of \onerst{} sequence $\orssym_{\widetilde{\crtind{k}}}$ and the first valuation of \onerst{} sequence $\orssym_{\widetilde{\crtind{k}}+1}$ can be arbitrary, as long as $\hat{v}_{0,\widetilde{\crtind{k}}+1}(z_{nr})$ satisfies the constraints of the guard of the transition taken in the run.
Hence, we need to show that we can define $\hat{v}_{n_{\widetilde{\crtind{k}}},\widetilde{\crtind{k}}}(z_{nr})$ so that we can suitably choose $\hat{\delta}_{n_{\widetilde{\crtind{k}}},\widetilde{\crtind{k}}}$ such that $\hat{v}_{0,\widetilde{\crtind{k}}+1}(z_{nr}) = \hat{v}_{n_{\widetilde{\crtind{k}}},\widetilde{\crtind{k}}}(z_{nr}) + \hat{\delta}_{n_{\widetilde{\crtind{k}}},\widetilde{\crtind{k}}}$ satisfies the induction hypothesis.

Let $\iota_{\widetilde{\crtind{k}},\bar{\param}} = \iota_{{v}_{n_{\widetilde{\crtind{k}}},\widetilde{\crtind{k}}}(z_{nr})}$ (resp., $\iota_{\widetilde{\crtind{k}}+1,\bar{\param}} = \iota_{{v}_{0,\widetilde{\crtind{k}}+1}(z_{nr})}$) be the interval to which the fractional part of the last (resp., first) valuation of $z_{nr}$ in \onerst{} sequence $\orssym_{\widetilde{\crtind{k}}}$ (resp., $\orssym_{\widetilde{\crtind{k}}+1}$) belongs, and let $\hat{\iota}_{\widetilde{\crtind{k}},\hat{\param}}$ (resp., $\hat{\iota}_{\widetilde{\crtind{k}}+1,\hat{\param}}$) be the corresponding interval referred to $\hat{\param}$.

\todoMR{}{Possibile idea: dare dei nomi mnemonici a ``primo/ultimo punto di \onerst{} sequence corrente'', ecc., al posto di $v_{0,\crtind{k}}$, ecc.}

\textbf{Point 9} of Figure \ref{fig:proofStructureGeneral}.
In the rest of the proof, it is useful to define $\hat{v}_{0,\widetilde{\crtind{k}}+1}(z_{nr})$ so that condition $\hat{\iota}_{\widetilde{\crtind{k}},\hat{\param}} \preceq {\iota}_{\widetilde{\crtind{k}},\bar{\param}}$ holds.
For example, we introduce and prove the following claim (which is exemplified in Figure \ref{fig:notInCompleteAgreement}), which will be useful in the rest of the proof:
\begin{claim}
\label{claim:notCompleteAgreement}
If ${v}_{0,\widetilde{\crtind{k}}+1}(z_{nr})$ and ${v}_{n_{\widetilde{\crtind{k}}},\widetilde{\crtind{k}}}(z_{nr})$ are not in complete agreement and $\hat{\iota}_{\widetilde{\crtind{k}},\hat{\param}} \preceq {\iota}_{\widetilde{\crtind{k}},\bar{\param}}$ holds, then 
$\hat{v}_{0,\widetilde{\crtind{k}}+1}(z_{nr})$ can be defined so that it satisfies the induction hypothesis.
\end{claim}
\begin{figure}[!tp]
\centering
\includegraphics[scale=0.55]{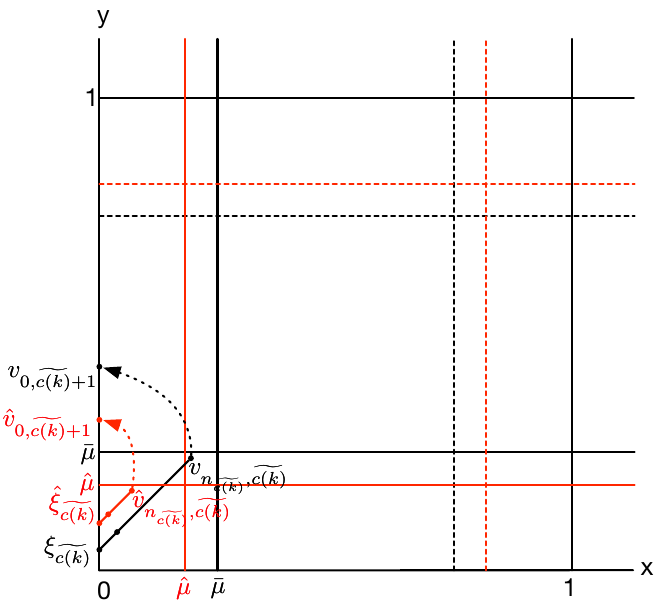}
\caption{If ${v}_{0,\widetilde{\crtind{k}}+1}(z_{nr})$ and ${v}_{n_{\widetilde{\crtind{k}}},\widetilde{\crtind{k}}}(z_{nr})$ are not in complete agreement (where $z_{nr} = y$ in this case), then $\hat{v}_{0,\widetilde{\crtind{k}}+1}(z_{nr})$ can be placed at an arbitrary position in $\hat{\iota}_{\widetilde{\crtind{k}}+1,\hat{\param}}$.}
\label{fig:notInCompleteAgreement}
\end{figure}
\begin{proof}[Proof of Claim \ref{claim:notCompleteAgreement}.]
First of all, notice that, when the conditions of Claim \ref{claim:notCompleteAgreement} hold, $\hat{v}_{0,\widetilde{\crtind{k}}+1}(z_{nr})$ and $\hat{v}_{n_{\widetilde{\crtind{k}}},\widetilde{\crtind{k}}}(z_{nr})$ cannot be in complete agreement, either, since the induction hypothesis requires that ${v}_{0,\widetilde{\crtind{k}}+1}(z_{nr})$ and $\hat{v}_{0,\widetilde{\crtind{k}}+1}(z_{nr})$ are (i.e., that ${\iota}_{\widetilde{\crtind{k}}+1,\bar{\param}} = \hat{\iota}_{\widetilde{\crtind{k}}+1,\hat{\param}}$ holds).
Then, it must hold that $\hat{v}_{0,\widetilde{\crtind{k}}+1}(z_{nr}) \geq \lfloor \hat{v}_{n_{\widetilde{\crtind{k}}},\widetilde{\crtind{k}}}(z_{nr}) \rfloor + \rightend{\hat{\iota}_{\widetilde{\crtind{k}},\hat{\param}}} \geq \hat{v}_{n_{\widetilde{\crtind{k}}},\widetilde{\crtind{k}}}(z_{nr})$, where at least one of the two inequalities is strict.
Hence, since $\hat{\iota}_{\widetilde{\crtind{k}},\hat{\param}} \preceq {\iota}_{\widetilde{\crtind{k}},\bar{\param}} \prec {\iota}_{\widetilde{\crtind{k}}+1,\bar{\param}}$ holds,
for any value of $\fract{\hat{v}_{n_{\widetilde{\crtind{k}}},\widetilde{\crtind{k}}}(z_{nr})} \in \hat{\iota}_{\widetilde{\crtind{k}},\hat{\param}}$, and for any value
$\hat{\varepsilon} \in \hat{\iota}_{\widetilde{\crtind{k}}+1,\hat{\param}} = {\iota}_{\widetilde{\crtind{k}}+1,\bar{\param}}$ (including of course all that satisfy the induction hypothesis) one can always define $\hat{\delta}_{n_{\widetilde{\crtind{k}}},\widetilde{\crtind{k}}}$
such that $\fract{{v}_{0,\widetilde{\crtind{k}}+1}(z_{nr})} = \hat{\varepsilon}$ holds.
\end{proof}

When the hypotheses of Claim \ref{claim:notCompleteAgreement} hold, this is enough to show that we can define $\hat{v}_{0,\widetilde{\crtind{k}}+1}(z_{nr})$ to satisfy the induction hypothesis.
However, the hypotheses of the claim do not hold in all configurations (in particular, when ${v}_{0,\widetilde{\crtind{k}}+1}(z_{nr})$ and ${v}_{n_{\widetilde{\crtind{k}}},\widetilde{\crtind{k}}}(z_{nr})$ are in complete agreement), so we will need to handle those other situations in a custom manner.
Both cases---to satisfy the hypotheses of Claim \ref{claim:notCompleteAgreement}, and in particular condition $\hat{\iota}_{\widetilde{\crtind{k}},\hat{\param}} \preceq {\iota}_{\widetilde{\crtind{k}},\bar{\param}}$, and to handle the other situations---require us to suitably define $\hat{v}_{n_{\widetilde{\crtind{k}}},\widetilde{\crtind{k}}}(z_{nr})$.
To this end, using Lemma \ref{lm:fracvalue} we investigate the constraints that hold on $\fract{\hat{v}_{n_{\widetilde{\crtind{k}}},\widetilde{\crtind{k}}}(z_{nr})}$ (i.e., at the end of \onerst{} sequence $\hat{\orssym}_{\widetilde{\crtind{k}}}$) with respect to $\fract{\hat{v}_{0,\widetilde{\crtind{k}}}(z_2)}$ (i.e., at the beginning of the same $\hat{\orssym}_{\widetilde{\crtind{k}}}$).
Since the hypotheses of Lemma \ref{lm:fracvalue} only consider the clock constraints that hold on $\hat{v}_{0,\widetilde{\crtind{k}}}$ and $\hat{v}_{n_{\widetilde{\crtind{k}}},\widetilde{\crtind{k}}}$, and these are the same as those that hold in  $v_{0,\widetilde{\crtind{k}}}$ and $v_{n_{\widetilde{\crtind{k}}},\widetilde{\crtind{k}}}$, we consider the relationship between $\fract{v_{n_{\widetilde{\crtind{k}}},\widetilde{\crtind{k}}}(z_{nr})}$ and $\fract{v_{0,\widetilde{\crtind{k}}}(z_2)}$ to draw conclusions on the relationship between $\fract{\hat{v}_{n_{\widetilde{\crtind{k}}},\widetilde{\crtind{k}}}(z_{nr})}$ and $\fract{\hat{v}_{0,\widetilde{\crtind{k}}}(z_2)}$.

We split the rest of the proof in cases \ref{case:i} and \ref{case:ii}.

\begin{figure}[tb]
\centering
\includegraphics[width=\columnwidth]{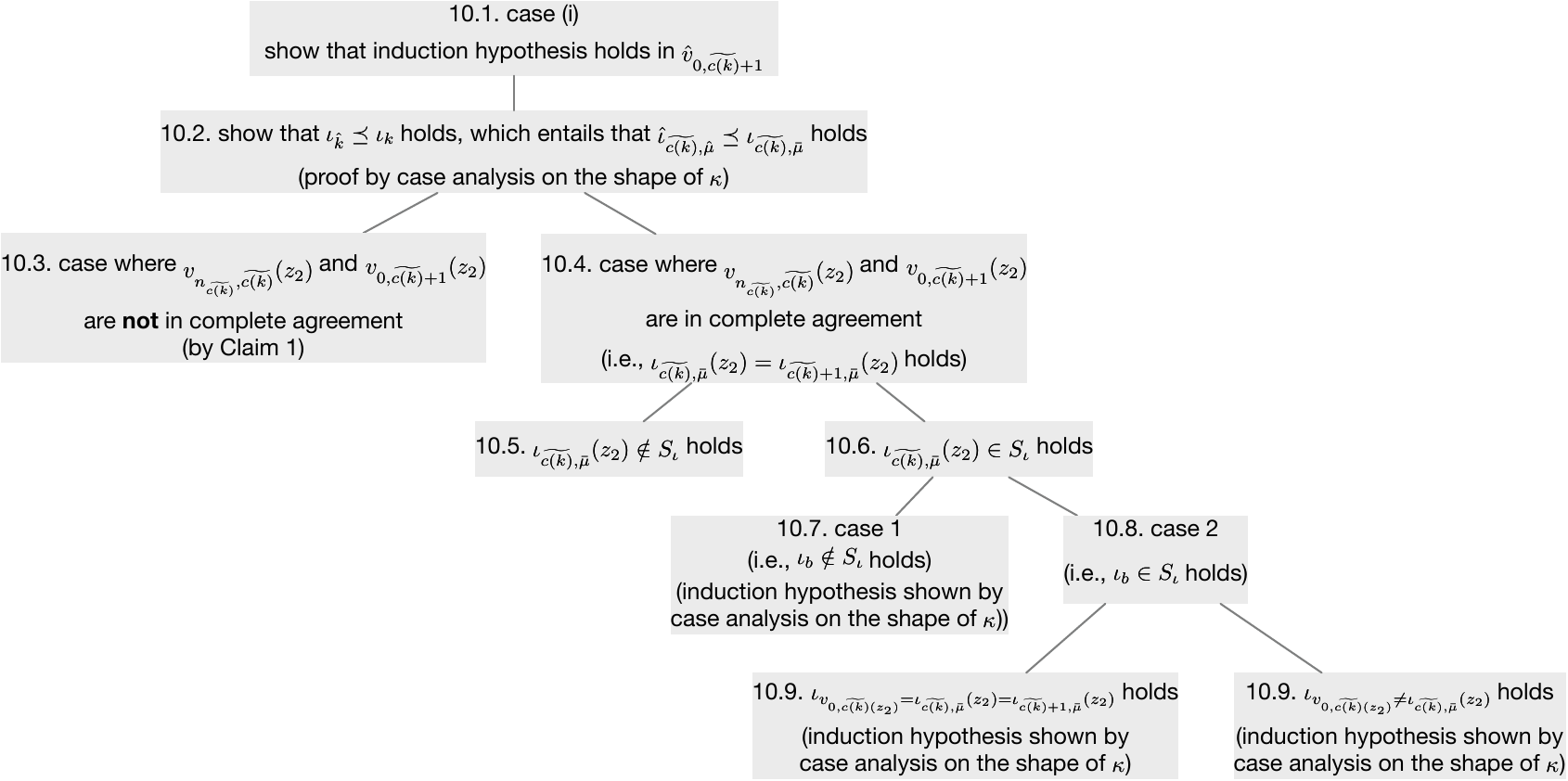}
\caption{Structure of the proof in case (i).}
\label{fig:proofStructureCasei}
\end{figure}

\textbf{Point 10} of Figure \ref{fig:proofStructureGeneral}, which is further detailed in Figure \ref{fig:proofStructureCasei}.
We need to show that the induction hypothesis holds in $v_{0,\widetilde{\crtind{k}}+1}$.

Let us consider {\bf case \ref{case:i}} (i.e., $z_r = z_1$ and $z_{nr} = z_2$).
Let $b = \fract{v_{0,\widetilde{\crtind{k}}}(z_2)}$ and $\hat{b} = \fract{\hat{v}_{0,\widetilde{\crtind{k}}}(z_2)}$.
By Lemma \ref{lm:fracvalue} (cases \ref{lm4case:1}-\ref{lm4case:4}) we have that $\fract{v_{n_{\widetilde{\crtind{k}}},\widetilde{\crtind{k}}}(z_2)}$ is of the form $\kappa + \epsilon$, where $\kappa$ is one of $\set{0, b, b+ \fract{\bar{\param}}, b- (1 - \fract{\bar{\param}})}$ and $\epsilon \geq 0$ holds, and similarly for $\fract{\hat{v}_{n_{\widetilde{\crtind{k}}},\widetilde{\crtind{k}}}(z_2)}$ with respect to $\hat{b}$ (hence $\hat{\kappa}$) and $\fract{\hat{\param}}$ (where $\kappa = b$ holds if, and only if, $\hat{\kappa} = \hat{b}$ does, and so on).
Since $\fract{v_{n_{\widetilde{\crtind{k}}},\widetilde{\crtind{k}}}(z_2)} = \kappa + \epsilon$ holds, this means that the value of $\fract{v_{ n_{\widetilde{\crtind{k}}},\widetilde{\crtind{k}}}(z_2)}$ is in $[\kappa, 1)$---or, dually, that $\epsilon$ is in $[0, 1-\kappa)$---and for any $\hat{\varepsilon} \in [\hat{\kappa}, 1)$ there is $\hat{\delta}_{\widetilde{\crtind{k}}} = \sum_{h = i'-1}^{n_{\widetilde{\crtind{k}}}-1} \hat{\delta}_{h,\widetilde{\crtind{k}}}$ such that $\hat{v}_{n_{\widetilde{\crtind{k}}},\widetilde{\crtind{k}}}(z_2) = \hat{v}_{i'-1,\widetilde{\crtind{k}}}(z_2)+\hat{\delta}_{\widetilde{\crtind{k}}}$ and $\fract{\hat{v}_{n_{\widetilde{\crtind{k}}},\widetilde{\crtind{k}}}(z_2)} = \hat{\varepsilon}$ hold.
\todoMR{}{questo è vero, ma non è detto che ci interessi un qualunque valore fino a 1, magari dobbiamo fermarci a $\fract{\hat{\param}}$.}
\todoMR{}{più che altro, non mi è più tanto chiaro perché abbiamo tirato in ballo questo discorso. Forse in realtà non serve e basta; ricontrollare quel che viene dopo, se mai questo fatto viene richiamato, se no toglierlo e basta.}

\textbf{Point \hypertarget{bmk:p10.2}{10.2}}
of Figure \ref{fig:proofStructureCasei}.
We first show that in all cases condition $\iota_{\hat{\kappa}} \preceq \iota_{\kappa}$ holds, that is, either $\kappa$ and $\hat{\kappa}$ are in complete agreement, or it must be that
$\iota_{\hat{\kappa}} \prec \iota_{\kappa}$ holds
(notice that, if $\iota_{\hat{\kappa}} \preceq \iota_{\kappa}$ holds, then for any $\epsilon \geq 0$ one can always define $\hat{\epsilon} \geq 0$ such that $\iota_{\hat{\kappa}+\hat{\epsilon}} = \hat{\iota}_{\widetilde{\crtind{k}},\hat{\param}} \preceq {\iota}_{\widetilde{\crtind{k}},\bar{\param}} = \iota_{{\kappa}+{\epsilon}}$ holds).
This is trivial if $\kappa = 0$ or $\kappa = b$, as $b$ and $\hat{b}$ are in complete agreement by induction hypothesis.
For the other two cases ($\kappa = b+ \fract{\bar{\param}}$ and $\kappa = b- (1 - \fract{\bar{\param}})$) we separate the proof depending on the polarity.\\
If the polarity is \emph{positive}, we separately consider the cases $\kappa = b+ \fract{\bar{\param}}$ and $\kappa = b - (1 - \fract{\bar{\param}})$.
If $\kappa = b+ \fract{\bar{\param}}$ holds, then it must be $\iota_b = \ZA_{\bar{\param}}$ (otherwise it would hold that $\kappa \geq 1$), hence $\iota_{\kappa} = \ZE_{\bar{\param}}$; the reasoning holds also for $\hat{\kappa}$ (since $b$ and $\hat{b}$ are in complete agreement), hence we have $\iota_{\hat{\kappa}} = \ZE_{\hat{\param}}$.
If, instead, $\kappa = b - (1 - \fract{\bar{\param}})$ holds, then we have $\ZC_{\bar{\param}} \preceq \iota_{b} \preceq \ZE_{\bar{\param}}$ (otherwise $\kappa \leq 0$ would hold) and $\ZA_{\bar{\param}} \preceq \iota_{\kappa} \preceq \ZC_{\bar{\param}}$.
In this case, we can show that $\iota_{\hat{\kappa}} = \ZA_{\hat{\param}}$ holds.
Indeed, we first remark that, by definition, it holds that $\frac{1}{2} < \fract{\hat{\param}} < \frac{2}{3}$ (i.e., $\fract{\hat{\param}} - \frac{1}{2}  < \frac{1}{6}$; in fact, the upper bound is stricter, but this is enough for our purposes).
Then, if $\ZC_{\hat{\param}} \preceq \iota_{\hat{b}} \preceq \ZD_{\hat{\param}}$ holds, it also holds that $\iota_{\hat{\kappa}} = \ZA_{\hat{\param}}$.
If, instead, $\iota_{\hat{b}} \in \ZE_{\bar{\param}}$ holds, then by induction hypothesis \ref{I3b}, it holds that $\hat{b} < 1 - 2(2\cmax - \lfloor \hat{v}_{0,\crtind{k'}}(z_2) \rfloor +1)(\fract{\hat{\param}}-\frac{1}{2})$ (that is, $\hat{b} < 1 - 2\cdot 2(\fract{\hat{\param}}-\frac{1}{2})$, since $2\cmax - \lfloor \hat{v}_{0,\crtind{k'}}(z_2) \rfloor \geq 1$ holds), hence also $\hat{b}- (1 - \fract{\hat{\param}}) < 1 - \fract{\hat{\param}}$, i.e., $\iota_{\hat{\kappa}} = \ZA_{\hat{\param}}$.
\todoMR{}{ %
Ho espanso la dimostrazione, ma è da ricontrollare.
}
\\
If the polarity is \emph{negative}, if $\kappa = b+ \fract{\bar{\param}}$ holds then it also holds that $\iota_b \preceq \ZC_{\bar{\param}}$
(or $b+ \fract{\bar{\param}} \geq 1$ would hold) and $\ZC_{\bar{\param}} \preceq \iota_{\kappa} \preceq \ZE_{\bar{\param}}$.
In this case, we show that it also holds that $\iota_{\hat{\kappa}} = \ZC_{\hat{\param}}$.
In fact, if $\iota_{\hat{b}} \preceq \ZB_{\hat{\param}}$ holds, since by definition we have that $\fract{\hat{\param}} < \frac{1}{3}$ when the polarity is negative (indeed, the bound is stricter, but this is enough for our current purposes),
we also have $\fract{\hat{\param}} < \hat{b} + \fract{\hat{\param}} < 1 - \fract{\hat{\param}}$ (i.e., $\iota_{\hat{\kappa}} = \ZC_{\hat{\param}}$).
If $\iota_{\hat{b}} = \ZC_{\hat{\param}}$ holds, then, if $\intprm = 0$ holds (hence $\fract{\hat{\param}} = \hat{\param}$), as discussed in Point
\hyperlink{bmk:condDn}{4}
above, it also holds that
$\hat{b} = \fract{\hat{v}_{0,\crtind{k}+R}(z_2)} = \fract{\hat{v}_{0,\crtind{k}}(z_2)} + \hat{\param}P + \hat{\varepsilon}' < 1-(2\cmax - \lfloor \hat{v}_{0,\crtind{k}}(z_2) \rfloor -1)|Q|\hat{\param} - 2\hat{\param}$ and $\fract{\hat{\param}} < \hat{\kappa} = \hat{b} + \fract{\hat{\param}} < 1 - \hat{\param}$ (i.e., $\iota_{\hat{\kappa}} = \ZC_{\hat{\param}}$) also hold.
\todoMR{}{aggiustata ipotesi}
If, instead, $\intprm > 0$ holds, then by induction hypothesis \ref{I3a} it must hold that $1-((2\cmax - \lfloor \hat{v}_{0,\crtind{k}}(z_2) \rfloor)\cdot|Q| +2)\fract{\hat{\param}} < \hat{b} < 1-((2\cmax - \lfloor \hat{v}_{0,\crtind{k}}(z_2) \rfloor)\cdot|Q|+1)\fract{\hat{\param}}$, so it also holds that $\fract{\hat{\param}} < \hat{b} + \fract{\hat{\param}} < 1 - \fract{\hat{\param}}$ (i.e., $\iota_{\hat{\kappa}} = \ZC_{\hat{\param}}$; recall that, as discussed in Point
\hyperlink{bmk:enoughintervals}{2} above, the length of interval $\ZC_{\hat{\param}}$ is greater than $(2\cmax \cdot |Q| +1)\chi_{\hat{\param}}$).
\todoMR{}{sistemato i conti}
If $\kappa = b - (1 -\fract{\bar{\param}})$ holds, then it must also hold that $\iota_b = \ZE_{\bar{\param}}$ (hence also $\iota_{\hat{b}} = \ZE_{\hat{\param}}$), or $b - (1 -\fract{\bar{\param}}) \leq 0$ would hold.
Then, $\iota_{\kappa} = \ZA_{\bar{\param}}$ and $\iota_{\hat{\kappa}} = \ZA_{\hat{\param}}$ hold.

We further split the proof in two cases, depending on whether $v_{n_{\widetilde{\crtind{k}}},\widetilde{\crtind{k}}}(z_2)$ and $v_{0,\widetilde{\crtind{k}}+1}(z_2)$ are in complete agreement or not, and for each case we show that the induction hypothesis holds.

\textbf{Point 10.3} of Figure \ref{fig:proofStructureCasei}.
If $v_{n_{\widetilde{\crtind{k}}},\widetilde{\crtind{k}}}(z_2)$ and $v_{0,\widetilde{\crtind{k}}+1}(z_2)$ are not in complete agreement, then, by Claim \ref{claim:notCompleteAgreement}, the induction hypothesis can be proved from the fact that $\hat{\iota}_{\widetilde{\crtind{k}}, \hat{\param}}(z_2) \preceq \iota_{\widetilde{\crtind{k}}, \bar{\param}}(z_2)$ holds.

\textbf{Point 10.4} of Figure \ref{fig:proofStructureCasei}.
Let us now consider the case in which $v_{n_{\widetilde{\crtind{k}}},\widetilde{\crtind{k}}}(z_2)$ and $v_{0,\widetilde{\crtind{k}}+1}(z_2)$ are in complete agreement.
We further separate two cases, depending on whether $\iota_{\widetilde{\crtind{k}}, \bar{\param}}(z_2) \in \SetZZ$ holds or not.

\textbf{Point 10.5} of Figure \ref{fig:proofStructureCasei}.
If $v_{n_{\widetilde{\crtind{k}}},\widetilde{\crtind{k}}}(z_2)$ and $v_{0,\widetilde{\crtind{k}}+1}(z_2)$ are in complete agreement (which entails $\iota_{\widetilde{\crtind{k}}, \bar{\param}}(z_2) = \iota_{\widetilde{\crtind{k}}+1, \bar{\param}}(z_2)$), but $\iota_{\widetilde{\crtind{k}}, \bar{\param}}(z_2) \notin \SetZZ$ holds, then the induction hypothesis (which in this case reduces to \ref{I2}, i.e., that $\iota_{\widetilde{\crtind{k}}+1, \bar{\param}}(z_2) = \hat{\iota}_{\widetilde{\crtind{k}}+1, \hat{\param}}(z_2)$ holds) can be easily proved by considering that, as shown above, $\hat{\iota}_{\widetilde{\crtind{k}}, \hat{\param}}(z_2) \preceq \iota_{\widetilde{\crtind{k}}, \bar{\param}}(z_2)$ holds, and the delay $\hat{\delta}_{n_{\widetilde{\crtind{k}}},\widetilde{\crtind{k}}}$ between $\hat{v}_{n_{\widetilde{\crtind{k}}},\widetilde{\crtind{k}}}$ and $\hat{v}_{0,\widetilde{\crtind{k}}+1}$ is arbitrary, since $\A$ is an nrtTA; hence, one can always choose $\hat{\delta}_{n_{\widetilde{\crtind{k}}},\widetilde{\crtind{k}}}$ so that $\iota_{\widetilde{\crtind{k}}+1, \bar{\param}}(z_2) = \hat{\iota}_{\widetilde{\crtind{k}}+1, \hat{\param}}(z_2)$ holds (see Figure \ref{fig:inCompleteAgreementNotInSj} for a graphical depiction).
\todoMR{}{ %
Forse un po' sbrigativa come spiegazione}

\begin{figure}[tb]
\centering
\includegraphics[scale=0.55]{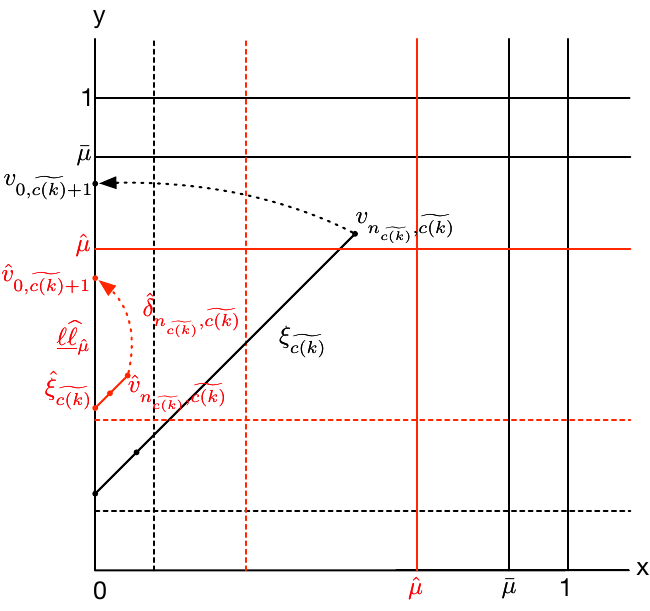}
\caption{Depiction of the case in which ${v}_{0,\widetilde{\crtind{k}}+1}(z_{2})$ and ${v}_{n_{\widetilde{\crtind{k}}},\widetilde{\crtind{k}}}(z_{2})$ are in complete agreement (where $z_{2} = y$ in this case), but $\iota_{\widetilde{\crtind{k}}, \bar{\param}}(z_2)$ does not belong to $\SetZZ$ (notice that in this case the polarity is positive, so $\ZC_{\bar{\param}} \notin \SetZZ$ holds).}
\label{fig:inCompleteAgreementNotInSj}
\end{figure}

\textbf{Point 10.6} of Figure \ref{fig:proofStructureCasei}.
If $v_{n_{\widetilde{\crtind{k}}},\widetilde{\crtind{k}}}(z_2)$ and $v_{0,\widetilde{\crtind{k}}+1}(z_2)$ are in complete agreement
and $\iota_{\widetilde{\crtind{k}}, \bar{\param}}(z_2) \in \SetZZ$ holds, then we separately consider the cases in which $b$ (hence also $\hat{b}$) belongs to an interval of $\SetZZ$ or not (i.e., cases \ref{case:2} and \ref{case:1} of Point 
\hyperlink{bmk:p4}{4}, respectively), and for each case we deal with the various shapes of $\kappa$.

\textbf{Point 10.7} of Figure \ref{fig:proofStructureCasei}.
Let us first consider case \ref{case:1}.
If $\kappa = 0$ or $\kappa = b$ hold, then it must also hold that $\kappa \leq \leftend{\iota_{\widetilde{\crtind{k}}, \bar{\param}}(z_2)}$ (hence also $\hat{\kappa} \leq \leftend{\hat{\iota}_{\widetilde{\crtind{k}}, \hat{\param}}(z_2)}$) since by hypothesis $\kappa$ is not in an interval of $\SetZZ$,
but $\kappa + \epsilon$ is; hence, for any $\hat{\varepsilon} \in \hat{\iota}_{\widetilde{\crtind{k}+1}, \hat{\param}}(z_2)$ (which must be such that $\hat{\varepsilon} > \hat{\kappa}$ holds), we can always define $\hat{\epsilon}$ and $\hat{\delta}_{n_{\widetilde{\crtind{k}}},\widetilde{\crtind{k}}}$ so that $\fract{\hat{\kappa} + \hat{\epsilon} + \hat{\delta}_{n_{\widetilde{\crtind{k}}},\widetilde{\crtind{k}}}} = \hat{\kappa} + \hat{\epsilon} + \hat{\delta}_{n_{\widetilde{\crtind{k}}}} = \hat{\varepsilon}$ holds, because $\A$ is an nrtTA, and $\hat{\delta}_{n_{\widetilde{\crtind{k}}},\widetilde{\crtind{k}}}$ is arbitrary (see Figure \ref{fig:inCompleteAgreementInSjCase1} for a graphical representation).
\todoMR{}{ %
Ho messo esplicitamente che la parte frazionaria è $\hat{\kappa} + \hat{\epsilon} + \hat{\delta}_{n_{\widetilde{\crtind{k}}}}$, ma forse è da spiegare (ultimo punto della \onerst{} sequence e quello successivo sono in complete agreement per ipotesi, non si può cambiare parte intera)}
\begin{figure}[tb]
\centering
\includegraphics[scale=0.55]{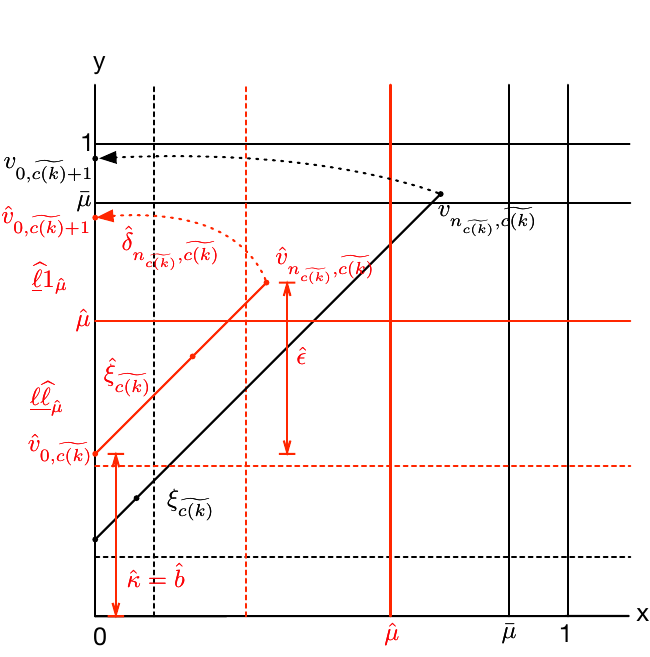}
\caption{Depiction of the case (with positive polarity) in which ${v}_{0,\widetilde{\crtind{k}}+1}(z_{2})$ and ${v}_{n_{\widetilde{\crtind{k}}},\widetilde{\crtind{k}}}(z_{2})$ are in complete agreement (with $z_{2} = y$), $\iota_{\widetilde{\crtind{k}}, \bar{\param}}(z_2)$ belongs to $\SetZZ$, but $\iota_{b}$ does not; in addition, in this case it holds that $\kappa = b$.}
\label{fig:inCompleteAgreementInSjCase1}
\end{figure}
If $\hat{\kappa} = \hat{b}+ \fract{\hat{\param}}$ or $\hat{\kappa} = \hat{b} - (1-\fract{\hat{\param}})$ holds, then it is easy to see that $\hat{\kappa} \leq 2\chi_{\hat{\param}}$ holds.
Indeed, if $\hat{\kappa} = \hat{b}+ \fract{\hat{\param}}$ holds, as discussed in Point 
\hyperlink{bmk:p10.2}{10.2} above, $\hat{b}$ is in $\ZA_{\hat{\param}}$ if the polarity is positive, or in one of $\ZA_{\hat{\param}}$, $\ZB_{\hat{\param}}$, $\ZC_{\hat{\param}}$ if the polarity is negative.
Since we are in case \ref{case:1}, the polarity must be negative, and $\hat{b}$ must be in one of $\ZA_{\hat{\param}}$, $\ZB_{\hat{\param}}$, hence $\hat{b}+ \fract{\hat{\param}} \leq 2\chi_{\hat{\param}}$ holds.
Similarly, if $\hat{\kappa} = \hat{b} - (1-\fract{\hat{\param}})$ holds and the polarity is negative, $\iota_{\hat{\kappa}} = \ZA_{\hat{\param}}$ holds, hence the result.
if $\hat{\kappa} = \hat{b} - (1-\fract{\hat{\param}})$ holds and the polarity is positive, $\hat{b}$ is in $\ZC_{\hat{\param}}$ or in $\ZD_{\hat{\param}}$ since we are in case 1, which entails $\hat{b} \leq 1 - \fract{\hat{\param}} + 2\chi_{\hat{\param}}$, hence the result.
In all these cases $\hat{\epsilon}$ can be chosen so that $\hat{\kappa}+\hat{\epsilon}$ trivially satisfies the induction hypothesis.
In particular, in all cases we can define $\hat{\varepsilon}$ and $\hat{\delta}_{n_{\widetilde{\crtind{k}}},\widetilde{\crtind{k}}}$ so that $\hat{v}_{0,\widetilde{\crtind{k}}+1}(z_2)$ falls in the desired interval. For example, if the polarity is negative, we can define the values so that $\LowK{\hat{v}_{n_{\widetilde{\crtind{k}}},\widetilde{\crtind{k}}}(z_2), \chi_{\hat{\param}}} = \LowK{\hat{v}_{0,\widetilde{\crtind{k}}+1}(z_2), \chi_{\hat{\param}}} = (2\cmax - \lfloor \hat{v}_{0,\widetilde{\crtind{k}}+1}(z_2) \rfloor)\cdot|Q|+1$ holds (where $(2\cmax - \lfloor \hat{v}_{0,\widetilde{\crtind{k}}+1}(z_2) \rfloor)\cdot|Q|+1$ is the desired interval for $\LowK{\hat{v}_{0,\widetilde{\crtind{k}}+1}(z_2), \chi_{\hat{\param}}}$ in this case, see \ref{I3a}).

\textbf{Point 10.8} of Figure \ref{fig:proofStructureCasei}.
Let us now consider case \ref{case:2}.
The case in which $\kappa = 0$ holds is handled in the same way as above.
We consider two further cases: $\iota_{{v}_{0,\widetilde{\crtind{k}}}(z_{2})} = \iota_{\widetilde{\crtind{k}}+1, \bar{\param}}(z_{2})$
and $\iota_{{v}_{0,\widetilde{\crtind{k}}}(z_{2})} \neq \iota_{\widetilde{\crtind{k}}+1, \bar{\param}}(z_{2})$.\\

\textbf{Point 10.9} of Figure \ref{fig:proofStructureCasei}.
If $\iota_{{v}_{0,\widetilde{\crtind{k}}}(z_{2})} = \iota_{\widetilde{\crtind{k}}+1, \bar{\param}}(z_{2})$ holds---hence also $\iota_{{v}_{0,\widetilde{\crtind{k}}}(z_{2})} = \iota_{\widetilde{\crtind{k}}, \bar{\param}}(z_{2})$---since all three intervals are, by hypothesis, in $\SetZZ$, but $\orssym_{\widetilde{\crtind{k}}+1}$ is a \crtonerst{} \onerst{} sequence, then it must hold that $\lfloor {v}_{0,\widetilde{\crtind{k}}}(z_{2}) \rfloor < \lfloor {v}_{0,\widetilde{\crtind{k}}+1}(z_{2}) \rfloor$ (see Figure \ref{fig:inCompleteAgreementInSjCase2a}).
\begin{figure}[tb]
\centering
\includegraphics[scale=0.55]{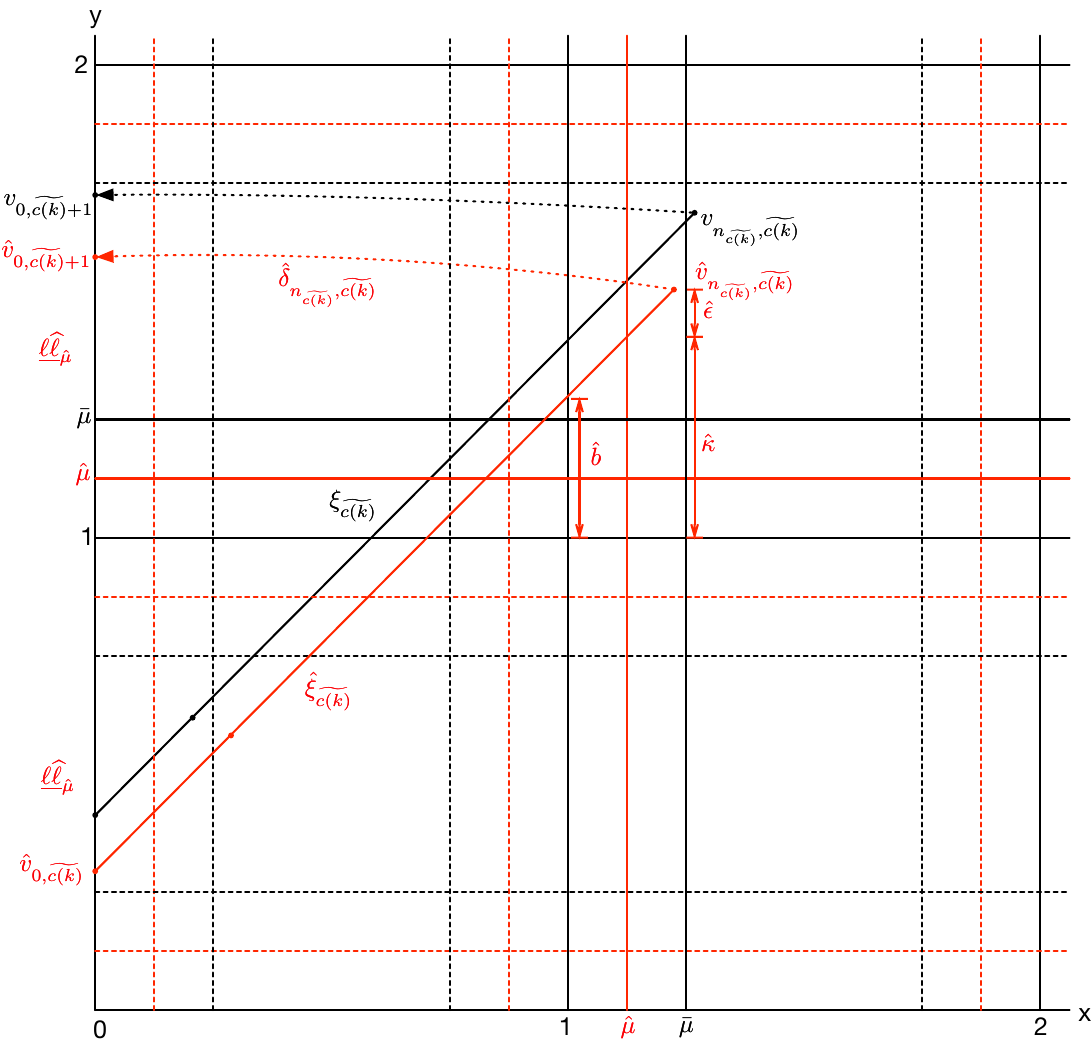}
\caption{Depiction of the case (with negative polarity) in which ${v}_{0,\widetilde{\crtind{k}}+1}(z_{2})$ and ${v}_{n_{\widetilde{\crtind{k}}},\widetilde{\crtind{k}}}(z_{2})$ are in complete agreement (with $z_{2} = y$), $\iota_{\widetilde{\crtind{k}}, \bar{\param}}(z_2)$ and $\iota_{b}$ belong to $\SetZZ$, and $\iota_{{v}_{0,\widetilde{\crtind{k}}}(z_{2})} = \iota_{\widetilde{\crtind{k}}, \bar{\param}}(z_{2}) = \iota_{\widetilde{\crtind{k}}+1, \bar{\param}}(z_{2})$ holds; in addition, in this case it holds that $\widetilde{\crtind{k}} = \crtind{k}$ and $\kappa = b + \fract{\bar{\param}}$.}
\label{fig:inCompleteAgreementInSjCase2a}
\end{figure}
If $\iota_{{v}_{0,\widetilde{\crtind{k}}}(z_{2})} = \ZC_{\bar{\param}}$ holds (i.e., the polarity is negative), then, as discussed in Point
\hyperlink{bmk:condDn}{4} above, $\hat{b} < 1-((2\cmax - \lfloor \hat{v}_{0,\crtind{k}}(z_2) \rfloor - 1)|Q| +1)\fract{\hat{\param}}$ holds (notice that, by hypothesis \ref{I3a}, this is true also when $m >0$ holds).
Similarly, if $\iota_{{v}_{0,\widetilde{\crtind{k}}}(z_{2})} = \ZA_{\bar{\param}}$ (resp., $\iota_{{v}_{0,\widetilde{\crtind{k}}}(z_{2})} = \ZE_{\bar{\param}}$) holds (i.e., the polarity is positive), then, by hypothesis \ref{I3b}, $\hat{b} < 1- \fract{\bar{\param}} - (2(2\cmax - \lfloor \hat{v}_{0,\crtind{k}}(z_2) \rfloor)-1)(\fract{\hat{\param}}-\frac{1}{2})$ (resp., $\hat{b} < 1- (2(2\cmax - \lfloor \hat{v}_{0,\crtind{k}}(z_2) \rfloor)-1)(\fract{\hat{\param}}-\frac{1}{2})$) holds.
In all cases, if $\kappa = b$ holds, then it is easy to define $\hat{\epsilon}$ such that $\hat{\kappa} + \hat{\epsilon}$ satisfies the induction hypothesis (e.g., \ref{I3a}, that is $\hat{\kappa} + \hat{\epsilon} < 1-((2\cmax - \lfloor \hat{v}_{0,\crtind{k+1}}(z_2) \rfloor)\cdot|Q| +1)\hat{\param}$ if the polarity is negative and \onerst{} sequence $\orssym_{\widetilde{\crtind{k}}+2}$ exists and is such that $v_{0, \widetilde{\crtind{k}}+2}(z_1) = 0$ and $v_{0, \widetilde{\crtind{k}}+2}(z_2) > 0$ hold),
since it holds that $\lfloor \hat{v}_{0,\crtind{k}}(z_2) \rfloor + 1 \leq \lfloor \hat{v}_{0,\crtind{k+1}}(z_2) \rfloor$ (recall that $\crtind{k+1} = \widetilde{\crtind{k}}+1$ holds).
If $\kappa = b + \fract{\bar{\param}}$ holds, then the polarity must be negative (otherwise it cannot hold that $\iota_{{v}_{0,\widetilde{\crtind{k}}}(z_{2})} = \iota_{\widetilde{\crtind{k}}, \bar{\param}}(z_{2})$) and, as mentioned above, $\lfloor {v}_{0,\widetilde{\crtind{k}}}(z_{2}) \rfloor < \lfloor {v}_{0,\widetilde{\crtind{k}}+1}(z_{2}) \rfloor$ holds, so from Lemma \ref{lm:fracvalue} it must also hold that $\intprm > 0$ (we are in case \ref{lm4case:3a}, with $\const_2 > z_{2,0}$, see Figure \ref{fig:inCompleteAgreementInSjCase2a} for a graphical depiction, where $\cmax = 1$ and $|Q| = 2$ hold);
\todoMR{}{dovrebbe essere questo il caso}
hence, in this case it holds that $P=0$ and $\hat{b} < 1-((2\cmax - \lfloor \hat{v}_{0,\crtind{k}}(z_2) \rfloor)\cdot|Q| +1)\fract{\hat{\param}}$, so it is easy to define $\hat{\epsilon}$ so that $\hat{\kappa} + \hat{\epsilon} =\hat{b} + \fract{\hat{\param}} + \hat{\epsilon} < 1-((2\cmax - \lfloor \hat{v}_{0,\crtind{k+1}}(z_2) \rfloor)\cdot|Q| +1)\fract{\hat{\param}}$ holds (which entails that we can also define $\hat{\epsilon}$ so that $1 - 2\fract{\hat{\param}} < \hat{\kappa} + \hat{\epsilon} < 1- \fract{\hat{\param}}$).
Finally, case $\kappa = b - (1- \fract{\bar{\param}})$ cannot occur if $\iota_{{v}_{0,\widetilde{\crtind{k}}}(z_{2})} = \iota_{\widetilde{\crtind{k}}, \bar{\param}}(z_{2})$ holds and $\iota_{\widetilde{\crtind{k}}, \bar{\param}}$ is either $\ZA_{\bar{\param}}$ or $\ZC_{\bar{\param}}$, as $b - (1 - \fract{\bar{\param}})$ would be negative in these cases.
If $\iota_{\widetilde{\crtind{k}}, \bar{\param}}(z_{2})$ is $\ZE_{\bar{\param}}$, instead, then the polarity is positive (since we are in case \ref{case:2}) and $\kappa$ is in $\ZA_{\bar{\param}}$, so, for any $\hat{\varepsilon}$ in $\ZE_{\hat{\param}}$, as before, one
can always define $\hat{\epsilon}$ and $\hat{\delta}_{n_{\widetilde{\crtind{k}}},\widetilde{\crtind{k}}}$ so that $\fract{\hat{\kappa} + \hat{\epsilon} + \hat{\delta}_{n_{\widetilde{\crtind{k}}},\widetilde{\crtind{k}}}} = \hat{\kappa} + \hat{\epsilon} + \hat{\delta}_{n_{\widetilde{\crtind{k}}}} = \hat{\varepsilon}$ holds, hence also one that satisfies the induction hypothesis.

\textbf{Point 10.10} of Figure \ref{fig:proofStructureCasei}.
If, instead, $\iota_{{v}_{0,\widetilde{\crtind{k}}}(z_{2})} \neq \iota_{\widetilde{\crtind{k}}+1, \bar{\param}}(z_{2})$
holds, then the polarity must be positive (because $|\SetZZ| > 1$ holds) and one of the two intervals is $\ZA_{\bar{\param}}$, while the other is $\ZE_{\bar{\param}}$ (see Figure \ref{fig:inCompleteAgreementInSjCase2b}).
If $\iota_{{v}_{0,\widetilde{\crtind{k}}}(z_{2})}$ is $\ZA_{\bar{\param}}$ and $\iota_{\widetilde{\crtind{k}}, \bar{\param}}(z_{2})$ (i.e., $\iota_{\widetilde{\crtind{k}}+1, \bar{\param}}(z_{2})$, since they are in complete agreement) is $\ZE_{\bar{\param}}$, then the only possible two cases are $\kappa = b$ and $\kappa = b + \fract{\bar{\param}}$ ($b - (1 - \fract{\bar{\param}})$ would be negative).
If $\kappa = b$ holds, then for any $\hat{\varepsilon}$ in $\ZE_{\hat{\param}}$ that satisfies the induction hypothesis one can always define
$\hat{\epsilon}$ and $\hat{\delta}_{n_{\widetilde{\crtind{k}}},\widetilde{\crtind{k}}}$ so that $\hat{\kappa} + \hat{\epsilon} + \hat{\delta}_{n_{\widetilde{\crtind{k}}},\widetilde{\crtind{k}}} = \hat{\varepsilon}$ holds.
If $\kappa = b + \fract{\bar{\param}}$ holds, instead, since (by induction hypothesis), $\hat{b} < 1- \fract{\hat{\param}} - (2(2\cmax - \lfloor \hat{v}_{0,\crtind{k}}(z_2) \rfloor)-1)(\fract{\hat{\param}}-\frac{1}{2})$ holds, then it also holds that $\hat{b} + \fract{\hat{\param}} < 1- (2(2\cmax - \lfloor \hat{v}_{0,\crtind{k}}(z_2) \rfloor)-1)(\fract{\hat{\param}}-\frac{1}{2})$, hence $\hat{\epsilon}$ can be chosen so that $\hat{\kappa} + \hat{\epsilon} < 1- (2(2\cmax - \lfloor \hat{v}_{0,\crtind{k}}(z_2) \rfloor)-1)(\fract{\hat{\param}}-\frac{1}{2})$ holds
(which entails that we can also define $\hat{\epsilon}$ so that $1 - (\fract{\hat{\param}}-\frac{1}{2}) < \hat{\kappa} + \hat{\epsilon} < 1$ holds)
 and then the induction hypothesis is satisfied.
If $\iota_{{v}_{0,\widetilde{\crtind{k}}}(z_{2})}$ is $\ZE_{\bar{\param}}$ and $\iota_{\widetilde{\crtind{k}}, \bar{\param}}(z_{2})$ is $\ZA_{\bar{\param}}$, then the only possible case is $\kappa = b - (1- \fract{\bar{\param}})$ (in the other cases $\kappa$ would be greater than $1 - \fract{\bar{\param}}$, hence $\kappa + \epsilon$ could not be in $\ZA_{\bar{\param}}$).
\begin{figure}[tb]
\centering
\includegraphics[scale=0.55]{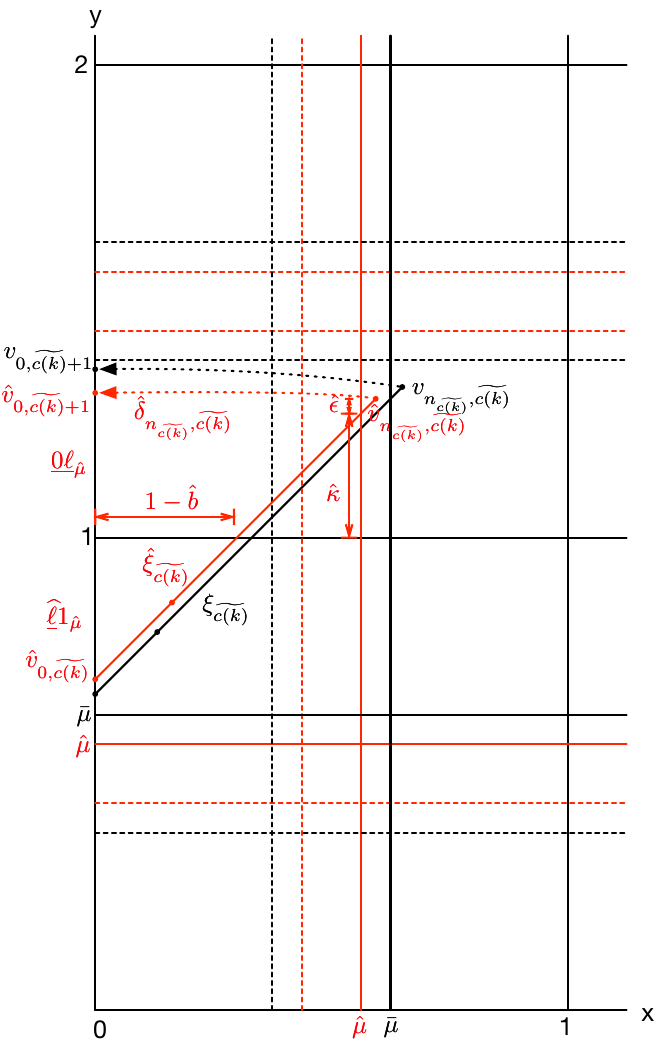}
\caption{Depiction of the case (with positive polarity) in which ${v}_{0,\widetilde{\crtind{k}}+1}(z_{2})$ and ${v}_{n_{\widetilde{\crtind{k}}},\widetilde{\crtind{k}}}(z_{2})$ are in complete agreement (with $z_{2} = y$), $\iota_{\widetilde{\crtind{k}}, \bar{\param}}(z_2)$ and $\iota_{b}$ belong to $\SetZZ$, and $\iota_{{v}_{0,\widetilde{\crtind{k}}}(z_{2})} \neq \iota_{\widetilde{\crtind{k}}+1, \bar{\param}}(z_{2})$ holds; in addition, in this case it holds that $\kappa = b - (1- \fract{\bar{\param}}) = \fract{\bar{\param}} - (1-b)$.}
\label{fig:inCompleteAgreementInSjCase2b}
\end{figure}
Notice that, in this case, $\lfloor {v}_{0,\widetilde{\crtind{k}}}(z_{2}) \rfloor < \lfloor {v}_{0,\widetilde{\crtind{k}}+1}(z_{2}) \rfloor$ holds, and (by induction hypothesis) it also holds that $\hat{b} < 1- (2(2\cmax - \lfloor \hat{v}_{0,\crtind{k}}(z_2) \rfloor)-1)(\fract{\hat{\param}}-\frac{1}{2})$ (Figure \ref{fig:inCompleteAgreementInSjCase2b} provides a graphical depiction of this case, where $\cmax = 1$ holds).
Hence $\hat{b} - (1 - \fract{\hat{\param}})< 1- \fract{\hat{\param}} - (2(2\cmax - (\lfloor \hat{v}_{0,\crtind{k}}(z_2) \rfloor +1))-1)(\fract{\hat{\param}}-\frac{1}{2})$ holds, and $\hat{\epsilon}$ can be chosen so that $\hat{\kappa} + \hat{\epsilon} < 1- \fract{\hat{\param}} - (2(2\cmax - (\lfloor \hat{v}_{0,\crtind{k}}(z_2) \rfloor +1))-1)(\fract{\hat{\param}}-\frac{1}{2})$ holds
(which entails that we can also define $\hat{\epsilon}$ so that $1 - \fract{\hat{\param}} - (\fract{\hat{\param}}-\frac{1}{2}) < \hat{\kappa} + \hat{\epsilon} < 1 - \fract{\hat{\param}}$ holds)
and then the induction hypothesis is satisfied.

\begin{figure}[tb]
\centering
\includegraphics[width=\columnwidth]{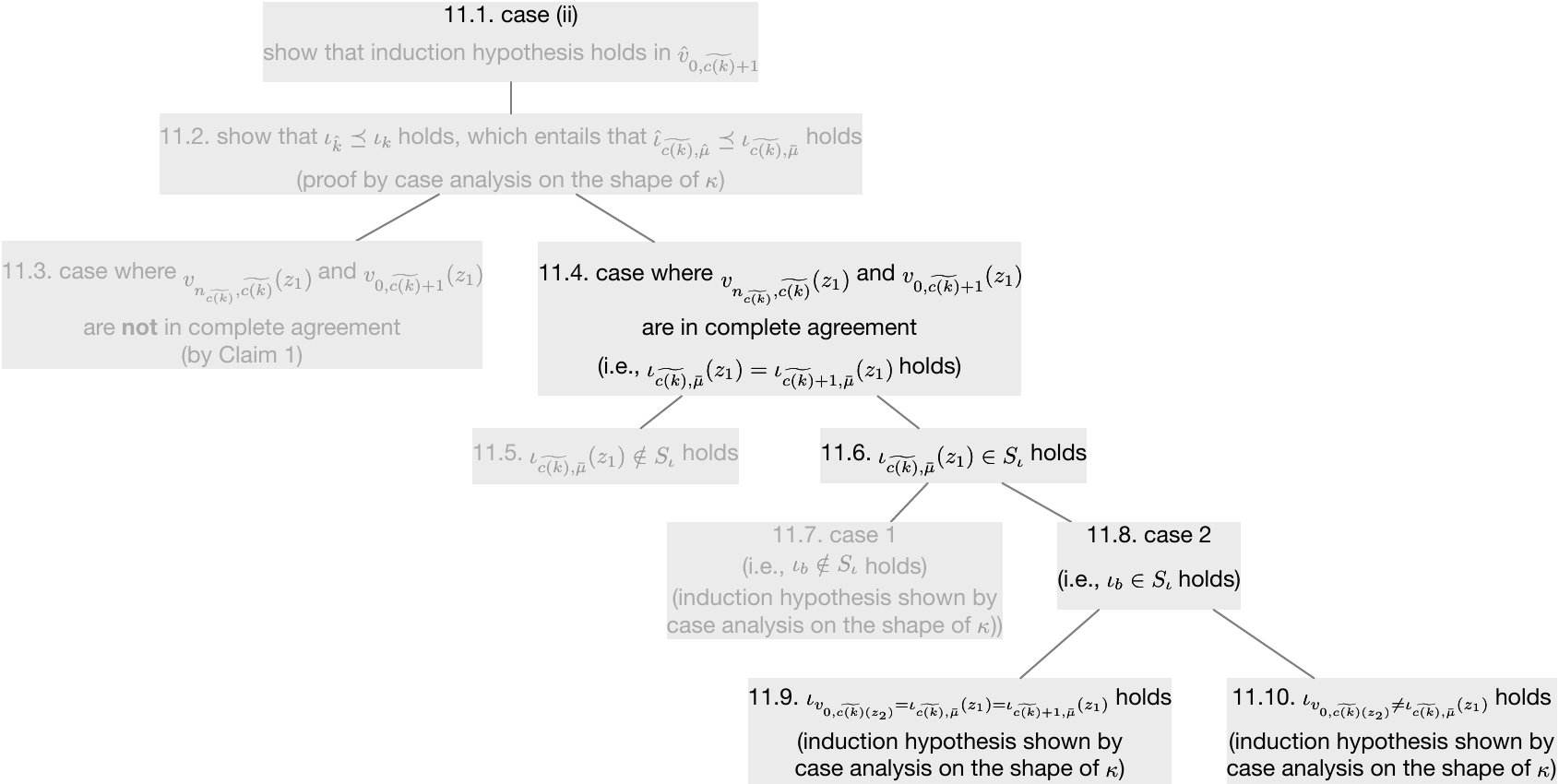}
\caption{Structure of the proof in case (ii).}
\label{fig:proofStructureCaseii}
\end{figure}

\textbf{Point 11} of Figure \ref{fig:proofStructureGeneral}, which is further detailed in Figure \ref{fig:proofStructureCaseii}.
In {\bf case \ref{case:ii}} ($z_r = z_2$ and $z_{nr} = z_1$) we need to investigate the constraint that must hold on $\fract{v_{n_{\widetilde{\crtind{k}}},\crtind{k}}(z_1)}$ with respect to $\fract{v_{0,\widetilde{\crtind{k}}}(z_2)} = b$.
By Lemma \ref{lm:fracvalue} (cases \ref{lm4case:5}-\ref{lm4case:8}) we have that $\fract{v_{n_{\widetilde{\crtind{k}}},\widetilde{\crtind{k}}}(z_1)}$ is of the form $\kappa + \epsilon$, where $\kappa \in \set{0, 1-b, 1-b+ \fract{\bar{\param}}, 1-b-(1 -\fract{\bar{\param}})}$ (similarly for $\fract{\hat{v}_{n_{\widetilde{\crtind{k}}},\widetilde{\crtind{k}}}(z_1)}$ with respect to $\hat{b}$, $\hat{\kappa}$ and $\fract{\hat{\param}}$).
Notice that $b$ and $1-b$ behave in a symmetric way with respect to intervals $\ZA_{\bar{\param}}, \ZB_{\bar{\param}}, \ZC_{\bar{\param}}, \ZD_{\bar{\param}}, \ZE_{\bar{\param}}$ (in fact, it holds that $\frac{1}{2} - b = 1 - b - \frac{1}{2}$). In fact, $b \in \ZA_{\bar{\param}}$ (resp., $b \in \ZB_{\bar{\param}}$) holds if, and only if, $1-b \in \ZE_{\bar{\param}}$ (resp., $1-b \in \ZD_{\bar{\param}}$) holds; also, $b \in \ZC_{\bar{\param}}$ holds if, and only if, $1-b \in \ZC_{\bar{\param}}$ also holds.
This also means that $b$ belongs to an interval of $\SetZZ$ if, and only if, $1-b$ does.
Then, the same reasoning applied to case (i) can essentially be repeated in case (ii) by considering $1-b$ instead of $b$ (this is true in particular for points 11.1, 11.2, 11.3, 11.5 and 11.7 of Figure \ref{fig:proofStructureCaseii}).

The only case where the proof requires modifications that are not completely trivial is the one (highlighted in Figure \ref{fig:proofStructureCaseii}) where $v_{n_{\widetilde{\crtind{k}}},\widetilde{\crtind{k}}}(z_1)$ and $v_{0,\widetilde{\crtind{k}}+1}(z_1)$ are in complete agreement, $\iota_{\widetilde{\crtind{k}}, \bar{\param}}(z_1) \in \SetZZ$ holds, and $b$ (hence also $\hat{b}$, $1-b$ and $1 - \hat{b}$) belongs to an interval of $\SetZZ$ (i.e., case \ref{case:2} above).
Then, let us consider this case separately and expand its proof.

The case in which $\kappa = 0$ holds is handled in the same way as above.
We consider two further cases: $\iota_{{v}_{0,\widetilde{\crtind{k}}}(z_{2})} =  \iota_{\widetilde{\crtind{k}}+1, \bar{\param}}(z_{1})$ holds and $\iota_{{v}_{0,\widetilde{\crtind{k}}}(z_{2})} \neq \iota_{\widetilde{\crtind{k}}+1, \bar{\param}}(z_{1})$ holds (recall that we have $\iota_{\widetilde{\crtind{k}}, \bar{\param}}(z_1) = \iota_{\widetilde{\crtind{k}}+1, \bar{\param}}(z_{1})$).\\

\textbf{Point \hypertarget{bmk:p11.9}{11.9}}
of Figure \ref{fig:proofStructureCaseii}.
Let us first consider case $\iota_{{v}_{0,\widetilde{\crtind{k}}}(z_{2})} = \iota_{\widetilde{\crtind{k}}, \bar{\param}}(z_1) = \iota_{\widetilde{\crtind{k}}+1, \bar{\param}}(z_{1})$ (see Figure \ref{fig:inCompleteAgreementInSjCaseiiCase2a}).
\begin{figure}[tb]
\centering
\includegraphics[scale=0.55]{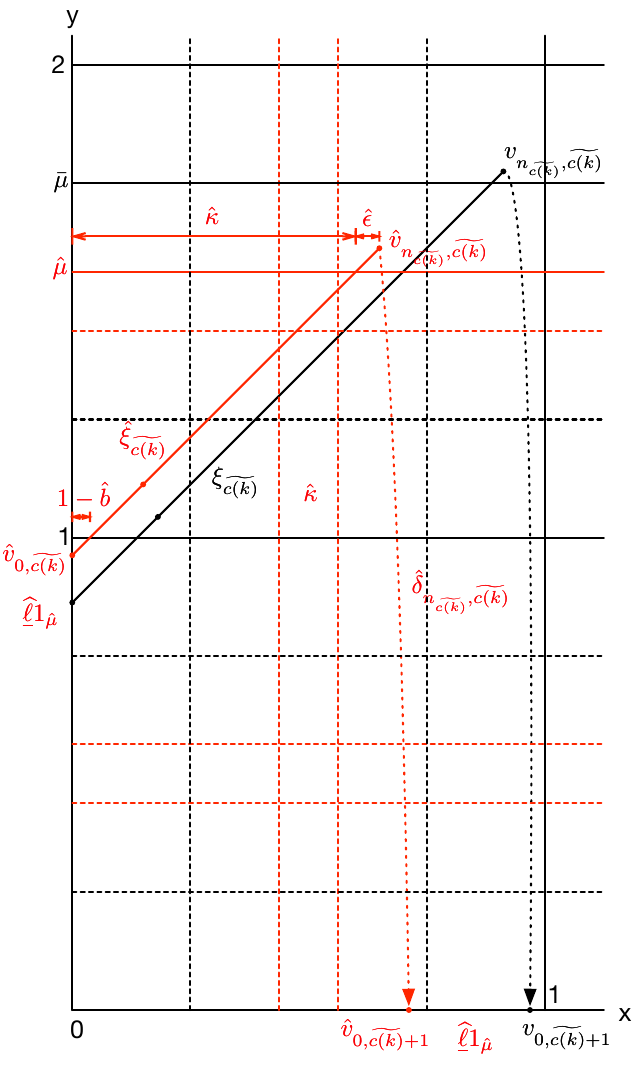}
\caption{Depiction of the case (with positive polarity) in which ${v}_{0,\widetilde{\crtind{k}}+1}(z_{1})$ and ${v}_{n_{\widetilde{\crtind{k}}},\widetilde{\crtind{k}}}(z_{1})$ are in complete agreement (with $z_{1} = x$), $\iota_{\widetilde{\crtind{k}}, \bar{\param}}(z_1)$ and $\iota_{b}$ belong to $\SetZZ$, and $\iota_{{v}_{0,\widetilde{\crtind{k}}}(z_{2})} = \iota_{\widetilde{\crtind{k}}, \bar{\param}}(z_{1}) = \iota_{\widetilde{\crtind{k}}+1, \bar{\param}}(z_{1})$ holds; in addition, in this case $\kappa = 1-b + \fract{\bar{\param}}$ holds.}
\label{fig:inCompleteAgreementInSjCaseiiCase2a}
\end{figure}
If $\iota_{{v}_{0,\widetilde{\crtind{k}}}(z_{2})} = \ZC_{\bar{\param}}$ holds (i.e., the polarity is negative), then, by induction hypothesis \ref{I4},
\todoMR{}{controllare ref}
$1 - 2\fract{\hat{\param}} < \hat{b} < 1-\fract{\hat{\param}}$ holds, hence also $\fract{\hat{\param}} < 1 - \hat{b} < 2\fract{\hat{\param}}$ (i.e., $\fract{\hat{\param}} < 1 - \hat{b} < \fract{\hat{\param}} + \chi_{\hat{\param}}$, since in this case it holds that $\fract{\hat{\param}} = \chi_{\hat{\param}}$).
If, instead, $\iota_{{v}_{0,\widetilde{\crtind{k}}}(z_{2})} = \ZA_{\bar{\param}}$ (resp., $\iota_{{v}_{0,\widetilde{\crtind{k}}}(z_{2})} = \ZE_{\bar{\param}}$) holds (i.e., the polarity is positive, hence it holds that $\fract{\hat{\param}} - \frac{1}{2} = \chi_{\hat{\param}}$) then, by induction hypothesis \ref{I4} again, $1 - \fract{\hat{\param}} - \chi_{\hat{\param}} < \hat{b} < 1 - \fract{\hat{\param}}$ (resp., $1 - \chi_{\hat{\param}} < \hat{b} < 1$) holds, hence also $\fract{\hat{\param}} < 1 - \hat{b} < \fract{\hat{\param}} + \chi_{\hat{\param}}$ (resp., $0 < 1 - \hat{b} < \chi_{\hat{\param}}$).
In all cases, if $\kappa = 1-b$ holds (notice that, in this case, it cannot be that $\iota_{{v}_{0,\widetilde{\crtind{k}}}(z_{2})} = \ZA_{\bar{\param}}$, or $\kappa > \fract{\bar{\param}}$ and $\kappa +\epsilon > \fract{\bar{\param}}$ hold, so it could not hold that $\iota_{{v}_{0,\widetilde{\crtind{k}}}(z_{2})} = \iota_{\widetilde{\crtind{k}}, \bar{\param}}(z_{1})$), then it is easy to define $\hat{\epsilon}$ such that $\hat{\kappa} + \hat{\epsilon}$ satisfies the induction hypothesis (e.g., $\hat{\kappa} + \hat{\epsilon} < 1-((2\cmax - \lfloor \hat{v}_{0,\crtind{k+1}}(z_2) \rfloor)\cdot|Q| +1)\hat{\param}$
if the polarity is negative and \onerst{} sequence $\orssym_{\widetilde{\crtind{k}}+2}$ exists and is such that $v_{0, \widetilde{\crtind{k}}+2}(z_2) = 0$ and $v_{0, \widetilde{\crtind{k}}+2}(z_1) > 0$ hold).
If $\kappa = 1-b + \fract{\bar{\param}}$ holds, then it is either $\iota_{{v}_{0,\widetilde{\crtind{k}}}(z_{2})} = \ZC_{\bar{\param}}$ (and the polarity is negative), or $\iota_{{v}_{0,\widetilde{\crtind{k}}}(z_{2})} = \ZE_{\bar{\param}}$ (and the polarity is positive), since in case $\iota_{{v}_{0,\widetilde{\crtind{k}}}(z_{2})} = \ZA_{\bar{\param}}$ $\kappa$ would be greater than 1.
If $\iota_{{v}_{0,\widetilde{\crtind{k}}}(z_{2})} = \ZC_{\bar{\param}}$ holds, then also $\hat{\kappa} < 3\fract{\hat{\param}} = \fract{\hat{\param}} + 2\chi_{\hat{\param}}$ holds;
hence, from the considerations of Point
\hyperlink{bmk:enoughintervals}{2} above regarding the number of intervals of length $\chi_{\hat{\param}}$ that there are in $\ZC_{\hat{\param}}$, it is easy to define
\todoMR{}{messo rimando}
$\hat{\epsilon}$ so that $\hat{\kappa} + \hat{\epsilon} = 1- \hat{b} + \fract{\hat{\param}} + \hat{\epsilon} < 1-((2\cmax - \lfloor \hat{v}_{0,\crtind{k+1}}(z_2) \rfloor)\cdot|Q| +1)\fract{\hat{\param}}$ holds (which entails that we can also define $\hat{\epsilon}$ so that $1 - 2\fract{\hat{\param}} = 1 - \fract{\hat{\param}} - \chi_{\hat{\param}}< \hat{\kappa} + \hat{\epsilon} < 1- \fract{\hat{\param}}$).
If, instead, $\iota_{{v}_{0,\widetilde{\crtind{k}}}(z_{2})} = \ZE_{\bar{\param}}$ holds (Figure \ref{fig:inCompleteAgreementInSjCaseiiCase2a} provides a graphical depiction of this case, where $\cmax = 1$ holds), then $\fract{\hat{\param}} < \hat{\kappa} < \fract{\hat{\param}} + \chi_{\hat{\param}}$ holds so, also for the considerations of Point
\hyperlink{bmk:enoughintervals}{2}, we have that
$\hat{\kappa} < 1-4\cmax\fract{\hat{\param}} \leq 1-2(2\cmax - \lfloor \hat{v}_{0,\crtind{k+1}}(z_2) \rfloor)\fract{\hat{\param}}$ holds, which means that we can define $\hat{\epsilon}$ so that $\hat{\kappa} + \hat{\epsilon}$ satisfies hypothesis \ref{I3b}, which in turn entails that we can also define $\hat{\epsilon}$ so that $1 - \chi_{\hat{\param}}< \hat{\kappa} + \hat{\epsilon} < 1$).
Finally, case $\kappa = 1-b - (1- \fract{\bar{\param}})$ cannot occur if $\iota_{{v}_{0,\widetilde{\crtind{k}}}(z_{2})} = \iota_{\widetilde{\crtind{k}}, \bar{\param}}(z_{1})$ holds and $\iota_{\widetilde{\crtind{k}}, \bar{\param}}(z_{1})$ is either $\ZC_{\bar{\param}}$ or $\ZE_{\bar{\param}}$, as $1 -b - (1 - \fract{\bar{\param}})$ would be negative in these cases.
If $\iota_{\widetilde{\crtind{k}}, \bar{\param}}(z_{1})$ is $\ZA_{\bar{\param}}$, instead, then the polarity is positive (since we are in case \ref{case:2}), $\kappa$ is in $\ZA_{\bar{\param}}$, and, since from induction hypothesis \ref{I4} it follows that $\fract{\hat{\param}} < 1 - \hat{b} < \fract{\hat{\param}} + \chi_{\hat{\param}}$ holds (see above), then $2(\fract{\hat{\param}} - \frac{1}{2}) = 2\chi_{\hat{\param}} < 1 -\hat{b} - (1 - \fract{\hat{\param}}) = \hat{\kappa} < 3(\fract{\hat{\param}} - \frac{1}{2}) = 3\chi_{\hat{\param}}$ holds.
Depending on what clock is reset at the beginning of \onerst{} sequence $\orssym_{\widetilde{\crtind{k}}+2}$, we need to define $\hat{v}_{0,\widetilde{\crtind{k}}+1}(z_1) = \hat{v}_{n_{\widetilde{\crtind{k}}},\widetilde{\crtind{k}}}(z_1) + \delta_{n_{\widetilde{\crtind{k}}},\widetilde{\crtind{k}}}$ so that either $\LowK{\hat{v}_{0,\widetilde{\crtind{k}}+1}(z_1), \chi_{\hat{\param}}} = 2(2\cmax - \lfloor \hat{v}_{0,\widetilde{\crtind{k}}+1}(z_1) \rfloor)$ holds \ref{I3b}, or $\LowK{\hat{v}_{0,\widetilde{\crtind{k}}+1}(z_1), \chi_{\hat{\param}}} = 1$ holds \ref{I4}.
By the considerations of Point
\hyperlink{bmk:enoughintervals}{2}, we have that $1 - \fract{\hat{\param}} = \frac{1}{2} - \chi_{\hat{\param}} > 8\cmax\chi_{\hat{\param}}$ holds, so also $3\chi_{\hat{\param}} < 1 - \fract{\hat{\param}} - 4\cmax$ (as we have $\cmax \geq 1$);
hence, no matter what clock is reset at the beginning of \onerst{} sequence $\orssym_{\widetilde{\crtind{k}}+2}$, we can define $\hat{\epsilon}$ and $\delta_{n_{\widetilde{\crtind{k}}},\widetilde{\crtind{k}}}$ so that the induction hypothesis holds.
\todoMR{}{messo riferimento a calcoli precisi, tolti da qui}

\textbf{Point 11.10} of Figure \ref{fig:proofStructureCaseii}.
If, instead, $\iota_{{v}_{0,\widetilde{\crtind{k}}}(z_{2})} \neq \iota_{\widetilde{\crtind{k}}, \bar{\param}}(z_{1})$ holds (see Figure \ref{fig:inCompleteAgreementInSjCaseiiCase2b}), then the polarity must be positive (because $|\SetZZ| > 1$ holds) and one of the two intervals is $\ZA_{\bar{\param}}$, while the other is $\ZE_{\bar{\param}}$ (recall that in this case $\frac{1}{2} - \chi_{\hat{\param}} = 1 - \fract{\hat{\param}}$ holds).
\begin{figure}[!tp]
\centering
\includegraphics[scale=0.55]{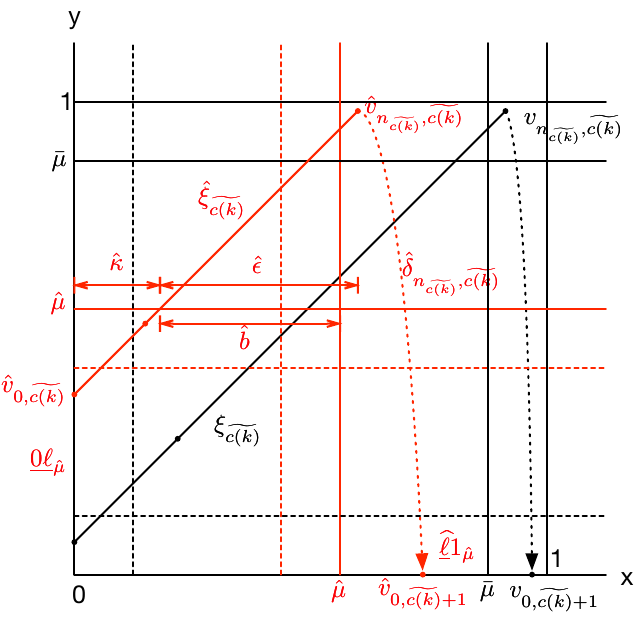}
\caption{Depiction of the case (with positive polarity) in which ${v}_{0,\widetilde{\crtind{k}}+1}(z_{1})$ and ${v}_{n_{\widetilde{\crtind{k}}},\widetilde{\crtind{k}}}(z_{1})$ are in complete agreement (with $z_{1} = x$), $\iota_{\widetilde{\crtind{k}}, \bar{\param}}(z_1)$ and $\iota_{b}$ belong to $\SetZZ$, and
$\iota_{{v}_{0,\widetilde{\crtind{k}}}(z_{2})} \neq \iota_{\widetilde{\crtind{k}}, \bar{\param}}(z_{1})$ holds; in addition, in this case $\kappa = 1-b + (1 -\fract{\bar{\param}}) = \fract{\bar{\param}} - b$ holds.}
\label{fig:inCompleteAgreementInSjCaseiiCase2b}
\end{figure}
If $\iota_{{v}_{0,\widetilde{\crtind{k}}}(z_{2})}$ is $\ZA_{\bar{\param}}$ and $\iota_{\widetilde{\crtind{k}}, \bar{\param}}(z_{1})$ is $\ZE_{\bar{\param}}$, then the only possible two cases are $\kappa = 1- b$ and $1 -b - (1 - \fract{\bar{\param}})$ ($\kappa = 1-b + \fract{\bar{\param}}$ would be greater than 1).
If $\kappa = 1-b$ holds, then since (by induction hypothesis \ref{I4}), $1- \fract{\hat{\param}} - \chi_{\hat{\param}} < \hat{b} < 1- \fract{\hat{\param}}$ holds, then it also holds that $\fract{\hat{\param}} < 1-\hat{b} < \fract{\hat{\param}} + \chi_{\hat{\param}}$, hence (as in Point \hyperlink{bmk:p11.9}{11.9} above) $\hat{\epsilon}$ can be chosen so that $\hat{\kappa} + \hat{\epsilon} < 1- 4\cmax\chi_{\hat{\param}}$ holds
(which entails that we can also define $\hat{\epsilon}$ so that $1 - \chi_{\hat{\param}} < \hat{\kappa} + \hat{\epsilon} < 1$ holds)
and then the induction hypothesis is satisfied.
If $\kappa = 1 -b - (1 - \fract{\bar{\param}})$ holds (hence also $\kappa < \fract{\bar{\param}}$), instead, for any $\hat{\varepsilon}$ in $\ZE_{\hat{\param}}$ that satisfies the induction hypothesis, one can always define
$\hat{\epsilon}$ and $\hat{\delta}_{n_{\widetilde{\crtind{k}}},\widetilde{\crtind{k}}}$ so that $\hat{\kappa} + \hat{\epsilon} + \hat{\delta}_{n_{\widetilde{\crtind{k}}}} = \hat{\varepsilon}$ holds (Figure \ref{fig:inCompleteAgreementInSjCaseiiCase2b} provides a graphical depiction of this case).
If $\iota_{{v}_{0,\widetilde{\crtind{k}}}(z_{2})}$ is $\ZE_{\bar{\param}}$ and $\iota_{\widetilde{\crtind{k}}, \bar{\param}}(z_{1})$ is $\ZA_{\bar{\param}}$, then the only possible case is $\kappa = 1-b$ ($1-b - (1 - \fract{\bar{\param}})$ is negative, and $1-b + \fract{\bar{\param}}$ is greater than $\fract{\bar{\param}}$, hence $\kappa + \epsilon$ could not be in $\ZA_{\bar{\param}}$).
Then, since (by induction hypothesis \ref{I4}) $1- \chi_{\hat{\param}} < \hat{b} < 1$ holds, it also holds that $0 < 1-\hat{b} < \chi_{\hat{\param}}$, hence $\hat{\epsilon}$ can be chosen so that $\hat{\kappa} + \hat{\epsilon} < 1-\fract{\hat{\param}}- 4\cmax\chi_{\hat{\param}}$ holds
(which entails that we can also define $\hat{\epsilon}$ so that $1 -\fract{\hat{\param}} - \chi_{\hat{\param}} < \hat{\kappa} + \hat{\epsilon} < 1 - \fract{\hat{\param}}$ holds)
and then the induction hypothesis is satisfied.

\end{proof}

The following result is a direct consequence of Lemma \ref{lm:fromTAtonrtTA2} and of Theorem \ref{thm:nrtTAdec}.

\begin{corollary}
The problem of deciding whether a parametric TA with only one clock and one parameter accepts an empty set of timed $\omega$-words is decidable.
\todoMR{}{Cambiato.}
\end{corollary}


\bibliographystyle{elsarticle-num}
\bibliography{bibliografia}

\end{document}